[ english version / português abaixo ]

# **Climate change denial and anti-science communities on brazilian Telegram:** climate disinformation as a gateway to broader conspiracy networks

*Ergon Cugler de Moraes Silva*

Brazilian Institute of Information in
Science and Technology (IBICT)
Brasília, Federal District, Brazil

contato@ergoncugler.com
www.ergoncugler.com

## Abstract

Conspiracy theories related to climate change denial and anti-science have found fertile ground on Telegram, particularly among Brazilian communities that distrust scientific institutions and oppose global environmental policies. This study seeks to answer the research question: **how are Brazilian conspiracy theory communities on climate change and anti-science themes characterized and articulated on Telegram?** It is worth noting that this study is part of a series of seven studies aimed at understanding and characterizing Brazilian conspiracy theory communities on Telegram. This series of studies is openly and originally available on arXiv from Cornell University, applying a mirrored method across all seven studies, changing only the thematic focus of analysis, and providing replicable investigation methods, including custom-developed and proprietary codes, contributing to the culture of open-source software. Regarding the main findings of this study, the following observations were made: Climate change denial and anti-science communities interact synergistically, creating a complex network that mutually reinforces disinformation narratives; Apocalyptic themes, such as Apocalypse and Survivalism, act as gateways to climate denial, with 5,057 links directed to these communities; Anti-science communities function as gatekeepers, distributing links evenly to theories such as the New World Order and Globalism, among others; During the COVID-19 pandemic, anti-science discussions experienced a significant peak, driven by vaccine disinformation; The intersection between anti-science narratives and esoteric beliefs reinforces the idea of a supposed alternative truth that challenges science; Since 2022, discussions on climate change have evolved to align with global domination theories; Additionally, the UN's 2030 Agenda is portrayed as part of a global conspiracy.

### Key findings

- ➔ Climate change denial and anti-science form a critical intersection in the spread of disinformation, creating a complex network that amplifies and mutually reinforces false narratives, making it more difficult to combat disinformation across broad sectors;
- ➔ Climate change is targeted by apocalyptic narratives, with Apocalypse and Survivalism communities directing 5,057 links to climate denial groups, exacerbating the rejection of science and increasing polarization;
- ➔ Anti-science communities act as distributors of disinformation for various conspiracy theories, functioning as gatekeepers and evenly distributing links to other themes like the New World Order and Globalism;

- ➔ There was significant growth in anti-science discussions during the COVID-19 pandemic, with a notable peak in 2021 driven by disinformation about vaccines and public health measures, reflecting the exploitation of global crises to expand conspiratorial beliefs;
- ➔ Climate change denial communities attract members from other conspiracy theories, with Globalism and New World Order sending 3,013 and 2,903 links, respectively, demonstrating the convergence of these narratives;
- ➔ Disinformation about climate change is strongly linked to geopolitical disputes, with terms like "Ukraine" and "Russia" frequently associated with climate debates, suggesting the geopolitical manipulation of environmental crises;
- ➔ Anti-science narratives integrate esoteric and conspiratorial beliefs, with Occultism and Esotericism communities frequently connected to anti-science discussions, reinforcing the idea of an "alternative truth" that challenges science;
- ➔ Discussions on climate change are evolving towards themes related to global domination, with increasing references to global control since 2022, linking to New World Order theories and other global conspiracies;
- ➔ Anti-science communities promote disinformation about public health based on conspiracy theories, using vaccine disinformation as a central pillar to challenge health policies and foster distrust in science;
- ➔ The UN's 2030 Agenda is portrayed as a conspiracy for population control, with denialist communities associating UN initiatives with global domination theories, undermining international sustainability efforts.

## 1. Introduction

After examining thousands of Brazilian conspiracy theory communities on Telegram and extracting tens of millions of pieces of content created and/or shared by millions of users within these communities, this study aims to contribute to a series of seven studies that address the phenomenon of conspiracy theories on Telegram, using Brazil as a case study. Through the identification approaches implemented, it was possible to identify a total of 36 Brazilian conspiracy theory communities on Telegram focused on climate change and anti-science themes. These communities collectively generated 1,287,938 pieces of content from May 2018 (the first publications) until August 2024 (the period of this study), with a total of 78,252 users across these communities. Thus, this study aims to understand and characterize the communities focused on climate change and anti-science themes within this identified Brazilian conspiracy network on Telegram.

To this end, a mirrored method will be applied across all seven studies, changing only the thematic object of analysis and providing investigation replicability. In this way, we will adopt technical approaches to observe the connections, temporal series, content, and overlaps of themes within the conspiracy theory communities. In addition to this study, the other six are openly and originally available on arXiv at Cornell University. This series paid particular attention to ensuring data integrity and respecting user privacy, as provided by Brazilian legislation (Law No. 13,709/2018 / Brazilian law from 2018).

Therefore, the research question is: **how are Brazilian conspiracy theory communities focused on climate change and anti-science themes characterized and structured on Telegram?**

## 2. Materials and methods

The methodology of this study is organized into three subsections: **2.1. Data extraction**, which describes the process and tools used to collect information from Telegram communities; **2.2. Data processing**, which discusses the criteria and methods applied to classify and anonymize the collected data; and **2.3. Approaches to data analysis**, which details the techniques used to investigate the connections, temporal series, content, and thematic overlaps within conspiracy theory communities.

### 2.1. Data extraction

This project began in February 2023 with the publication of the first version of TelegramScrap (Silva, 2023), a proprietary, free, and open-source tool that utilizes Telegram's Application Programming Interface (API) by Telethon library and organizes data extraction cycles from groups and open channels on Telegram. Over the months, the database was expanded and refined using four approaches to identifying conspiracy theory communities:

**(i) Use of keywords:** at the project's outset, keywords were listed for direct identification in the search engine of Brazilian groups and channels on Telegram, such as "apocalypse", "survivalism", "climate change", "flat earth", "conspiracy theory", "globalism", "new world order", "occultism", "esotericism", "alternative cures", "qAnon" "reptilians", "revisionism", "aliens", among others. This initial approach provided some communities whose titles and/or descriptions of groups and channels explicitly contained terms related to conspiracy theories. However, over time, it was possible to identify many other communities that the listed keywords did not encompass, some of which deliberately used altered characters to make it difficult for those searching for them on the network.

**(ii) Telegram channel recommendation mechanism:** over time, it was identified that Telegram channels (except groups) have a recommendation tab called "similar channels", where Telegram itself suggests ten channels that have some similarity with the channel being observed. Through this recommendation mechanism, it was possible to find more Brazilian conspiracy theory communities, with these being recommended by the platform itself.

**(iii) Snowball approach for invitation identification:** after some initial communities were accumulated for extraction, a proprietary algorithm was developed to identify URLs containing "t.me/", the prefix for any invitation to Telegram groups and channels. Accumulating a frequency of hundreds of thousands of links that met this criterion, the unique addresses were listed, and their repetitions counted. In this way, it was possible to investigate new Brazilian groups and channels mentioned in the messages of those already investigated, expanding the network. This process was repeated periodically to identify new communities aligned with conspiracy theory themes on Telegram.

**(iv) Expansion to tweets published on X mentioning invitations:** to further diversify the sources of Brazilian conspiracy theory communities on Telegram, a proprietary search query was developed to identify conspiracy theory-themed keywords using tweets published on X (formerly Twitter) that, in addition to containing one of the keywords, also included "t.me/", the prefix for any invitation to Telegram groups and channels, "https://x.com/search?q=lang%3Apt%20%22t.me%2F%22%20SEARCH-TERM".

With the implementation of community identification approaches for conspiracy theories developed over months of investigation and method refinement, it was possible to build a project database encompassing a total of 855 Brazilian conspiracy theory communities on Telegram (including other themes not covered in this study). These communities have collectively published 27,227,525 pieces of content from May 2016 (the first publications) to August 2024 (the period of this study), with a combined total of 2,290,621 users across the Brazilian communities. It is important to note that this volume of users includes two elements: first, it is a variable figure, as users can join and leave communities daily, so this value represents what was recorded on the publication extraction date; second, it is possible that the same user is a member of more than one group and, therefore, is counted more than once. In this context, while the volume remains significant, it may be slightly lower when considering the deduplicated number of citizens within these Brazilian conspiracy theory communities.

### 2.2. Data processing

With all the Brazilian conspiracy theory groups and channels on Telegram extracted, a manual classification was conducted considering the title and description of the community. If there was an explicit mention in the title or description of the community related to a specific theme, it was classified into one of the following categories: (i) "Anti-Science"; (ii) "Anti-Woke and Gender"; (iii) "Antivax"; (iv) "Apocalypse and Survivalism"; (v) "Climate Changes"; (vi) "Flat Earth"; (vii) "Globalism"; (viii) "New World Order"; (ix) "Occultism and Esotericism"; (x) "Off Label and Quackery"; (xi) "QAnon"; (xii) "Reptilians and Creatures"; (xiii) "Revisionism and Hate Speech"; (xiv) "UFO and Universe". If there was no explicit mention related to the themes in the title or description of the community, it was classified as (xv) "General Conspiracy". In the following table, we can observe the metrics related to the classification of these conspiracy theory communities in Brazil.

**Table 01.** Conspiracy theory communities in Brazil (metrics up to August 2024)

| Categories | Groups | Users | Contents | Comments | Total |
|---|---|---|---|---|---|
| **Anti-Science** | 22 | 58,138 | 187,585 | 784,331 | 971,916 |
| **Anti-*Woke* and Gender** | 43 | 154,391 | 276,018 | 1,017,412 | 1,293,430 |
| **Antivax** | 111 | 239,309 | 1,778,587 | 1,965,381 | 3,743,968 |
| **Apocalypse and Survivalism** | 33 | 109,266 | 915,584 | 429,476 | 1,345,060 |
| **Climate Changes** | 14 | 20,114 | 269,203 | 46,819 | 316,022 |
| **Flat Earth** | 33 | 38,563 | 354,200 | 1,025,039 | 1,379,239 |
| **General Conspiracy** | 127 | 498,190 | 2,671,440 | 3,498,492 | 6,169,932 |
| **Globalism** | 41 | 326,596 | 768,176 | 537,087 | 1,305,263 |
| **NWO** | 148 | 329,304 | 2,411,003 | 1,077,683 | 3,488,686 |
| **Occultism and Esotericism** | 39 | 82,872 | 927,708 | 2,098,357 | 3,026,065 |
| **Off Label and Quackery** | 84 | 201,342 | 929,156 | 733,638 | 1,662,794 |
| **QAnon** | 28 | 62,346 | 531,678 | 219,742 | 751,420 |
| **Reptilians and Creatures** | 19 | 82,290 | 96,262 | 62,342 | 158,604 |
| **Revisionism and Hate Speech** | 66 | 34,380 | 204,453 | 142,266 | 346,719 |
| **UFO and Universe** | 47 | 58,912 | 862,358 | 406,049 | 1,268,407 |
| **Total** | **855** | **2,296,013** | **13,183,411** | **14,044,114** | **27,227,525** |

Source: Own elaboration (2024).

With this volume of extracted data, it was possible to segment and present in this study only the communities and content related to climate change denial and anti-science themes. In parallel, other themes within Brazilian conspiracy theory communities were also analyzed in separate studies aimed at characterizing the scope and dynamics of these networks, which are openly and originally available on arXiv by Cornell University.

Additionally, it should be noted that only open communities were extracted, meaning those that are not only publicly identifiable but also do not require any request to access the content, being available to any user with a Telegram account who needs to join the group or channel. Furthermore, in compliance with Brazilian legislation, particularly the General Data Protection Law (LGPD), or Law No. 13,709/2018 (Brazilian law from 2018), which deals with privacy control and the use/treatment of personal data, all extracted data were anonymized for the purposes of analysis and investigation. Therefore, not even the identification of the communities is possible through this study, thus extending the user's privacy by anonymizing their data beyond the community itself to which they submitted by being in a public and open group or channel on Telegram.

### 2.3. Approaches to data analysis

Totaling 36 selected communities within the themes of climate change denial and anti-science, comprising 1,287,938 publications and 78,252 users, four approaches will be employed to investigate the conspiracy theory communities selected for this study. These metrics are detailed in the following table:

**Table 02.** Selected communities for analysis (metrics up to August 2024)

| Categories | Groups | Users | Publications | Comments | Total |
|---|---|---|---|---|---|
| **Anticiência** | 22 | 58.138 | 187.585 | 784.331 | 971.916 |
| **Mudanças Climáticas** | 14 | 20.114 | 269.203 | 46.819 | 316.022 |
| **Total** | **36** | **78.252** | **456.788** | **831.150** | **1.287.938** |

Source: Own elaboration (2024).

**(i) Network:** by developing a proprietary algorithm to identify messages containing the term "t.me/" (inviting users to join other communities), we propose to present volumes and connections observed on how **(a)** communities focused on climate change denial and anti-science circulate invitations to encourage users to join more groups and channels within the same theme, with shared belief systems; and how **(b)** these same communities circulate invitations for their users to explore groups and channels dealing with other conspiracy theories, distinct from their explicit purpose. This approach is valuable for observing whether these communities use themselves as a source of legitimation and reference and/or rely on other conspiracy theory themes, even opening doors for their users to explore other conspiracies. Furthermore, it is worth mentioning the study by Rocha *et al.* (2024), where a network identification approach was also applied in Telegram communities, but by observing similar content based on an ID generated for each unique message and its similar ones;

**(ii) Time series:** the "Pandas" library (McKinney, 2010) is used to organize the investigation data frames, observing **(a)** the volume of publications over the months; and **(b)** the volume of engagement over the months, considering metadata of views, reactions, and comments collected during extraction. In addition to volumetry, the "Plotly" library (Plotly Technologies Inc., 2015) enabled the graphical representation of this variation;

**(iii) Content analysis:** in addition to the general word frequency analysis, time series are applied to the variation of the most frequent words over the semesters—observing from May 2018 (initial publications) to August 2024 (when this study was conducted). With the "Pandas" (McKinney, 2010) and "WordCloud" (Mueller, 2020) libraries, the results are presented both volumetrically and graphically;

**(iv) Thematic agenda overlap:** following the approach proposed by Silva & Sátiro (2024) for identifying thematic agenda overlap in Telegram communities, we used the "BERTopic" model (Grootendorst, 2020). BERTopic is a topic modeling algorithm that facilitates the generation of thematic representations from large amounts of text. First, the algorithm extracts document embeddings using sentence transformer models, such as "all-MiniLM-L6-v2". These embeddings are then reduced in dimensionality using techniques like "UMAP", facilitating the clustering process. Clustering is performed using "HDBSCAN", a density-based technique that identifies clusters of different shapes and sizes, as well as outliers. Subsequently, the documents are tokenized and represented in a bag-of-words structure, which is normalized (L1) to account for size differences between clusters. The topic representation is refined using a modified version of "TF-IDF", called "Class-TF-IDF", which considers the importance of words within each cluster (Grootendorst, 2020). It is important to note that before applying BERTopic, we cleaned the dataset by removing Portuguese "stopwords" using "NLTK" (Loper & Bird, 2002). For topic modeling, we used the "loky" backend to optimize performance during data fitting and transformation.

In summary, the methodology applied ranged from data extraction using the own tool TelegramScrap (Silva, 2023) to the processing and analysis of the collected data, employing various approaches to identify and classify Brazilian conspiracy theory communities on Telegram. Each stage was carefully implemented to ensure data integrity and respect for user privacy, as mandated by Brazilian legislation. The results of this data will be presented below, aiming to reveal the dynamics and characteristics of the studied communities.

## 3. Results

The results are detailed below in the order outlined in the methodology, beginning with the characterization of the network and its sources of legitimation and reference, progressing to the time series, incorporating content analysis, and concluding with the identification of thematic agenda overlap among the conspiracy theory communities.

### 3.1. Network

The following figures provide a detailed view of the dynamics connecting communities around the themes of anti-science and climate change within the conspiracy ecosystem. These analyses reveal how anti-science narratives and climate change discussions not only coexist but also reinforce each other through complex networks of interactions and cross-invitations. It is observed that these networks function both to attract new followers and to radicalize existing members, promoting a continuous cycle of disinformation spreading.

[ english version / português abaixo ]

**Figure 01.** Internal network between climate changes and anti-science communities

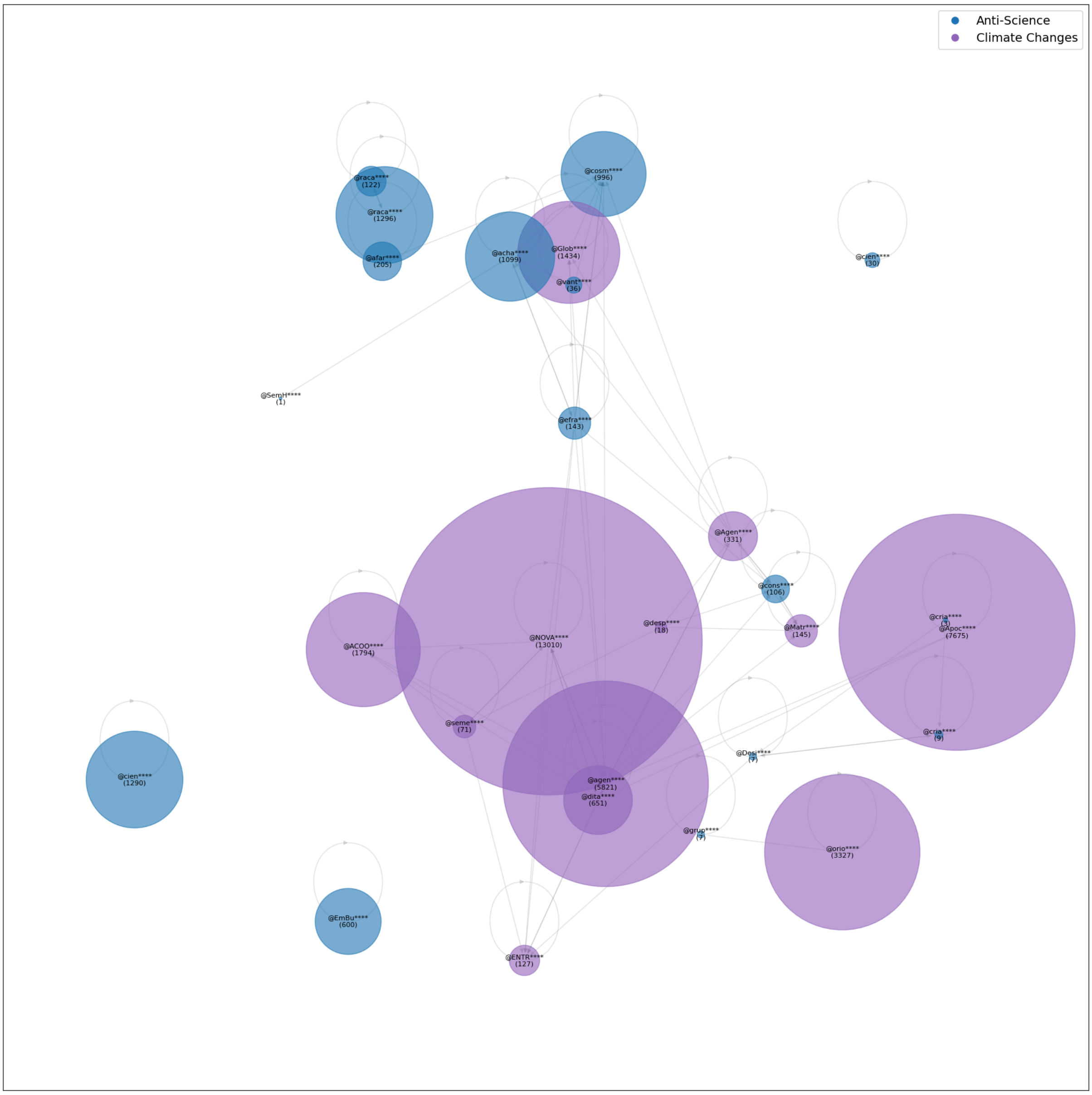

Source: Own elaboration (2024).

This figure explores the internal network connecting communities focused on anti-science and climate change, highlighting the interconnection of these themes. The connections reveal a strong correlation between anti-science rhetoric and climate change denial, suggesting that these narratives often reinforce one another. The size and centrality of the nodes indicate which communities play key roles in spreading these ideas, evidencing a feedback pattern that strengthens disinformation among participants. Notably, climate change denial not only occupies the center of this interconnection with anti-scientific content but also links to other anti-science communities, leaving only a few without direct connection to the conspiracy theory network.

**Figure 02.** Network of communities that open doors to the theme (gateway)

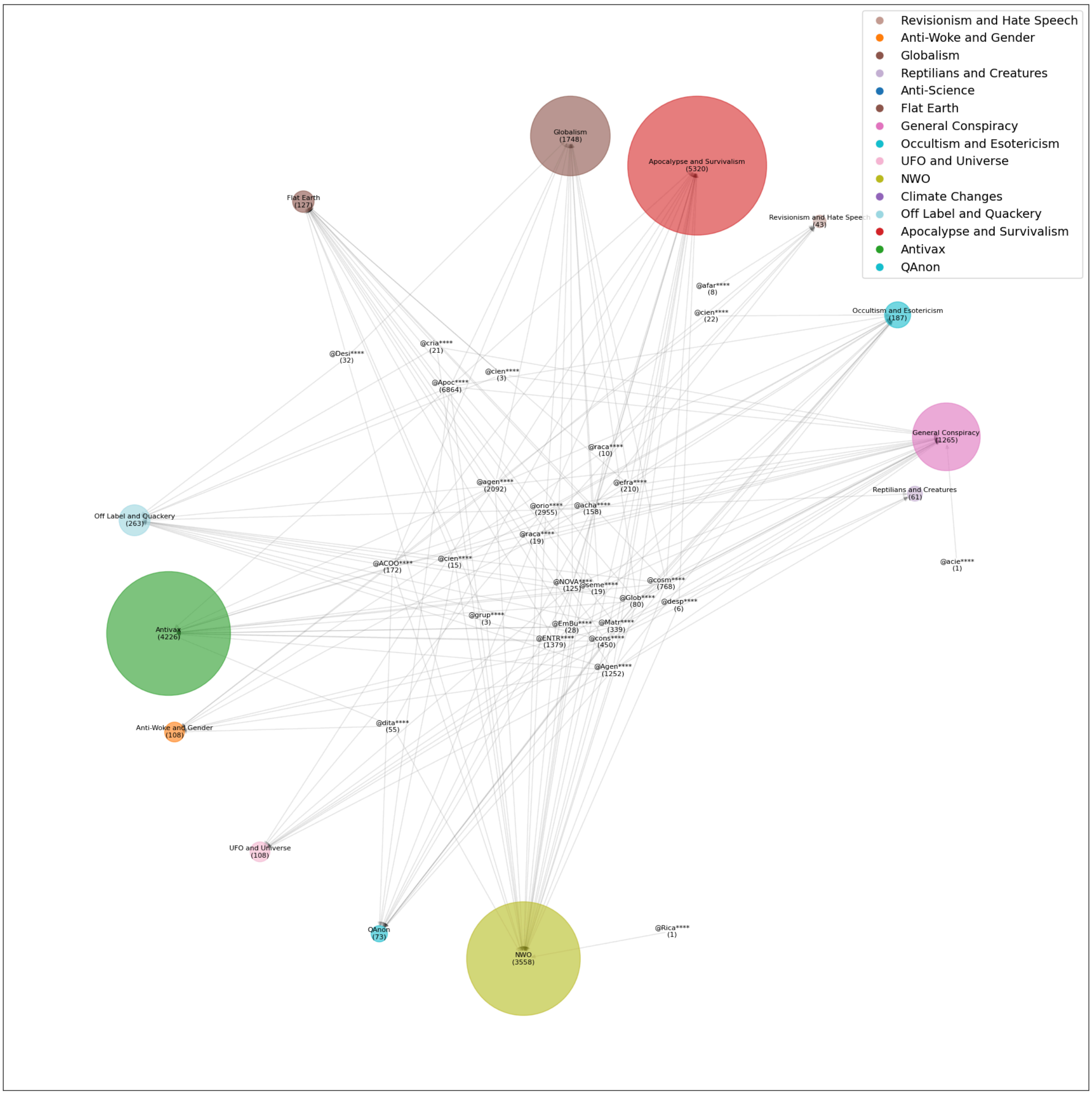

Source: Own elaboration (2024).

In this figure, we analyze the communities that serve as entry points for discussions on anti-science and climate change. The arrangement of nodes and their connections demonstrates that larger and more influential groups, such as those discussing global and apocalyptic conspiracies, act as initial gateways for new members. These central communities play a crucial role in introducing and spreading disinformation, linking various themes and presenting them as part of a cohesive, distorted worldview. They rely on alarmism and even intertwine biblical verses in attempts to discredit the necessary and important fight against climate change.

[ english version / português abaixo ]

**Figure 03.** Network of communities whose theme opens doors (exit point)

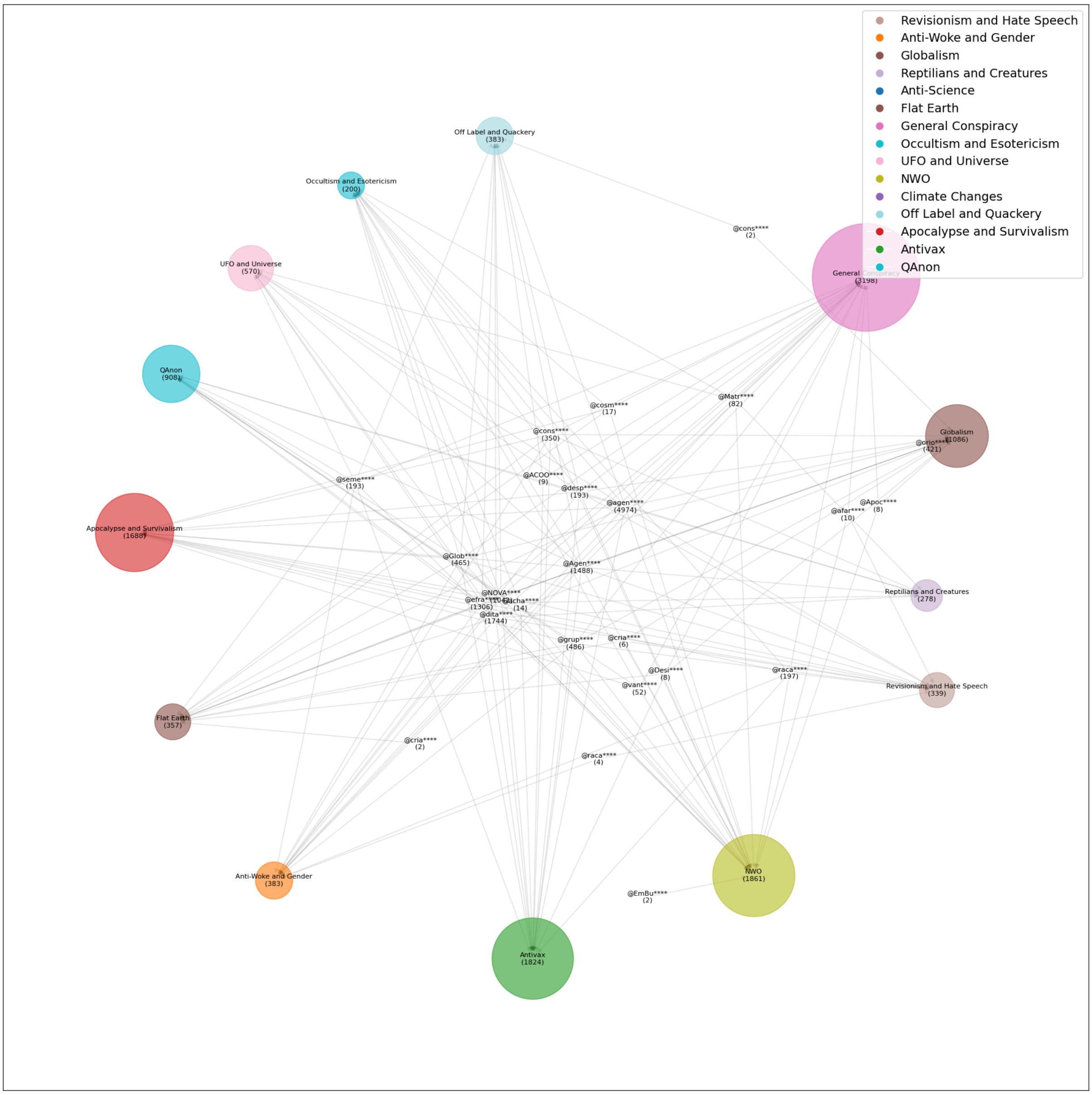

Source: Own elaboration (2024).

Here, the figure illustrates how certain communities, once explored, encourage members to transition to other topics within the conspiracy ecosystem. Groups centralizing discussions on climate change or anti-science often guide members toward related themes, such as the New World Order or General Conspiracy. This outbound movement not only diversifies members' beliefs but also deepens their involvement in broader conspiracy theories, intensifying the cycle of radicalization. An interesting insight from this observation is how climate change denial and anti-science communities seem to function as common distributors for other themes. Unlike other observations where one or two topics receive more links, these communities exhibit a balanced and hegemonic distribution, indicating that climate change denial and anti-science communities may act as gatekeepers for other conspiracy theories.

[ english version / português abaixo ]

**Figure 04.** Flow of invitation links between climate changes communities

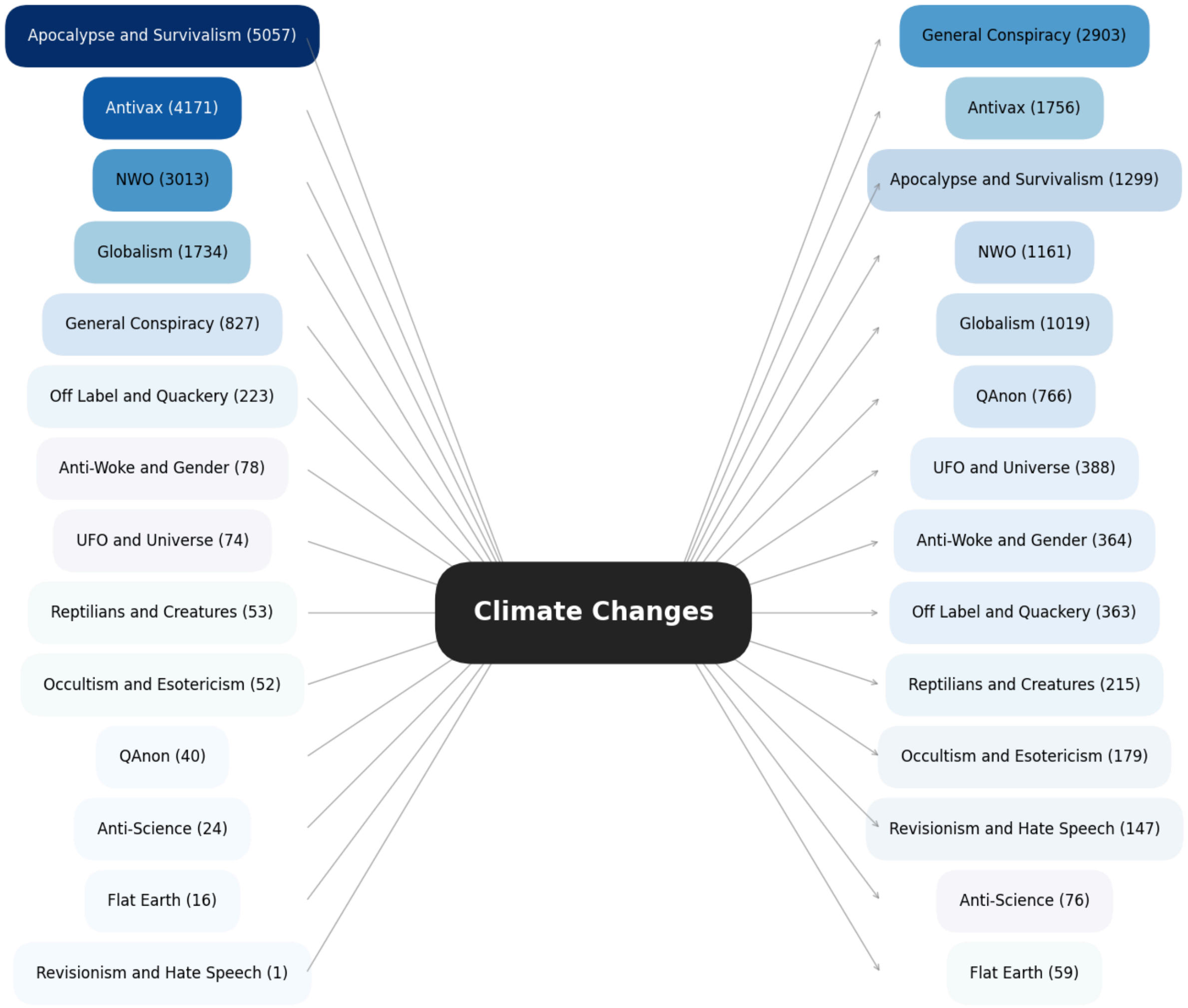

Source: Own elaboration (2024).

The figure shows the flow of invitation links originating from communities focused on climate change and directing their members to other conspiracy themes, as well as the reverse flow where these communities receive invitations from other themes. On the left, it is observed that the largest number of links (5,057) come from Apocalypse and Survivalism communities towards the climate change theme, followed by Anti-Vaccine with 4,171 links and New World Order with 3,013 links. These numbers indicate that discussions on climate change attract adherents from apocalyptic narratives and anti-vaccine theories, suggesting a significant intersection between these themes. On the right, it is notable that climate change communities redirect invitations quite evenly across a variety of other themes, particularly towards General Conspiracy (2,903 links), Anti-Vaccine (1,756 links), and Apocalypse and Survivalism (1,299 links). This uniform distribution suggests that climate change denial communities act as gatekeepers, distributing disinformation equitably across various other conspiracy narratives, strengthening their position within the conspiracy ecosystem.

[ english version / português abaixo ]

**Figure 05.** Flow of invitation links between anti-science communities

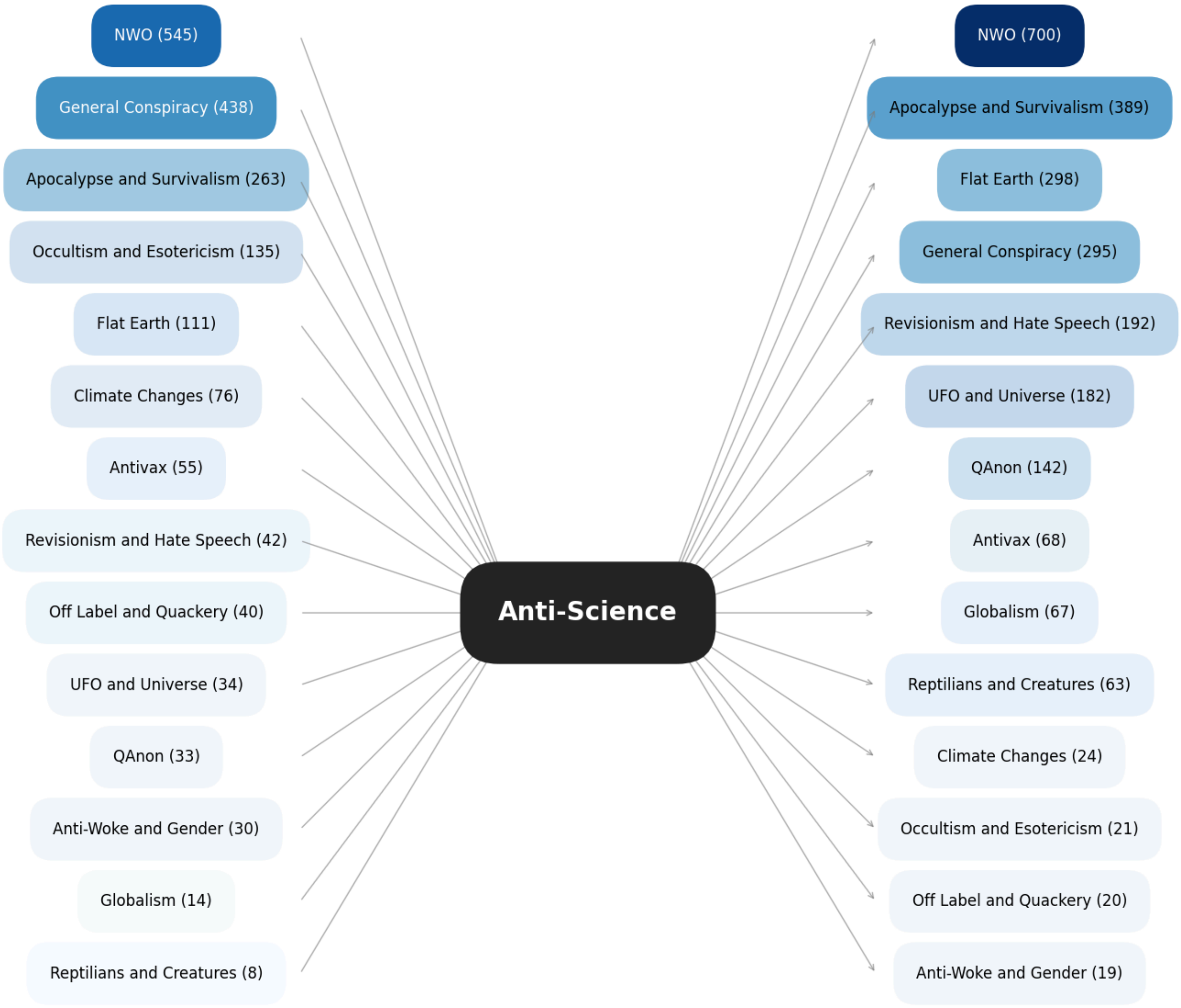

Source: Own elaboration (2024).

This figure focuses on the flow of invitation links originating from communities centered on anti-science to other themes and vice versa. On the left, the New World Order (NWO) stands out with 545 links leading to the anti-science theme, followed by General Conspiracy with 438 links and Apocalypse and Survivalism with 263 links. This reveals that the anti-scientific narrative appeals to those already immersed in global control and apocalyptic survivalism theories. On the right, the anti-science theme also distributes invitations, with NWO (700 links), Apocalypse and Survivalism (389 links), and Flat Earth (298 links) as the main destinations. Unlike climate change, where the distribution of invitations is more balanced, here we see a higher concentration of flows towards specific themes. This suggests that anti-science communities serve as catalysts for deepening members' beliefs in more focused areas of disinformation, consolidating beliefs in well-defined theories such as NWO and Flat Earth.

### 3.2. Time series

The following figures explore the temporal series of communities focused on anti-science and climate change over the years. Through different graphical representations, it is possible to observe how the evolution of these discussions behaved over time, revealing patterns of growth, decline, and the relationship between these themes.

**Figure 06.** Line graph over the period

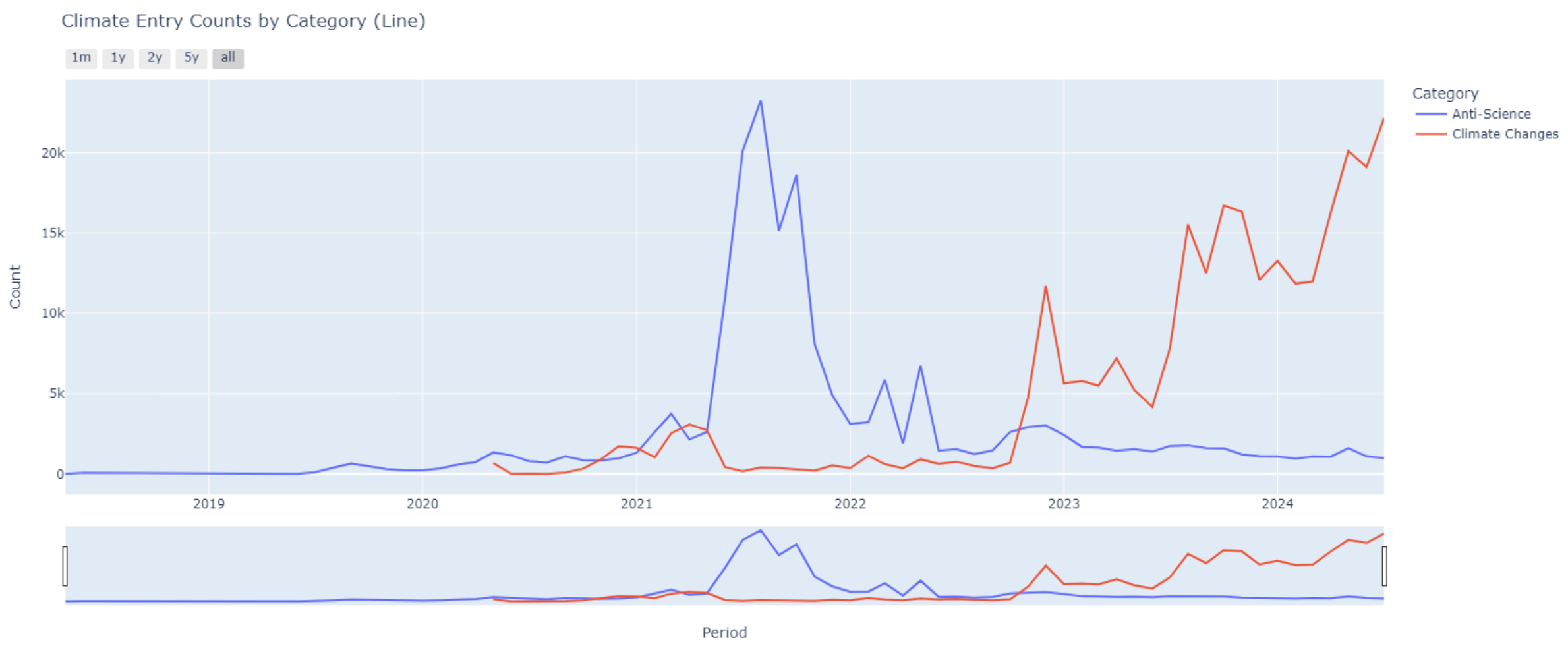

Source: Own elaboration (2024).

The line graph of the period highlights the temporal evolution of discussions on anti-science and climate change. We notice a significant peak in 2021 for the anti-science category, coinciding with the height of the COVID-19 pandemic, where debates about the efficacy of vaccines and public health measures generated intense activity in these communities. This marked increase reflects the polarization and amplification of conspiracy theories during moments of global crisis, where disinformation found fertile ground to grow. On the other hand, discussions about climate change showed more gradual but consistent growth, with notable increases from 2022 onwards, as extreme weather events and global political discussions, such as UN climate conferences, fueled denialist narratives.

[ english version / português abaixo ]

**Figure 07.** Absolute area chart over the period

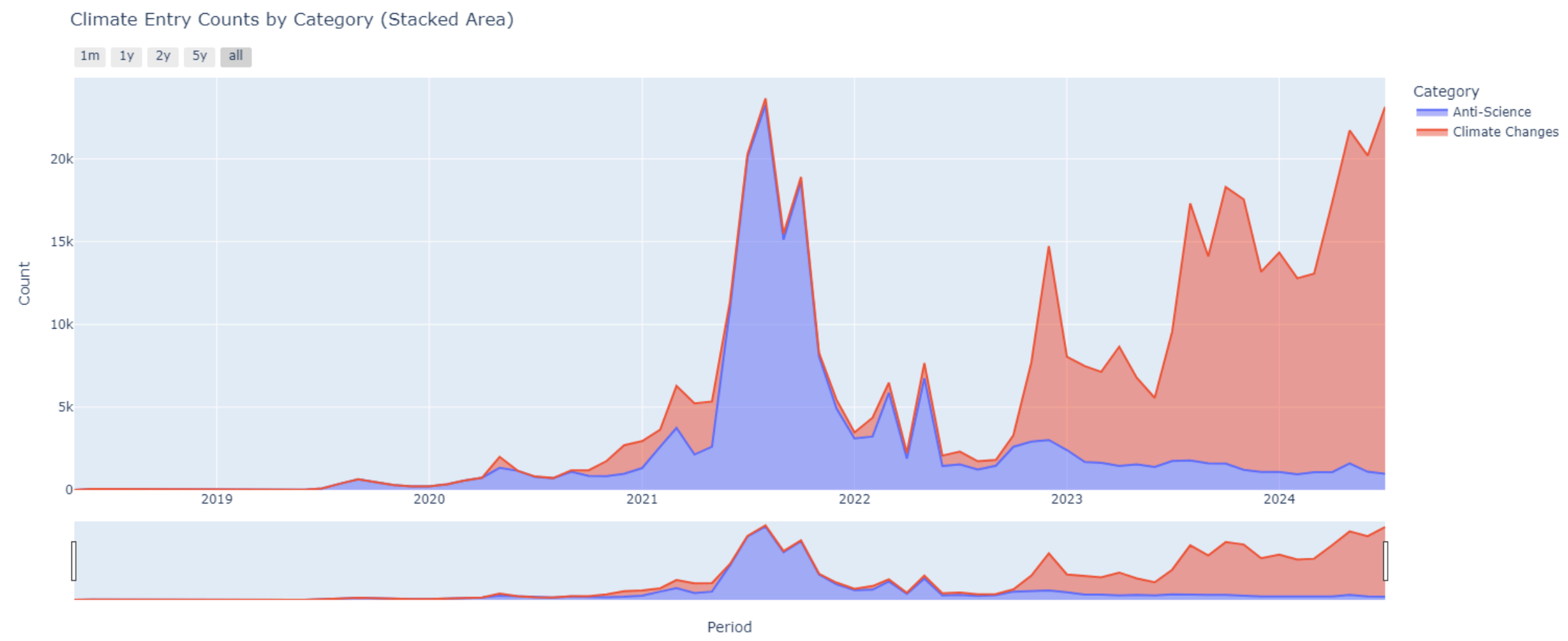

Source: Own elaboration (2024).

In the absolute area graph, the predominance of discussions on anti-science in 2021 becomes evident, occupying a large part of the graphical space during the pandemic peak. This is followed by a gradual increase in discussions on climate change, which gain more prominence from 2022 and continue to rise until 2024. This pattern suggests that as the pandemic was brought under control and new global concerns emerged, such as the climate crisis, the focus of conspiracy communities diversified. The transition from the absolute dominance of anti-science to a balance with climate change reflects the adaptation of conspiracy narratives to political agendas.

**Figure 08.** Relative area chart over the period

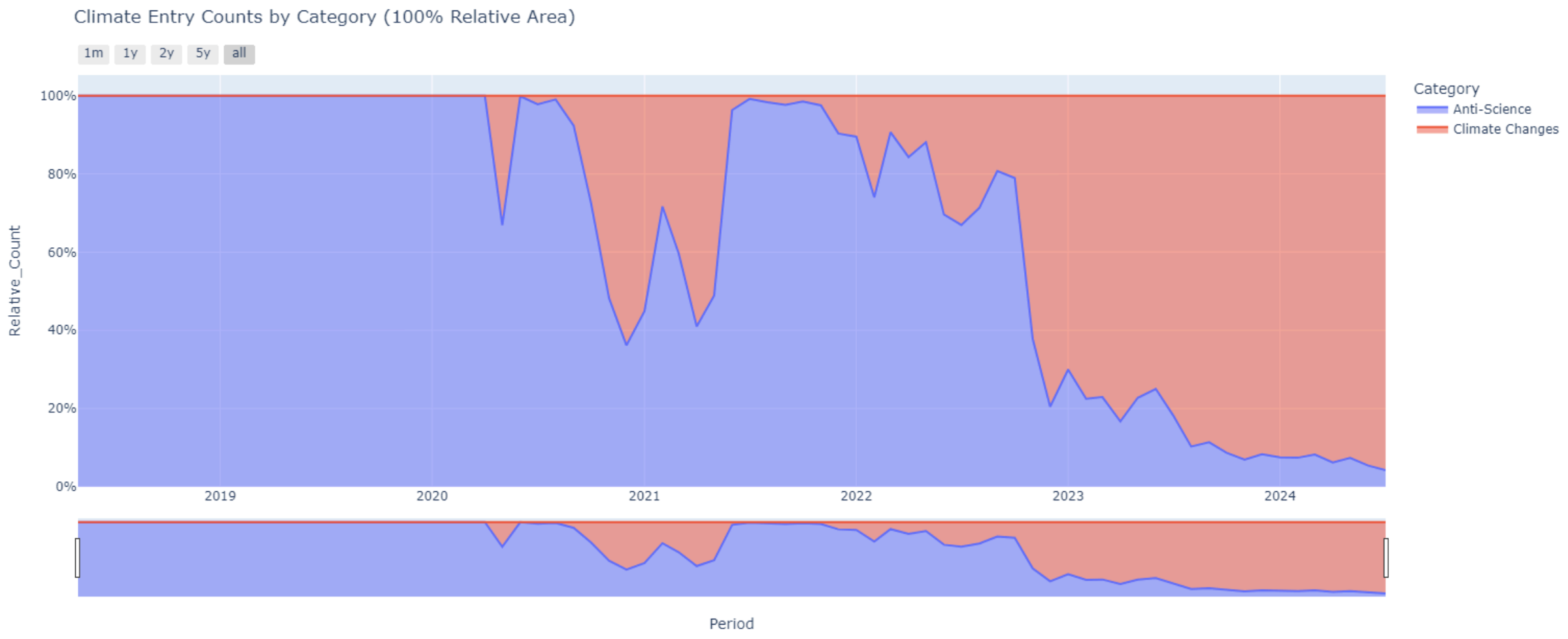

Source: Own elaboration (2024).

The relative area graph provides a comparative perspective between the two categories over time, highlighting the variation in the dominance of each theme. It is observed that during 2021, the anti-science theme almost completely dominated the discussions, reflecting the pandemic context as previously mentioned. However, from 2022 onwards, there is a gradual shift, where discussions on climate change begin to occupy more relative space, eventually surpassing anti-science. This shift suggests that over time and with the reduction of the pandemic's urgency, climate denial narratives gained prominence, possibly in response to growing environmental challenges and global climate policies. The change in dominance between the themes also reflects the ability of conspiracy communities to redirect the focus of their narratives according to the global context.

### 3.3. Content analysis

The following figures present a consolidated and temporal analysis of the most recurring words in discussions about climate change and anti-science. These word clouds allow for the identification of the most addressed arguments and topics, as well as the predominant narratives, offering insight into how certain themes emerge and transform over time, reflecting the dynamics of discussions within these communities.

**Figure 09.** Consolidated word cloud of climate change and anti-science

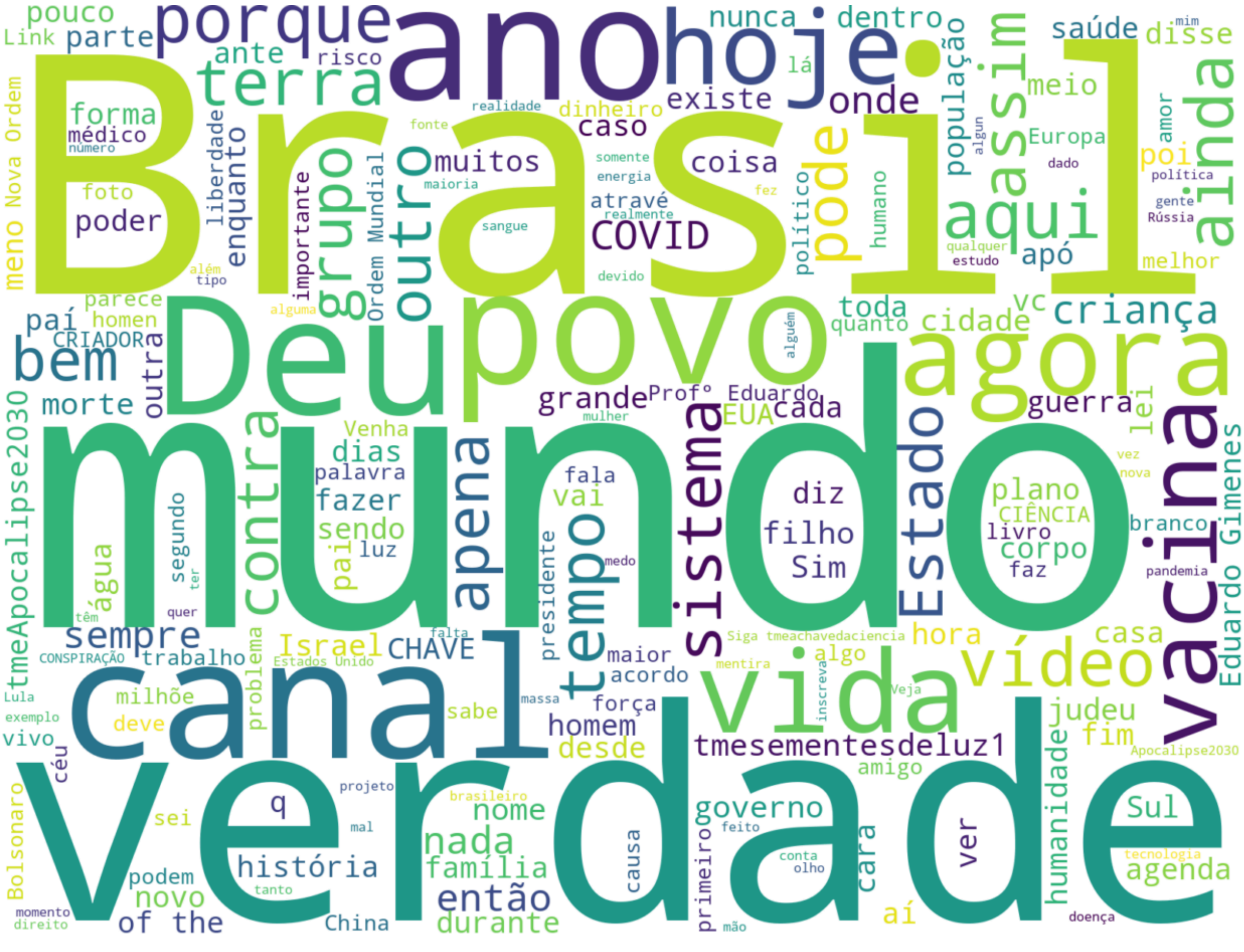

Source: Own elaboration (2024).

The consolidated word cloud reveals the intersection of narratives between climate change and anti-science. Words like "truth", "Brazil", "world", and "people" stand out, and by observing the publications, we can notice a strong presence of discourses that seek to imprint a nationalist bias, attempting to contest established scientific narratives. Terms like “vaccine” and “COVID” also appear prominently, reflecting the continuity of anti-science discussions during and after the pandemic.

**Chart 01.** Temporal word cloud of climate change

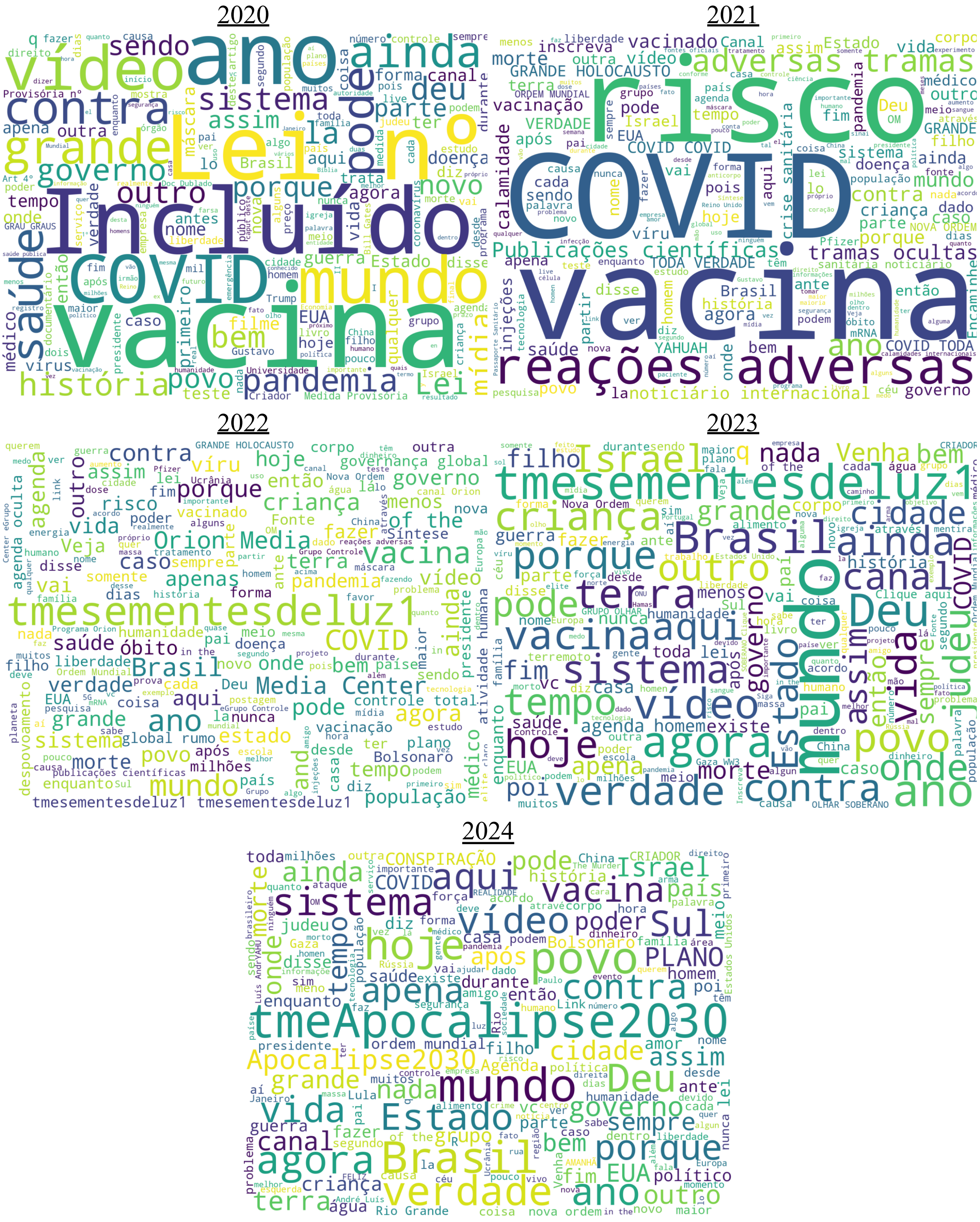

Source: Own elaboration (2024).

The temporal table on climate change shows an evolution of the topics discussed over the years. In 2020, the focus was clearly on “vaccines” and “COVID”, highlighting the overlap between climate and pandemic discussions. In 2021 and 2022, terms like “system”, “world” and “government” gained prominence, reflecting a transition to a broader discussion about global control and institutional distrust. In 2023 and 2024, we observe a diversification of narratives, with the introduction of terms like “Apocalypse2030”, along with critiques directed at the United Nations’ (UN) Agenda 2030 and the Sustainable Development Goals (SDGs), indicating a growing connection between climate denialism and apocalyptic theories.

**Chart 02.** Temporal word cloud of anti-science

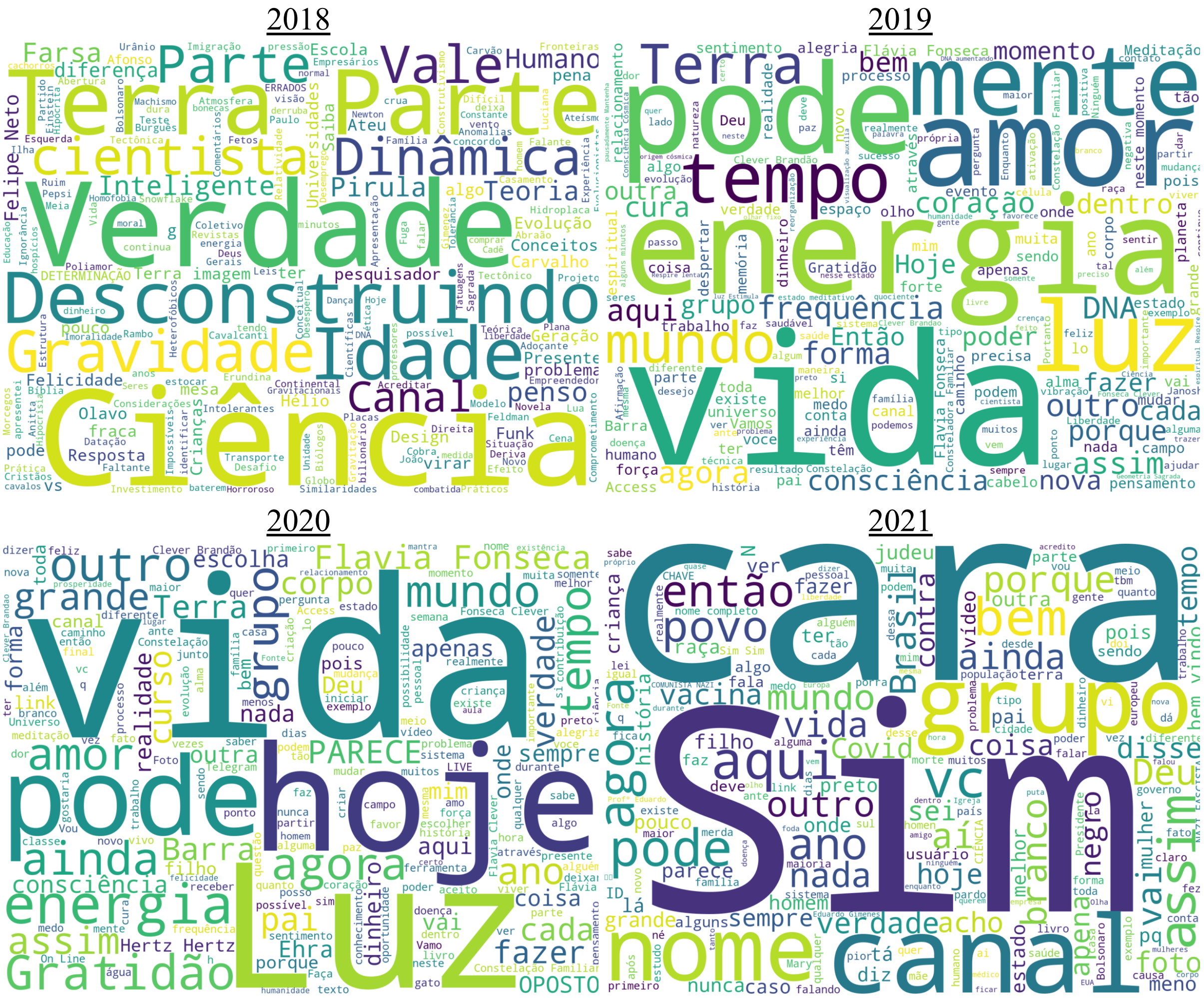

2022

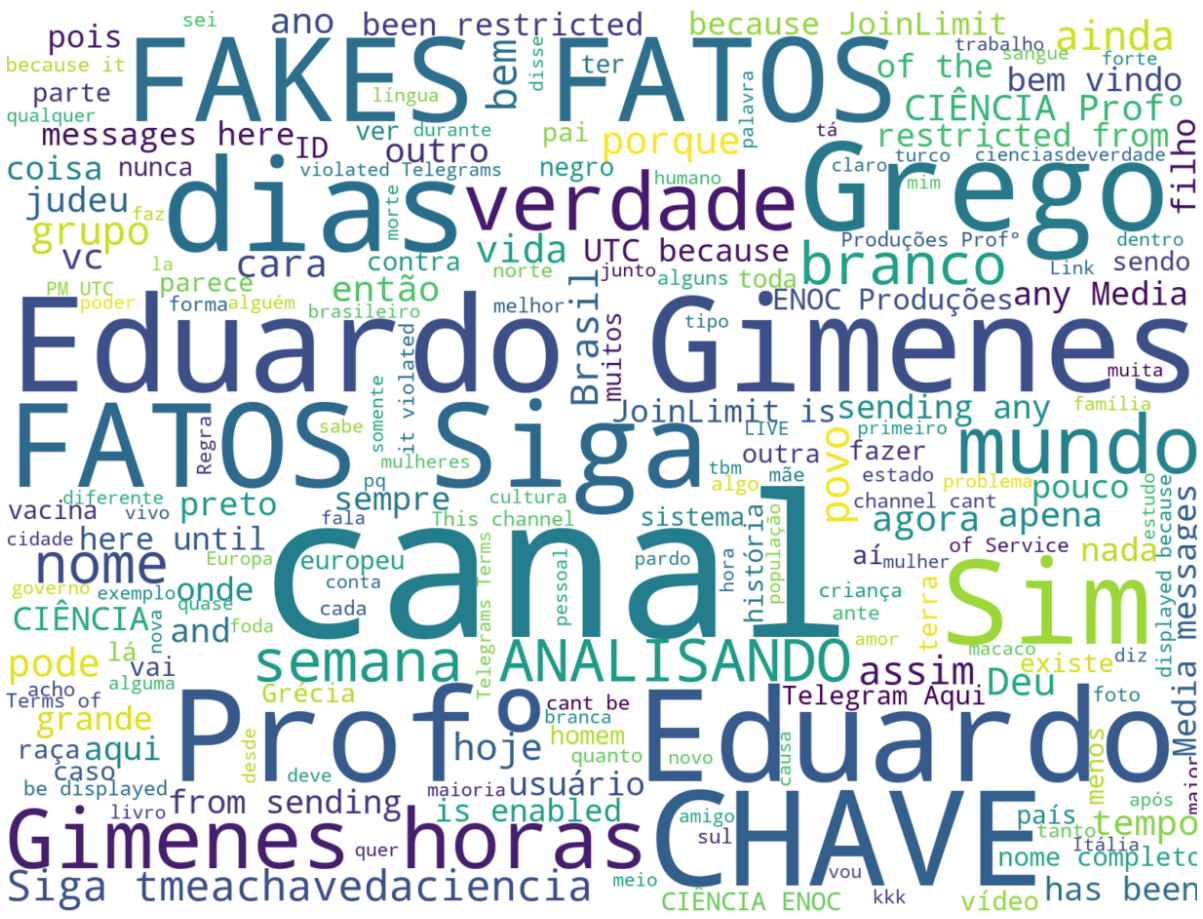

2023

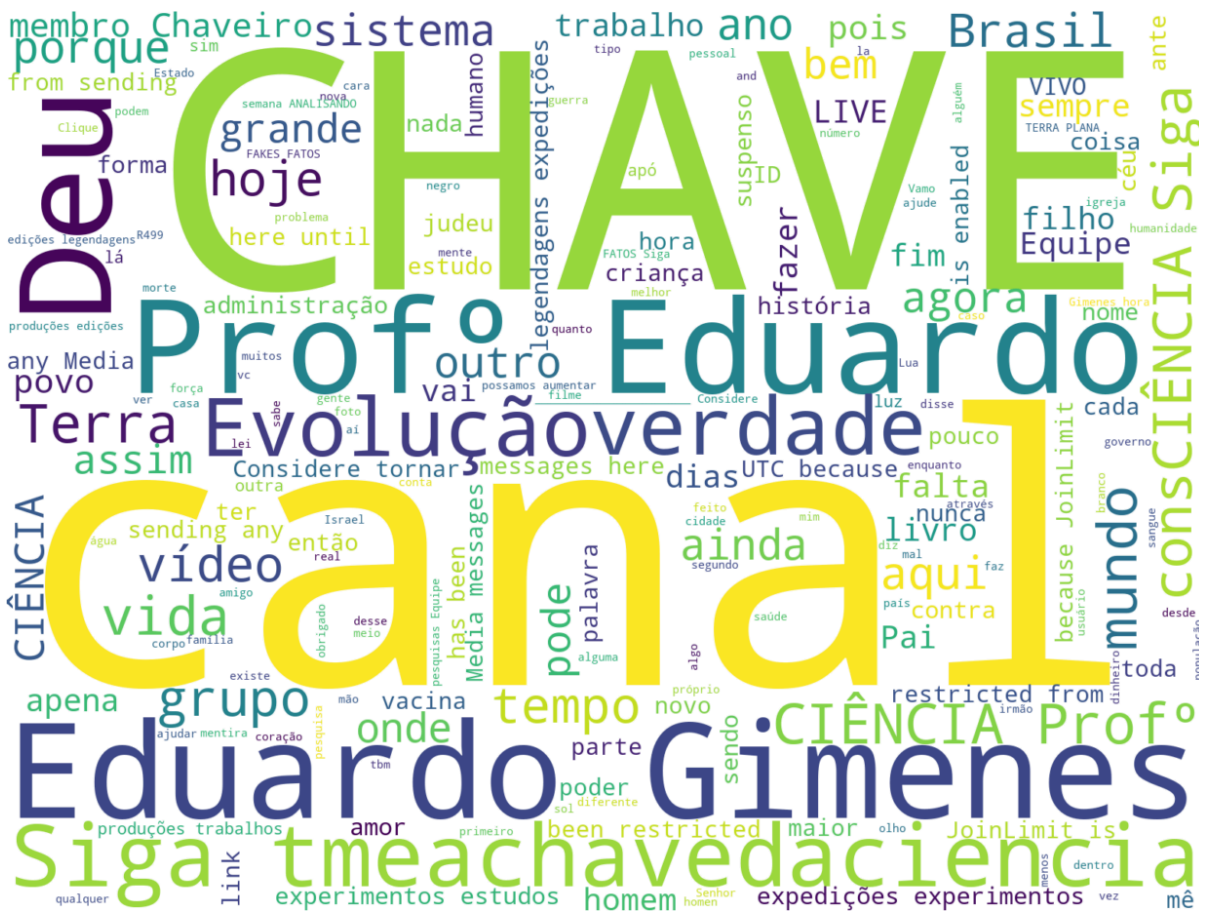

2024

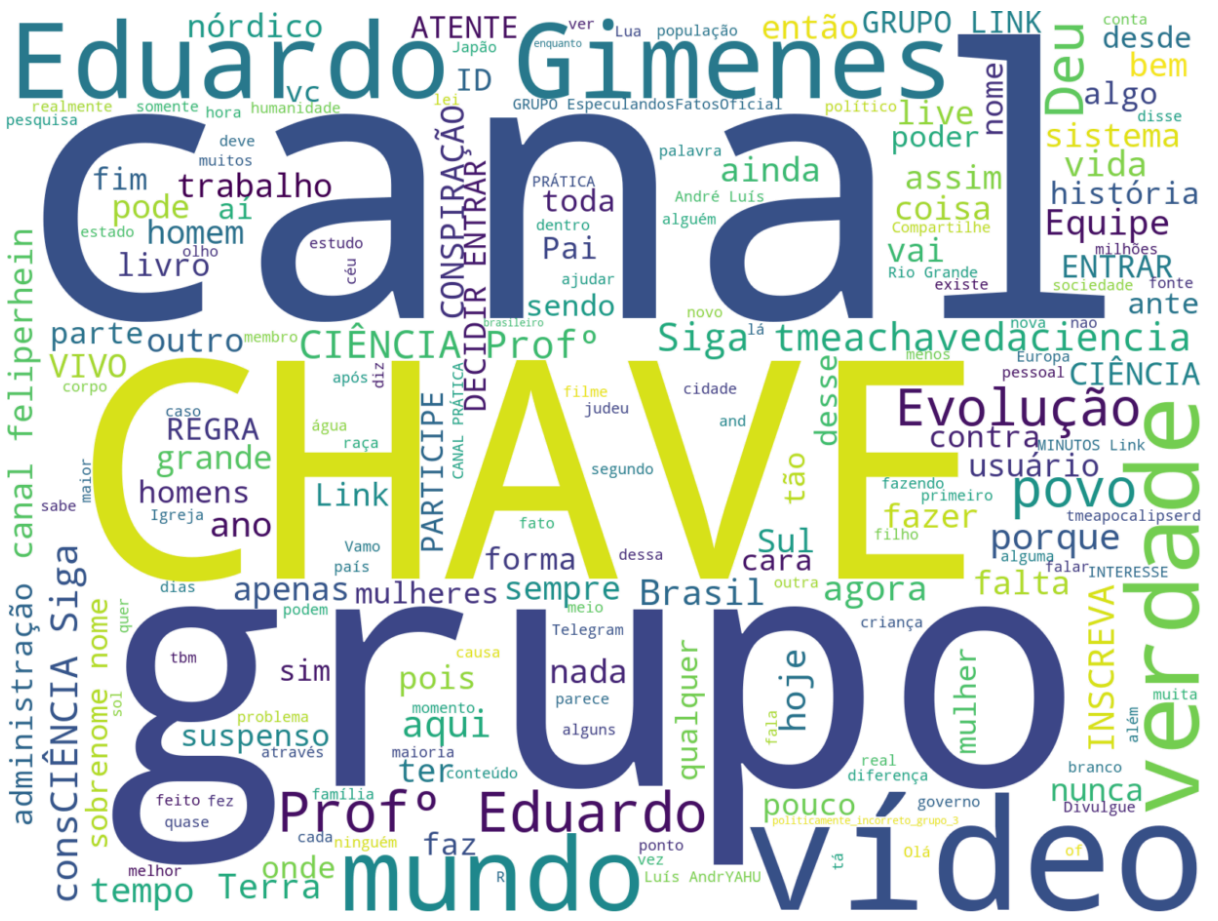

Source: Own elaboration (2024).

In the temporal analysis of anti-science, we observe consistency in the use of certain terms over the years, such as “life”, “channel”, “truth” and “light”. Words like “Eduardo Gimenes” and “CHAVE” emerge in 2023 and 2024, indicating the influence of specific figures or channels in propagating these narratives. The persistence of terms like “life” and “light” suggests an enduring narrative that blends esotericism and science, promoting a worldview where conventional science is discredited in favor of an alternative “truth”, often associated with charismatic or influential figures within these communities.

### 3.4. Thematic agenda overlap

The following figures explore various themes within conspiracy theory communities on Telegram, focusing on topics such as faith and religion, geopolitical disputes, vaccine disinformation and public health, the New World Order and globalism, cultural wars against the Anti-Woke movement, and disinformation surrounding the UN Agenda 2030 and the SDGs. These figures illustrate how different narratives interconnect, revealing the complexity of the discussions within these communities. The visual analysis of these connections allows for the identification of interaction patterns and thematic overlaps, highlighting how conspiracy beliefs are reinforced and disseminated through an ecosystem that blends science, religion, politics, and identity. These graphs provide a deep understanding of how seemingly distinct themes intertwine to create a cohesive network of disinformation and resistance to scientific and institutional information.

**Figure 10.** Themes of faith and religion

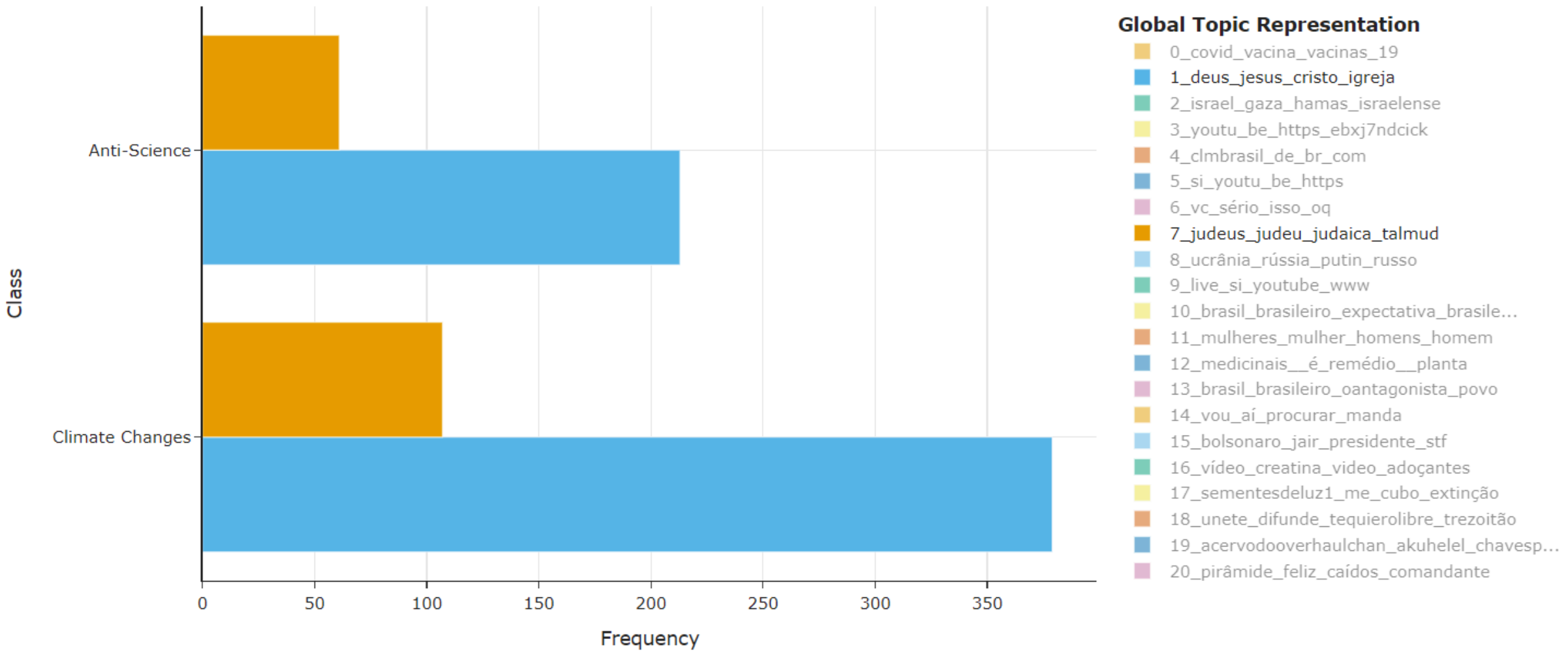

Source: Own elaboration (2024).

This graph illustrates how themes of faith and religion are interconnected with discussions on climate change and anti-science within conspiracy theory communities on Telegram. A strong presence of topics related to "God", "Jesus", and "Christ" is observed, particularly in discussions on anti-science. In many observed cases, biblical verses are used to supposedly support conspiratorial arguments. These connections suggest that faith is often mobilized to challenge scientific narratives, promoting a worldview in which science is seen as contradictory or threatening. On the other hand, climate change appears associated with topics like "Israel" and "Judaism", reflecting an overlap with conspiracy theories that link global crises to alleged religious domination plans. This interaction between religion and science in the discussions reveals the depth with which personal beliefs and conspiracy theories intertwine, strengthening resistance against scientific information.

[ english version / português abaixo ]

**Figure 11.** Themes of geopolitical disputes

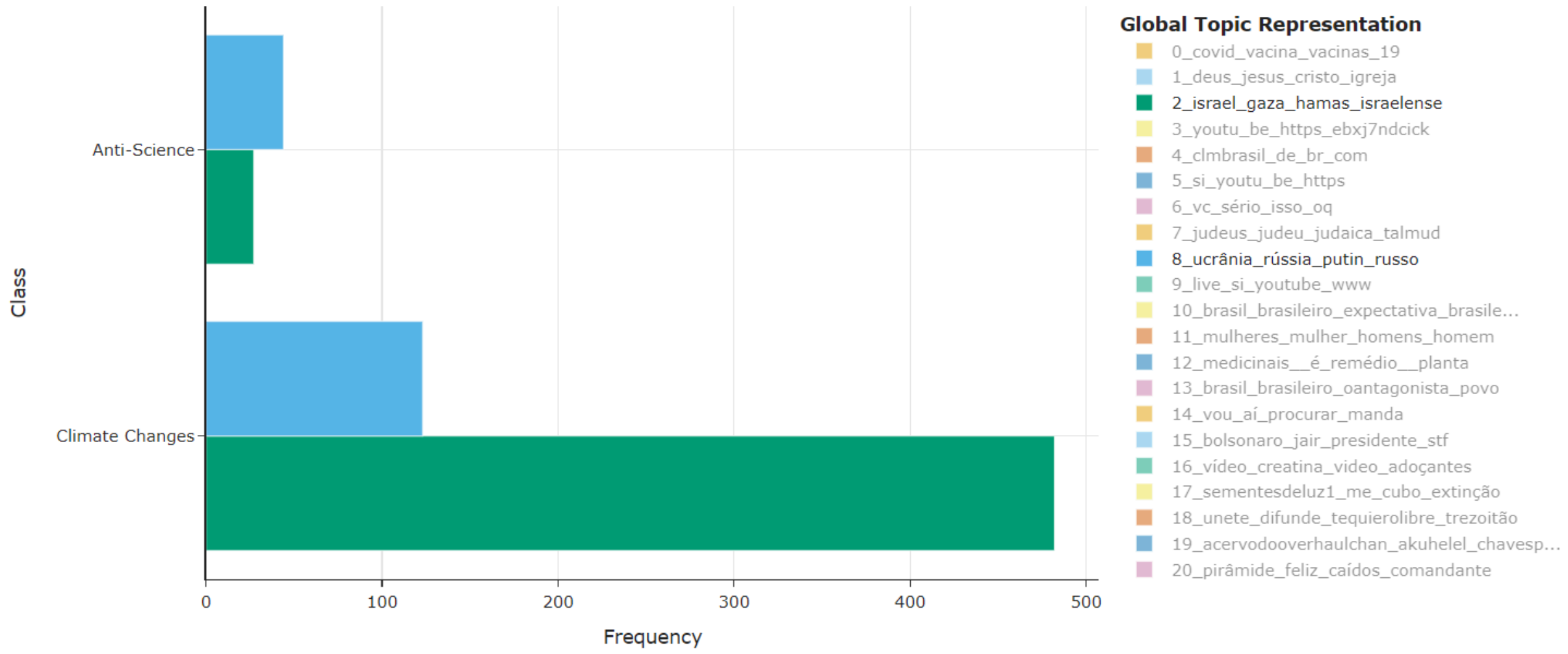

Source: Own elaboration (2024).

This graph highlights the themes of geopolitical disputes that permeate discussions on climate change and anti-science. Climate change, in particular, is strongly associated with topics like "Ukraine", "Russia", and "Putin", evidencing the narrative that extreme climate events and environmental policies are manipulated or used as a pretext in international conflicts. The significant presence of these themes suggests that climate discussions are often contextualized within a broader geopolitical framework, where climate change is seen not just as a natural phenomenon, but as a tool for global control. This reflects a strategy by these communities to insert scientific issues into debates about power and sovereignty, amplifying distrust towards international agendas and global climate policies.

**Figure 12.** Themes of vaccine disinformation and public health

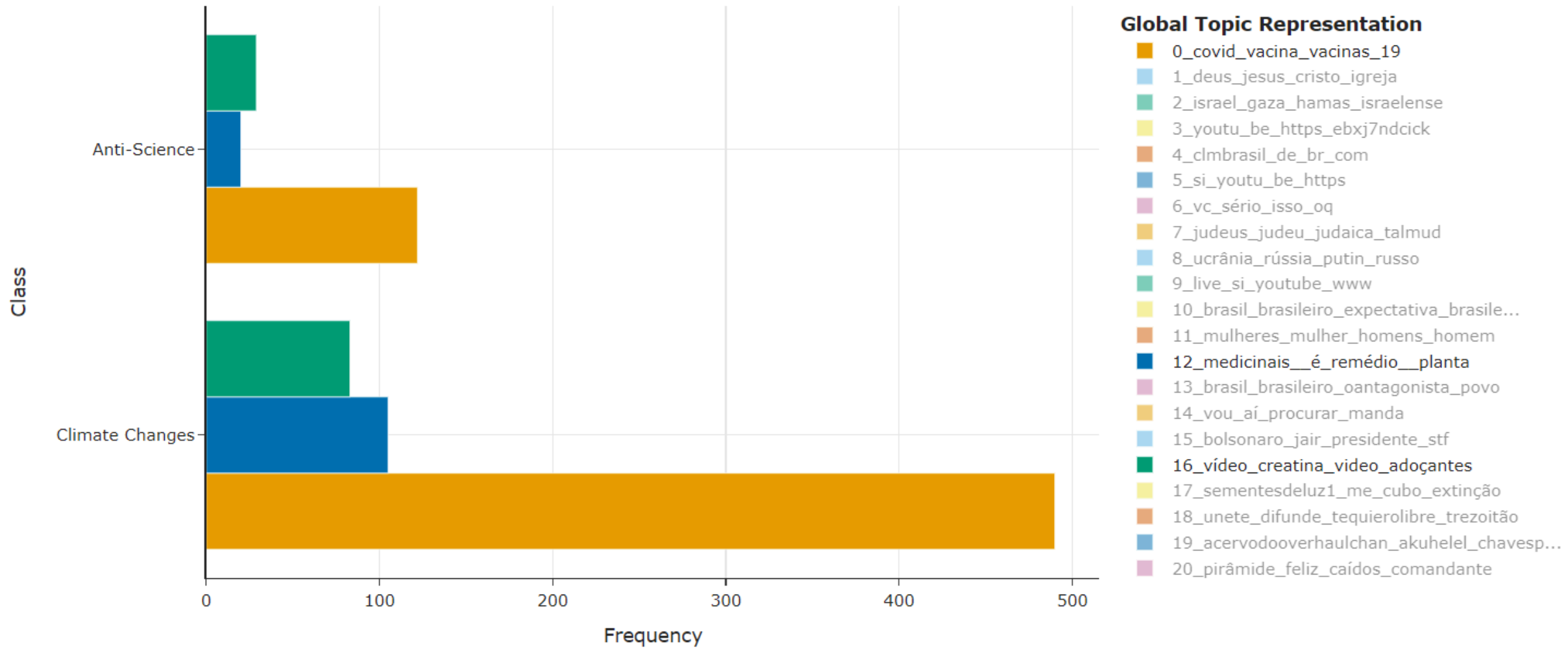

Source: Own elaboration (2024).

This graph examines how themes of vaccine disinformation and public health intertwine with climate change and anti-science. Anti-science, in particular, is closely linked to vaccine disinformation narratives, with topics like "vaccines" and "COVID" dominating the discussions. This reinforces the idea that communities denying science use vaccine disinformation as a central pillar in their disinformation campaigns. Meanwhile, climate change is connected to public health themes through discussions that question the validity of climate mitigation and adaptation policies, suggesting that these policies are part of a broader population control plan. The intersection of these topics reflects a strategy to unify forms of disinformation into a coherent narrative that challenges both science and health policies.

**Figure 13.** Themes of New World Order and globalism conspiracies

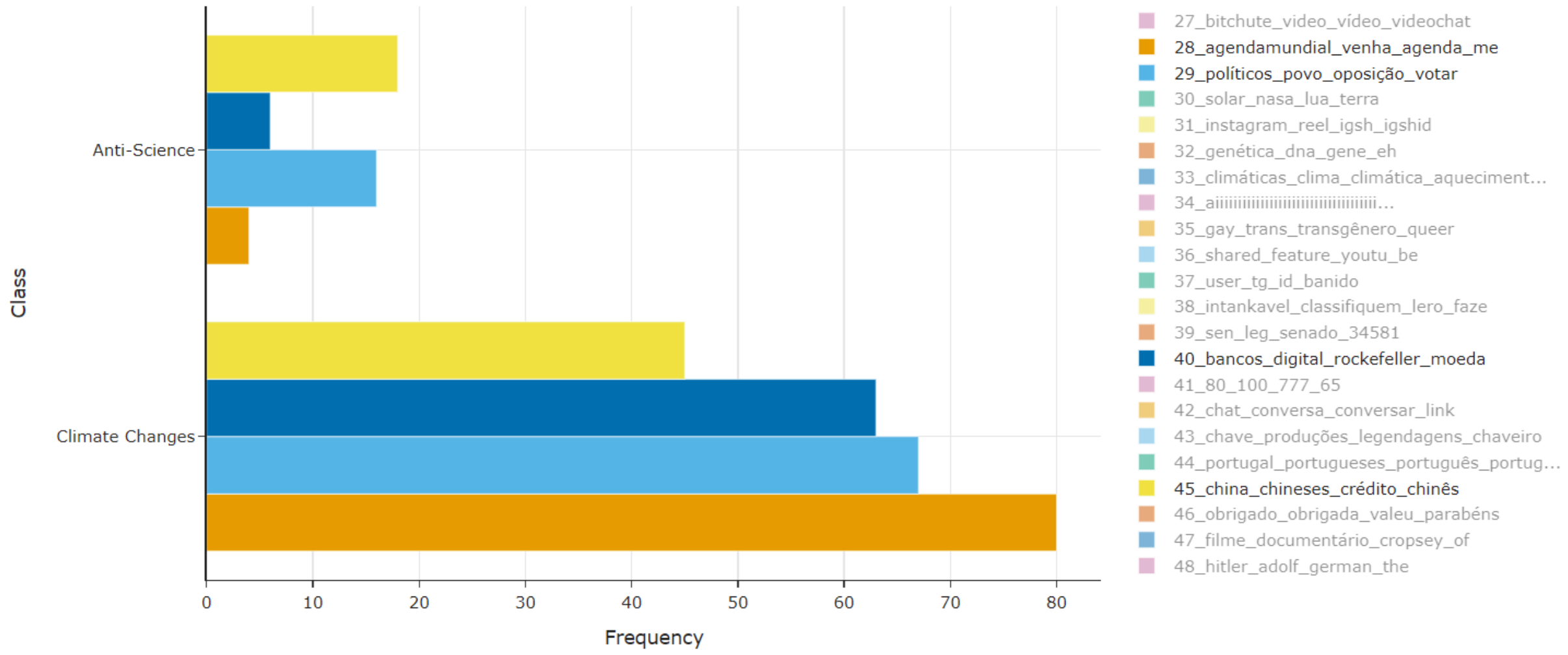

Source: Own elaboration (2024).

The graph reveals how themes related to the New World Order and globalism are present in discussions on climate change and anti-science. Climate change is frequently associated with topics like “world agenda”, “global”, and “digital currency”, indicating that climate events are seen as part of a global plan to establish a single government or global financial control. Anti-science is also linked to these theories, though less prominently, suggesting that while scientific disinformation is central, it is complemented by broader global conspiracy narratives. This overlap of themes indicates that climate and anti-scientific discussions are shaped by a common fear of loss of sovereignty and control by global elites, expanding the reach and persistence of these conspiratorial narratives.

[ english version / português abaixo ]

**Figure 14.** Themes of Anti-Woke cultural wars

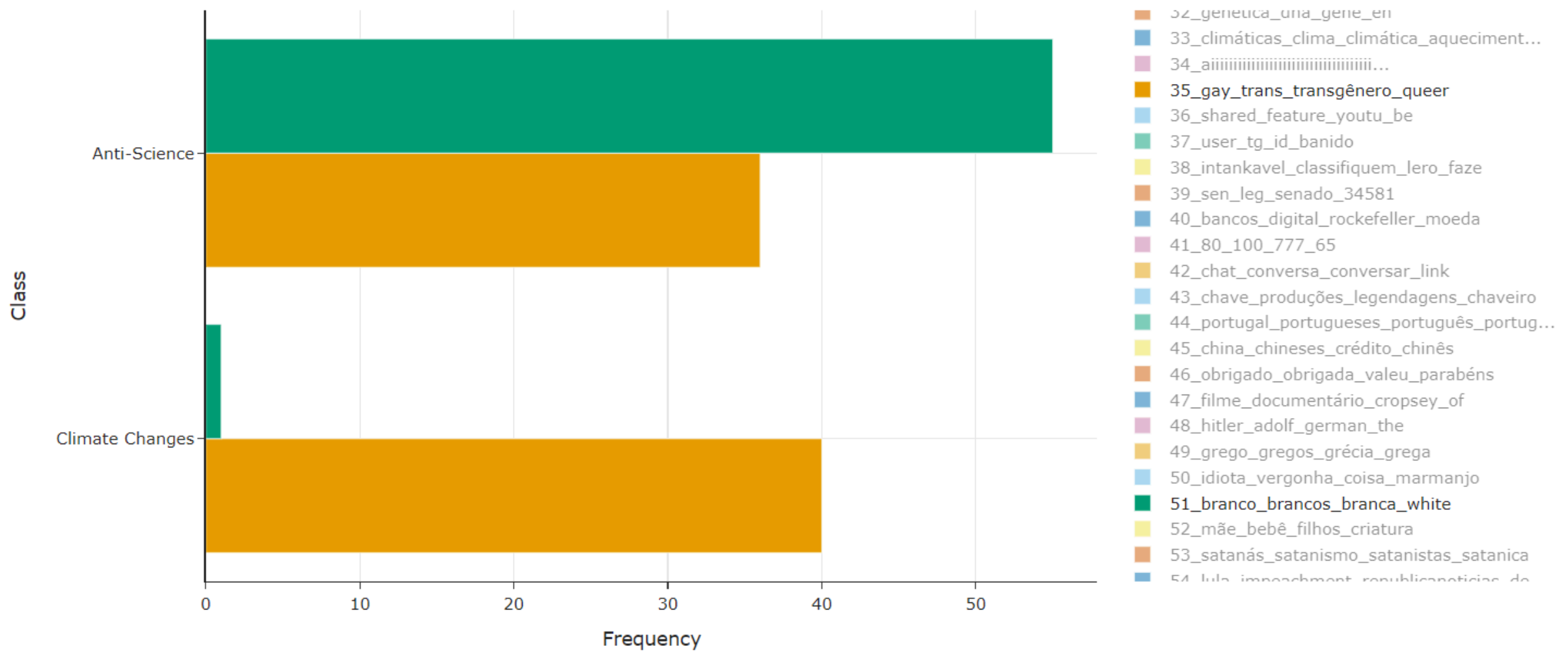

Source: Own elaboration (2024).

This graph observes the intersection between discussions on climate change, anti-science, and Anti-Woke cultural wars. Anti-science, in particular, is strongly associated with topics like “transgender”, “gay”, “queer”, and “whites”, reflecting the narrative that modern science and the fight for social justice are part of an ideological movement that threatens traditional values. Climate change also appears related to these cultural wars, but less intensely, suggesting that while climate discussions can be mobilized within debates on identity and culture, anti-science is more directly linked to these ideological battles. Furthermore, conspiracy theory communities suggest that climate change is part of a domination plan to masculinize women and feminize men through an alleged Anti-Woke agenda. This interaction highlights how conspiratorial narratives use cultural and identity themes to reinforce resistance against social and scientific changes.

**Figure 15.** Themes of disinformation about the UN Agenda 2030 and the SDGs

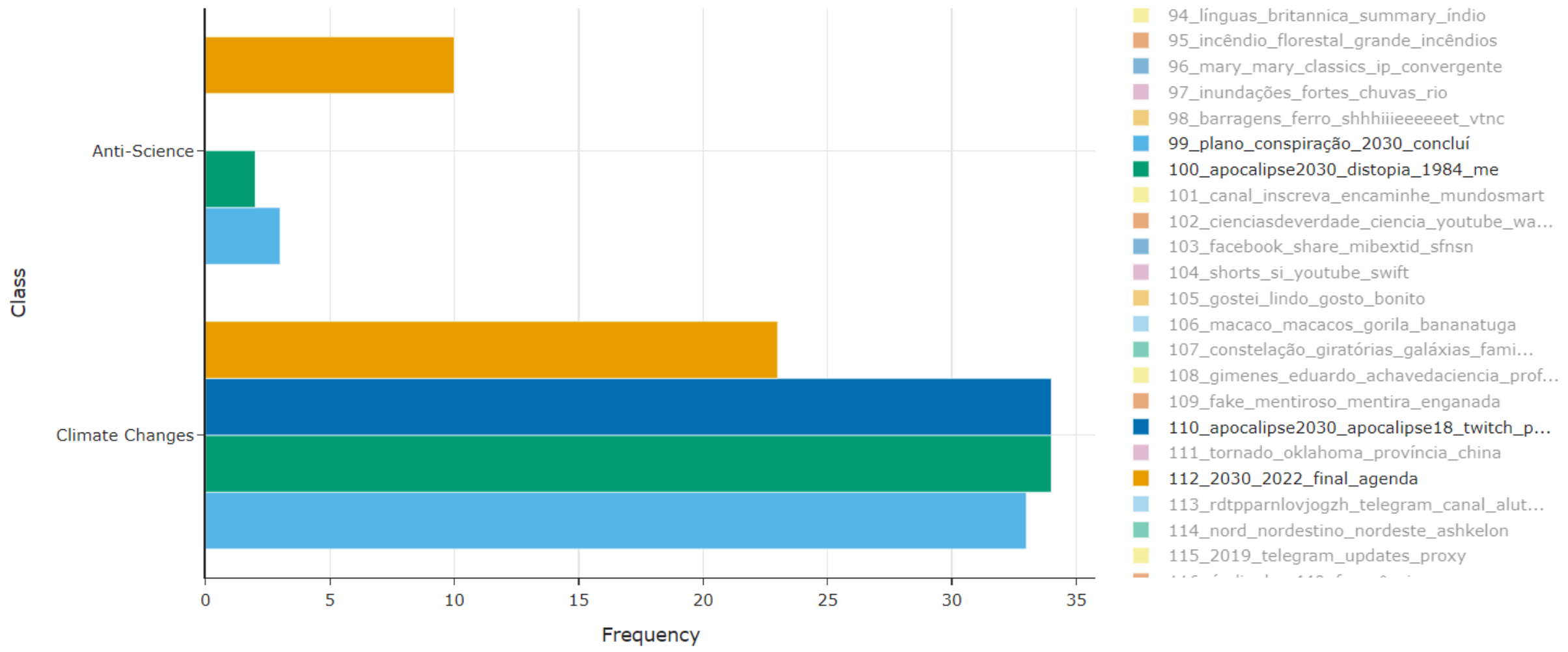

Source: Own elaboration (2024).

This graph explores how themes of disinformation about the UN Agenda 2030 and the Sustainable Development Goals (SDGs) relate to climate change and anti-science. Climate change, in particular, is strongly linked to conspiracy theories that describe Agenda 2030 as a plan to supposedly establish a global dictatorship or a fabricated crisis to justify population control. Topics like "Apocalypse 2030" and "conspiracy plan" appear frequently, suggesting that these communities view international climate policies as part of a globalist strategy. Anti-science is also present, indicating that scientific discussions are secondary to global conspiracy narratives. This overlap of themes reveals how climate disinformation is instrumentalized to challenge initiatives and promote the idea of a global conspiracy.

## 4. Reflections and future works

To answer the research question, **"How are Brazilian conspiracy theory communities on climate change and anti-science characterized and articulated on Telegram?"**, this study adopted techniques mirrored in a series of seven publications that seek to characterize and describe the phenomenon of conspiracy theories on Telegram, using Brazil as a case study. After months of investigation, it was possible to extract a total of 36 Brazilian conspiracy theory communities on Telegram related to climate change and anti-science, totaling 1,287,938 contents published between May 2018 (first publications) and August 2024 (when this study was conducted), with 78,252 users among the communities.

Four main approaches were adopted: **(i)** Network, which involved the creation of an algorithm to map connections between communities through invitations circulated among groups and channels; **(ii)** Time series, which used libraries like "Pandas" (McKinney, 2010) and "Plotly" (Plotly Technologies Inc., 2015) to analyze the evolution of publications and engagements over time; **(iii)** Content analysis, where textual analysis techniques were applied to identify patterns and word frequencies in the communities over the semesters; and **(iv)** Thematic agenda overlap, which utilized the BERTopic model (Grootendorst, 2020) to group and interpret large volumes of text, generating coherent topics from the analyzed publications. The main reflections are detailed below, followed by suggestions for future works.

### 4.1. Main reflections

**Climate change denial and anti-science form a critical intersection in the spread of disinformation:** Brazilian climate change denial and anti-science communities on Telegram work in synergy, creating a complex network of disinformation. These communities interact in a way that climate change denial narratives not only coexist with anti-scientific discourse but also reinforce each other. This phenomenon suggests that these conspiratorial themes take advantage of pre-existing distrust in scientific institutions, amplifying and spreading false information. The combined effect of these narratives is particularly potent as it operates on multiple fronts simultaneously, making it difficult to combat disinformation and expanding the reach of these beliefs;

**Climate change is targeted by apocalyptic narratives:** Communities discussing apocalypse and survival directed 5,057 links to climate change groups, demonstrating that these apocalyptic theories are a significant gateway to climate change denial. These apocalyptic narratives often associate extreme climate events with predictions of imminent catastrophes, which are then used to question the scientific validity of climate change. By promoting this view, these communities capture the attention of people already predisposed to believing in apocalyptic theories and redirect them to a broader rejection of climate science, exacerbating polarization around this issue;

**Anti-science communities function as distributors of disinformation for various conspiracy theories:** Links originating from these communities are evenly distributed to other topics, such as the New World Order and globalism, suggesting that anti-science acts as a central gatekeeper. These communities are not limited to a single conspiratorial narrative but serve as convergence points for various theories, where members are exposed to a wide range of disinformation. This creates an environment where different conspiracy narratives can reinforce each other, creating a complex and cohesive web of disinformation that is difficult to dismantle as it operates on multiple levels and themes simultaneously;

**Significant growth of anti-science discussions during the COVID-19 pandemic:** There was a notable peak in 2021 in anti-science discussions, driven primarily by disinformation about vaccines and other public health measures. During this period, anti-science communities on Telegram expanded rapidly, taking advantage of the fear and uncertainty generated by the pandemic to disseminate false information. This growth indicates how global crises are exploited by these communities to strengthen and expand their follower base. The explosion of these discussions reveals not only an increase in content but also an intensification of anti-scientific beliefs, which began to mix with other conspiracy theories, forming an even more robust network of disinformation;

**Climate change denial communities attract members from other conspiracy theories:** Groups focused on globalism and the New World Order sent 3,013 and 2,903 links, respectively, to climate change communities, indicating a convergence between these conspiracy narratives. This suggests that for adherents of these theories, climate change is seen not just as a natural phenomenon but as part of a larger global control plan. This intersection between different theories allows climate denial communities to expand and integrate new ideas, attracting members from other conspiratorial groups who share a skeptical or hostile view of authorities and traditional scientific explanations. Thus, climate change becomes another battleground in the fight against supposed global domination;

**Climate change disinformation is strongly linked to geopolitical disputes:** Terms like “Ukraine” and “Russia” are frequently associated with climate debates, suggesting that the climate crisis is framed as part of a geopolitical manipulation strategy. In these debates, extreme climate events and environmental policies are portrayed as tools used by global powers to influence and control other nations. This narrative is particularly powerful because it exploits real geopolitical tensions, adding a layer of distrust and conspiracism to climate change discussions. The result is a distorted view of environmental issues, where each climate event is seen not as a symptom of a global crisis but as a power tool used by nations or global elites to advance agendas;

**Anti-science narratives integrate esoteric and conspiratorial beliefs:** Communities that promote esotericism and occultism are often linked to anti-scientific discussions, reinforcing the idea of an "alternative truth" that challenges conventional science. These communities offer a space where the boundaries between science, spirituality, and conspiracy theory become blurred. Science is often portrayed as an oppressive force that hides the "true" nature of reality, supposedly revealed through esoteric knowledge. This not only strengthens beliefs within these communities but also creates a cohesive narrative that integrates different forms of disinformation, making them more attractive and convincing to those already predisposed to questioning conventional scientific explanations;

**Climate change discussions evolve into themes related to global domination:** From 2022 onwards, climate discussions began to include more references to supposed global control, connecting with New World Order theories and other conspiracies. This evolution reflects how climate change narratives are shaped to align with a broader view of global conspiracies, where climate events are seen as pretexts for implementing authoritarian and centralized policies. This approach resonates with those already suspicious of global elites and international organizations, reinforcing the idea that climate change is an invention or exaggeration created to justify a loss of freedom and sovereignty;

**Anti-science communities promote disinformation about public health based on conspiracy theories:** disinformation about vaccines is central to anti-science discussions, being used as a pillar to challenge public health policies and reinforce other conspiracy narratives. In these communities, vaccines are often presented not as a protective measure but as a tool for intentional control or harm. This distorted view feeds on fears and uncertainties already present in society, amplifying resistance to public health policies and creating a foundation to reject other scientific initiatives. From these discussions, narratives emerge that intertwine with global control theories, where science is seen as a tool in the hands of elites to manipulate and control the population;

**The UN Agenda 2030 is portrayed as a conspiracy for population control:** Climate change denial and anti-science communities often associate UN initiatives with conspiracy theories that claim a global domination plan. Within these narratives, Agenda 2030 and the Sustainable Development Goals (SDGs) are presented not as an effort to mitigate climate change and promote sustainable development but as a facade for a more sinister agenda of population control and loss of national sovereignty. This distorted view is fueled by fears of a world government and is widely spread among members of these communities, strengthening resistance to international policies and undermining global efforts to address environmental and social challenges.

### 4.2. Future works

Considering the findings detailed in this study, several promising research directions emerge to further explore the dynamics of climate change denial and the anti-science movement. One particularly rich area for investigation is the analysis of how the convergence between apocalyptic narratives and climate change denial contributes to the formation of highly polarized communities resistant to opinion change. Future studies could focus on examining the evolution of these narratives, especially in response to extreme climate events, and how these communities use global crises to reinforce and legitimize their beliefs among their members and beyond.

Another relevant direction would be to explore the role of geopolitical disputes in the spread of climate change disinformation. Future research could investigate how these connections are established, expanded, and how they affect public perception of climate change in different regional and sociopolitical contexts. Additionally, a deeper analysis could examine the impact of these geopolitical narratives on the formulation and implementation of environmental policies.

Furthermore, the interaction between esotericism and anti-science deserves special and detailed attention. Studies could focus on how communities that promote esoteric and occult beliefs contribute significantly to the spread of disinformation about climate change and science in general. Understanding the underlying psychology of these communities, the motivations that lead to adherence to these beliefs, and how they remain cohesive and attractive to new members can provide crucial insights for developing more effective strategies to combat disinformation, specifically tailored to these environments.

Future research could also delve deeper into the evolution of climate change discussions that are increasingly aligning with global control theories. This trend of connecting climate events to narratives of global domination and loss of individual freedom could have significant implications not only for public policy formulation but also for scientific communication in general. Investigating how these narratives develop, consolidate, and gain strength in different digital communities could offer valuable insights to interrupt the spread of disinformation before it becomes entrenched and influential on a large scale.

Finally, a more detailed analysis of the narratives that associate the UN Agenda 2030 with conspiracy theories would be essential to understand how these ideas are disseminated and amplified in climate change denial and anti-science communities. Studies could explore how this specific narrative is constructed, propagated, and reinforced within these communities, as well as evaluate the potential impact of these beliefs on the implementation of global sustainability policies. Identifying the main entry points for these narratives, understanding the motivations behind their dissemination, and developing innovative approaches to neutralize them may be crucial to protecting progress on climate and environmental issues, as well as ensuring adherence to essential international agreements and policies for the planet's future.

## 6. Author biography

**Ergon Cugler de Moraes Silva** has a Master's degree in Public Administration and Government (FGV), Postgraduate MBA in Data Science & Analytics (USP) and Bachelor's degree in Public Policy Management (USP). He is associated with the Bureaucracy Studies Center (NEB FGV), collaborates with the Interdisciplinary Observatory of Public Policies (OIPP USP), with the Study Group on Technology and Innovations in Public Management (GETIP USP) with the Monitor of Political Debate in the Digital Environment (Monitor USP) and with the Working Group on Strategy, Data and Sovereignty of the Study and Research Group on International Security of the Institute of International Relations of the University of Brasília (GEPSI UnB). He is also a researcher at the Brazilian Institute of Information in Science and Technology (IBICT), where he works for the Federal Government on strategies against disinformation. Brasília, Federal District, Brazil. Web site: https://ergoncugler.com/.

[ versão português / english above ]

# **Comunidades de negacionismo climático e anticiência no Telegram brasileiro:** desinformação climática como porta de entrada para redes conspiratórias mais amplas

*Ergon Cugler de Moraes Silva*

Instituto Brasileiro de Informação
em Ciência e Tecnologia (IBICT)
Brasília, Distrito Federal, Brasil

contato@ergoncugler.com
www.ergoncugler.com

## Resumo

As teorias da conspiração relacionadas ao negacionismo climático e anticiência têm encontrado terreno fértil no Telegram, especialmente entre comunidades brasileiras que desconfiam das instituições científicas e se opõem às políticas ambientais globais. Dessa forma, esse estudo busca responder à pergunta de pesquisa: **como são caracterizadas e articuladas as comunidades de teorias da conspiração brasileiras sobre temáticas de mudanças climáticas e anticiência no Telegram?** Vale ressaltar que este estudo faz parte de uma série de um total de sete estudos que possuem como objetivo principal compreender e caracterizar as comunidades brasileiras de teorias da conspiração no Telegram. Esta série de sete estudos está disponibilizada abertamente e originalmente no arXiv da Cornell University, aplicando um método espelhado nos sete estudos, mudando apenas o objeto temático de análise e provendo uma replicabilidade de investigação, inclusive com códigos próprios e autorais elaborados, somando-se à cultura de software livre e de código aberto. No que diz respeito aos principais achados deste estudo, observa-se: Comunidades de negacionismo climático e anticiência interagem em sinergia, criando uma rede complexa que reforça mutuamente as narrativas de desinformação; Temáticas apocalípticas, como Apocalipse e Sobrevivência, atuam como portas de entrada para o negacionismo climático, com 5.057 links direcionados para essas comunidades; Comunidades de anticiência funcionam como *gatekeepers*, distribuindo links de forma equilibrada para teorias como Nova Ordem Mundial e Globalismo, dentre outras; Durante a Pandemia da COVID-19, as discussões anticientíficas experimentaram um pico significativo, impulsionadas por desinformação sobre vacinas; A interseção entre narrativas anticientíficas e crenças esotéricas reforça a ideia de uma suposta verdade alternativa que desafia a ciência; Desde 2022, houve uma evolução nas discussões sobre mudanças climáticas, que passaram a se alinhar com teorias de dominação global; Além disso, a Agenda 2030 da ONU é retratada como parte de uma conspiração global.

### Principais descobertas

➔ Negacionismo climático e anticiência formam uma interseção crítica na propagação da desinformação, criando uma rede complexa que amplifica e reforça mutuamente as narrativas falsas, dificultando o combate à desinformação em setores amplos da população;

➔ Mudanças climáticas são alvo de narrativas apocalípticas, com comunidades de Apocalipse e Sobrevivência direcionando 5.057 links para grupos de negacionismo climático, exacerbando a rejeição da ciência e aumentando a polarização;

- ➔ Comunidades de anticiência funcionam como distribuidores de desinformação para diversas teorias conspiratórias, agindo como *gatekeepers* e distribuindo links de forma equilibrada para outras temáticas como Nova Ordem Mundial e Globalismo;
- ➔ Crescimento significativo de discussões anticientíficas durante a Pandemia da COVID-19, com um pico notável em 2021, impulsionado pela desinformação sobre vacinas e medidas de saúde pública, refletindo a exploração de crises globais para ampliar crenças conspiratórias;
- ➔ Comunidades negacionistas das mudanças climáticas atraem membros de outras teorias conspiratórias, com Globalismo e Nova Ordem Mundial enviando 3.013 e 2.903 links, respectivamente, demonstrando a convergência dessas narrativas;
- ➔ A desinformação sobre mudanças climáticas está fortemente vinculada a disputas geopolíticas, com termos como "Ucrânia" e "Rússia" frequentemente associados a debates climáticos, sugerindo a manipulação geopolítica das crises ambientais;
- ➔ Narrativas anticientíficas integram crenças esotéricas e conspiratórias, com comunidades de Ocultismo e Esoterismo frequentemente ligadas a discussões anticientíficas, reforçando a ideia de uma "verdade alternativa" que desafia a ciência;
- ➔ Discussões sobre mudanças climáticas evoluem para temas relacionados à dominação global, com referências crescentes a controle global desde 2022, conectando-se a teorias da Nova Ordem Mundial e outras conspirações globais;
- ➔ Comunidades anticientíficas promovem desinformação sobre saúde pública com base em teorias conspiratórias, usando a desinformação sobre vacinas como pilar central para desafiar as políticas de saúde e alimentar a desconfiança na ciência;
- ➔ A Agenda 2030 da ONU é retratada como uma conspiração para controle populacional, com comunidades negacionistas associando as iniciativas da ONU a teorias de dominação global, minando os esforços internacionais de sustentabilidade.

## 1. Introdução

Após percorrer milhares de comunidades brasileiras de teorias da conspiração no Telegram, extrair dezenas de milhões de conteúdos dessas comunidades, elaborados e/ou compartilhados por milhões de usuários que as compõem, este estudo tem o objetivo de compor uma série de um total de sete estudos que tratam sobre o fenômeno das teorias da conspiração no Telegram, adotando o Brasil como estudo de caso. Com as abordagens de identificação implementadas, foi possível alcançar um total de 36 comunidades de teorias da conspiração brasileiras no Telegram sobre temáticas de mudanças climáticas e anticiência, estas somando 1.287.938 de conteúdos publicados entre maio de 2018 (primeiras publicações) até agosto de 2024 (realização deste estudo), com 78.252 usuários somados dentre as comunidades. Dessa forma, este estudo tem como objetivo compreender e caracterizar as comunidades sobre temáticas de mudanças climáticas e anticiência presentes nessa rede brasileira de teorias da conspiração identificada no Telegram.

Para tal, será aplicado um método espelhado em todos os sete estudos, mudando apenas o objeto temático de análise e provendo uma replicabilidade de investigação. Assim, abordaremos técnicas para observar as conexões, séries temporais, conteúdos e sobreposições temáticas das comunidades de teorias da conspiração. Além desse estudo, é possível encontrar os seis demais disponibilizados abertamente e originalmente no arXiv da Cornell University. Essa série contou com a atenção redobrada para garantir a integridade dos dados e o respeito à privacidade dos usuários, conforme a legislação brasileira prevê (Lei nº 13.709/2018).

Portanto questiona-se: **como são caracterizadas e articuladas as comunidades de teorias da conspiração brasileiras sobre temáticas de mudanças climáticas e anticiência no Telegram?**

## 2. Materiais e métodos

A metodologia deste estudo está organizada em três subseções, sendo: **2.1. Extração de dados**, que descreve o processo e as ferramentas utilizadas para coletar as informações das comunidades no Telegram; **2.2. Tratamento de dados**, onde são abordados os critérios e métodos aplicados para classificar e anonimizar os dados coletados; e **2.3. Abordagens para análise de dados**, que detalha as técnicas utilizadas para investigar as conexões, séries temporais, conteúdos e sobreposições temáticas das comunidades de teorias da conspiração.

### 2.1. Extração de dados

Este projeto teve início em fevereiro de 2023, com a publicação da primeira versão do TelegramScrap (Silva, 2023), uma ferramenta própria e autoral, de software livre e de código aberto, que faz uso da Application Programming Interface (API) da plataforma Telegram por meio da biblioteca Telethon e organiza ciclos de extração de dados de grupos e canais abertos no Telegram. Ao longo dos meses, a base de dados pôde ser ampliada e qualificada fazendo uso de quatro abordagens de identificação de comunidades de teorias da conspiração:

**(i) Uso de palavras chave:** no início do projeto, foram elencadas palavras-chave para identificação diretamente no buscador de grupos e canais brasileiros no Telegram, tais como "apocalipse", "sobrevivencialismo", "mudanças climáticas", "terra plana", "teoria da conspiração", "globalismo", "nova ordem mundial", "ocultismo", "esoterismo", "curas alternativas", "qAnon", "reptilianos", "revisionismo", "alienígenas", dentre outras. Essa primeira abordagem forneceu algumas comunidades cujos títulos e/ou descrições dos grupos e canais contassem com os termos explícitos relacionados a teorias da conspiração. Contudo, com o tempo foi possível identificar outras diversas comunidades cujas palavras-chave elencadas não davam conta de abarcar, algumas inclusive propositalmente com caracteres trocados para dificultar a busca de quem a quisesse encontrar na rede;

**(ii) Mecanismo de recomendação de canais do Telegram:** com o tempo, foi identificado que canais do Telegram (exceto grupos) contam com uma aba de recomendação chamada de "canais similares", onde o próprio Telegram sugere dez canais que tenham alguma similaridade com o canal que se está observando. A partir desse mecanismo de recomendação do próprio Telegram, foi possível encontrar mais comunidades de teorias da conspiração brasileiras, com estas sendo recomendadas pela própria plataforma;

**(iii) Abordagem de bola de neve para identificação de convites:** após algumas comunidades iniciais serem acumuladas para a extração, foi elaborado um algoritmo próprio autoral de identificação de urls que contivessem “t.me/”, sendo o prefixo de qualquer convite para grupos e canais do Telegram. Acumulando uma frequência de centenas de milhares de links que atendessem a esse critério, foram elencados os endereços únicos e contabilizadas as suas repetições. Dessa forma, foi possível fazer uma investigação de novos grupos e canais brasileiros mencionados nas próprias mensagens dos já investigados, ampliando a rede. Esse processo foi sendo repetido periodicamente, buscando identificar novas comunidades que tivessem identificação com as temáticas de teorias da conspiração no Telegram;

**(iv) Ampliação para tweets publicados no X que mencionassem convites:** com o objetivo de diversificar ainda mais a fonte de comunidades de teorias da conspiração brasileiras no Telegram, foi elaborada uma query de busca própria que pudesse identificar as palavras-chave de temáticas de teorias da conspiração, porém usando como fonte tweets publicados no X (antigo Twitter) e que, além de conter alguma das palavras-chave, contivesse também o “t.me/”, sendo o prefixo de qualquer convite para grupos e canais do Telegram, “https://x.com/search?q=lang%3Apt%20%22t.me%2F%22%20TERMO-DE-BUSCA”.

Com as abordagens de identificação de comunidades de teorias da conspiração implementadas ao longo de meses de investigação e aprimoramento de método, foi possível construir uma base de dados do projeto com um total de 855 comunidades de teorias da conspiração brasileiras no Telegram (considerando as demais temáticas também não incluídas nesse estudo), estas somando 27.227.525 de conteúdos publicados entre maio de 2016 (primeiras publicações) até agosto de 2024 (realização deste estudo), com 2.290.621 usuários somados dentre as comunidades brasileiras. Há de se considerar que este volume de usuários conta com dois elementos, o primeiro é que trata-se de uma variável, pois usuários podem entrar e sair diariamente, portanto este valor representa o registrado na data de extração de publicações da comunidade; além disso, é possível que um mesmo usuário esteja em mais de um grupo e, portanto, é contabilizado mais de uma vez. Nesse sentido, o volume ainda sinaliza ser expressivo, mas pode ser levemente menor quando considerado o volume de cidadãos deduplicados dentro dessas comunidades brasileiras de teorias da conspiração.

### 2.2. Tratamento de dados

Com todos os grupos e canais brasileiros de teorias da conspiração no Telegram extraídos, foi realizada uma classificação manual considerando o título e a descrição da comunidade. Caso houvesse menção explícita no título ou na descrição da comunidade a alguma temática, esta foi classificada entre: (i) “Anticiência”; (ii) “Anti-Woke e Gênero”; (iii) “Antivax”; (iv) “Apocalipse e Sobrevivencialismo”; (v) “Mudanças Climáticas”; (vi) Terra Plana; (vii) “Globalismo”; (viii) “Nova Ordem Mundial”; (ix) “Ocultismo e Esoterismo”; (x) “Off Label e Charlatanismo”; (xi) “QAnon”; (xii) “Reptilianos e Criaturas”; (xiii) “Revisionismo e Discurso de Ódio”; (xiv) “OVNI e Universo”. Caso não houvesse nenhuma menção explícita relacionada às temáticas no título ou na descrição da comunidade, esta foi classificada como (xv) “Conspiração Geral”. Na Tabela a seguir, podemos observar as métricas relacionadas à classificação dessas comunidades de teorias da conspiração no Brasil.

**Tabela 01.** Comunidades de teorias da conspiração no Brasil (métricas até agosto de 2024)

| Categorias | Grupos | Usuários | Publicações | Comentários | Total |
|---|---|---|---|---|---|
| **Anticiência** | 22 | 58.138 | 187.585 | 784.331 | 971.916 |
| **Anti-Woke e Gênero** | 43 | 154.391 | 276.018 | 1.017.412 | 1.293.430 |
| **Antivacinas (*Antivax*)** | 111 | 239.309 | 1.778.587 | 1.965.381 | 3.743.968 |
| **Apocalipse e Sobrevivência** | 33 | 109.266 | 915.584 | 429.476 | 1.345.060 |
| **Mudanças Climáticas** | 14 | 20.114 | 269.203 | 46.819 | 316.022 |
| **Terraplanismo** | 33 | 38.563 | 354.200 | 1.025.039 | 1.379.239 |
| **Conspirações Gerais** | 127 | 498.190 | 2.671.440 | 3.498.492 | 6.169.932 |
| **Globalismo** | 41 | 326.596 | 768.176 | 537.087 | 1.305.263 |
| **Nova Ordem Mundial (NOM)** | 148 | 329.304 | 2.411.003 | 1.077.683 | 3.488.686 |
| **Ocultismo e Esoterismo** | 39 | 82.872 | 927.708 | 2.098.357 | 3.026.065 |
| **Medicamentos *off label*** | 84 | 201.342 | 929.156 | 733.638 | 1.662.794 |
| **QAnon** | 28 | 62.346 | 531.678 | 219.742 | 751.420 |
| **Reptilianos e Criaturas** | 19 | 82.290 | 96.262 | 62.342 | 158.604 |
| **Revisionismo e Ódio** | 66 | 34.380 | 204.453 | 142.266 | 346.719 |
| **OVNI e Universo** | 47 | 58.912 | 862.358 | 406.049 | 1.268.407 |
| **Total** | **855** | **2.296.013** | **13.183.411** | **14.044.114** | **27.227.525** |

Fonte: Elaboração própria (2024).

Com esse volume de dados extraídos, foi possível segmentar para apresentar neste estudo apenas comunidades e conteúdos referentes às temáticas de negacionismo sobre mudanças climáticas e anticiência. Em paralelo, as demais temáticas de comunidades brasileiras de teorias da conspiração também contaram com estudos elaborados para a caracterização da extensão e da dinâmica da rede, estes sendo disponibilizados abertamente e originalmente no arXiv da Cornell University.

Além disso, cabe citar que apenas foram extraídas comunidades abertas, isto é, não apenas identificáveis publicamente, mas também sem necessidade de solicitação para acessar ao conteúdo, estando aberto para todo e qualquer usuário com alguma conta do Telegram sem que este necessite ingressar no grupo ou canal. Além disso, em respeito à legislação brasileira e especialmente da Lei Geral de Proteção de Dados Pessoais (LGPD), ou Lei nº 13.709/2018, que trata do controle da privacidade e do uso/tratamento de dados pessoais, todos os dados extraídos foram anonimizados para a realização de análises e investigações. Dessa forma, nem mesmo a identificação das comunidades é possível por meio deste estudo, estendendo aqui a privacidade do usuário ao anonimizar os seus dados para além da própria comunidade à qual ele se submeteu ao estar em um grupo ou canal público e aberto no Telegram.

### 2.3. Abordagens para análise de dados

Totalizando 36 comunidades selecionadas nas temáticas de mudanças climáticas e anticiência, contendo 1.287.938 publicações e 78.252 usuários somados, quatro abordagens serão utilizadas para investigar as comunidades de teorias da conspiração selecionadas para o escopo do estudo. Tais métricas são detalhadas na Tabela a seguir:

**Tabela 02.** Comunidades selecionadas para análise (métricas até agosto de 2024)

| Categorias | Grupos | Usuários | Publicações | Comentários | Total |
|---|---|---|---|---|---|
| **Anticiência** | 22 | 58.138 | 187.585 | 784.331 | 971.916 |
| **Mudanças Climáticas** | 14 | 20.114 | 269.203 | 46.819 | 316.022 |
| **Total** | **36** | **78.252** | **456.788** | **831.150** | **1.287.938** |

Fonte: Elaboração própria (2024).

**(i) Rede:** com a elaboração de um algoritmo próprio para a identificação de mensagens que contenham o termo de “t.me/” (de convite para entrarem em outras comunidades), propomos apresentar volumes e conexões observadas sobre como **(a)** as comunidades da temática de negacionismo de mudanças climáticas e anticiência circulam convites para que os seus usuários conheçam mais grupos e canais da mesma temática, reforçando os sistemas de crença que comungam; e como **(b)** essas mesmas comunidades circulam convites para que os seus usuários conheçam grupos e canais que tratem de outras teorias da conspiração, distintas de seu propósito explícito. Esta abordagem é interessante para observar se essas comunidades utilizam a si próprias como fonte de legitimação e referência e/ou se embasam-se em demais temáticas de teorias da conspiração, inclusive abrindo portas para que seus usuários conheçam outras conspirações. Além disso, cabe citar o estudo de Rocha *et al.* (2024) em que uma abordagem de identificação de rede também foi aplicada em comunidades do Telegram, porém observando conteúdos similares a partir de um ID gerado para cada mensagem única e suas similares;

**(ii) Séries temporais:** utiliza-se da biblioteca “Pandas” (McKinney, 2010) para organizar os data frames de investigação, observando **(a)** o volume de publicações ao longo dos meses; e **(b)** o volume de engajamento ao longo dos meses, considerando metadados de visualizações, reações e comentários coletados na extração; Além da volumetria, a biblioteca “Plotly” (Plotly Technologies Inc., 2015) viabilizou a representação gráfica dessa variação;

**(iii) Análise de conteúdo:** além da análise geral de palavras com identificação das frequências, são aplicadas séries temporais na variação das palavras mais frequentes ao longo dos semestres — observando entre maio de 2018 (primeiras publicações) até agosto de 2024 (realização deste estudo). E com as bibliotecas “Pandas” (McKinney, 2010) e “WordCloud” (Mueller, 2020), os resultados são apresentados tanto volumetricamente quanto graficamente;

**(iv) Sobreposição de agenda temática:** seguindo a abordagem proposta por Silva & Sátiro (2024) para identificação de sobreposição de agenda temática em comunidades do Telegram, utilizamos o modelo “BERTopic” (Grootendorst, 2020). O BERTopic é um algoritmo de modelagem de tópicos que facilita a geração de representações temáticas a partir de grandes quantidades de textos. Primeiramente, o algoritmo extrai embeddings dos documentos usando modelos transformadores de sentenças, como o “all-MiniLM-L6-v2”. Em seguida, essas embeddings têm sua dimensionalidade reduzida por técnicas como “UMAP”, facilitando o processo de agrupamento. A clusterização é realizada usando “HDBSCAN”, uma técnica baseada em densidade que identifica clusters de diferentes formas e tamanhos, além de detectar outliers. Posteriormente, os documentos são tokenizados e representados em uma estrutura de bag-of-words, que é normalizada (L1) para considerar as diferenças de tamanho entre os clusters. A representação dos tópicos é refinada usando uma versão modificada do “TF-IDF”, chamada “Class-TF-IDF”, que considera a importância das palavras dentro de cada cluster (Grootendorst, 2020). Cabe destacar que, antes de aplicar o BERTopic, realizamos a limpeza da base removendo “stopwords” em português, por meio da biblioteca “NLTK” (Loper & Bird, 2002). Para a modelagem de tópicos, utilizamos o backend “loky” para otimizar o desempenho durante o ajuste e a transformação dos dados.

Em síntese, a metodologia aplicada compreendeu desde a extração de dados com a ferramenta própria autoral TelegramScrap (Silva, 2023), até o tratamento e a análise de dados coletados, utilizando diversas abordagens para identificar e classificar comunidades de teorias da conspiração brasileiras no Telegram. Cada uma das etapas foi cuidadosamente implementada para garantir a integridade dos dados e o respeito à privacidade dos usuários, conforme a legislação brasileira prevê. A seguir, serão apresentados os resultados desses dados, com o intuito de revelar as dinâmicas e as características das comunidades estudadas.

## 3. Resultados

A seguir, os resultados são detalhados na ordem prevista na metodologia, iniciando com a caracterização da rede e suas fontes de legitimação e referência, avançando para as séries temporais, incorporando a análise de conteúdo e concluindo com a identificação de sobreposição de agenda temática dentre as comunidades de teorias da conspiração.

### 3.1. Rede

As figuras a seguir proporcionam uma visão detalhada das dinâmicas que conectam as comunidades em torno das temáticas de anticiência e mudanças climáticas dentro do ecossistema conspiratório. Essas análises revelam como as narrativas anticientíficas e as discussões sobre mudanças climáticas não apenas coexistem, mas se reforçam mutuamente através de redes de interações complexas e convites cruzados. Observa-se que estas redes atuam tanto na atração de novos adeptos quanto na radicalização dos membros existentes, promovendo um ciclo contínuo de disseminação de desinformação.

**Figura 01.** Rede interna entre anticiência e mudanças climáticas

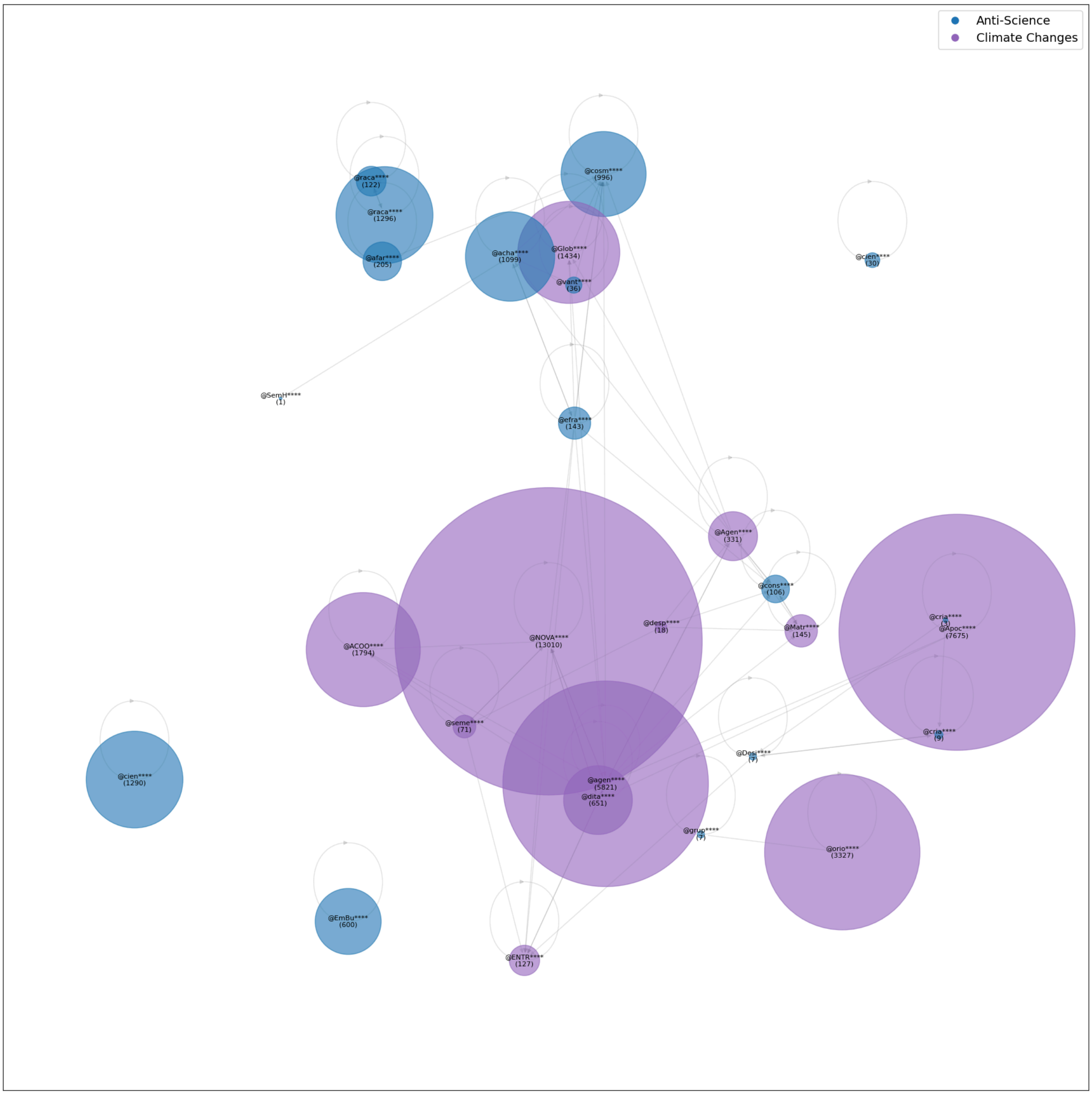

Fonte: Elaboração própria (2024).

Esta figura explora a rede interna que conecta comunidades focadas em anticiência e mudanças climáticas, destacando a interconexão dessas temáticas. As conexões revelam uma forte correlação entre discursos anticientíficos e a negação das mudanças climáticas, sugerindo que essas narrativas são frequentemente reforçadas umas pelas outras. O tamanho e a centralidade dos nós indicam quais comunidades desempenham papéis principais na propagação dessas ideias, evidenciando um padrão de retroalimentação que fortalece a desinformação entre os participantes. Há de se destacar como a temática de negacionismo das mudanças climáticas não apenas ocupa o centro dessa interconexão com conteúdos anticientíficos, mas também interconecta demais comunidades de anticiência, deixando apenas poucas sem conexão direta com a rede de teorias da conspiração.

**Figura 02.** Rede de comunidades que abrem portas para a temática (porta de entrada)

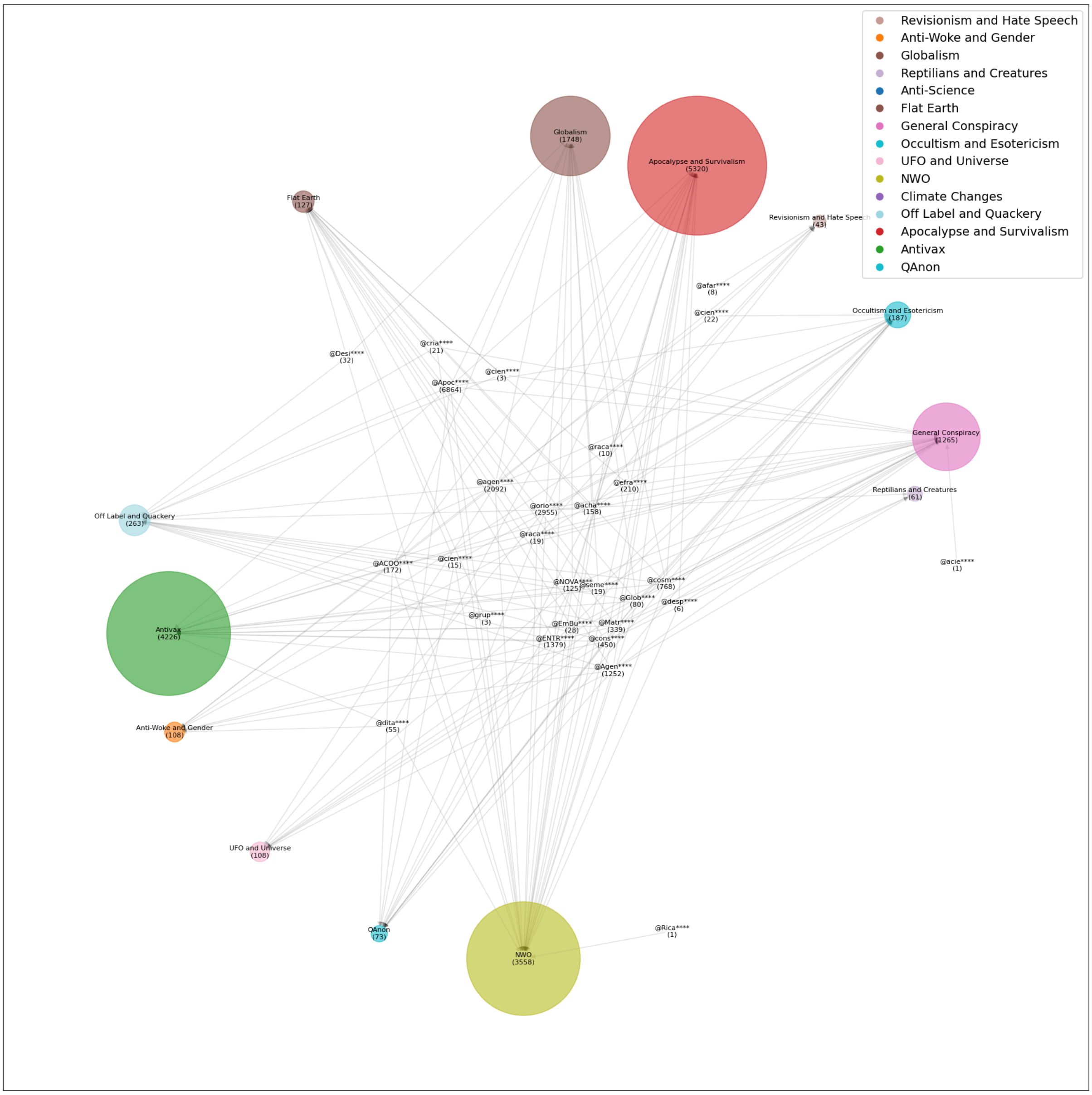

Fonte: Elaboração própria (2024).

Nesta figura, analisamos as comunidades que servem como portas de entrada para discussões sobre anticiência e mudanças climáticas. A disposição dos nós e suas conexões demonstram que grupos maiores e mais influentes, como os que discutem conspirações globais e apocalípticas, atuam como portais iniciais para novos membros. Essas comunidades centrais têm um papel crucial na introdução e disseminação de desinformação, conectando temas variados e apresentando-os como uma visão coesa de realidade distorcida. Embasam-se em alarmismo e inclusive interrelacionam versos bíblicos para tentar desmoralizar o necessário e importante enfrentamento às mudanças climáticas.

**Figura 03.** Rede de comunidades cuja temática abre portas (porta de saída)

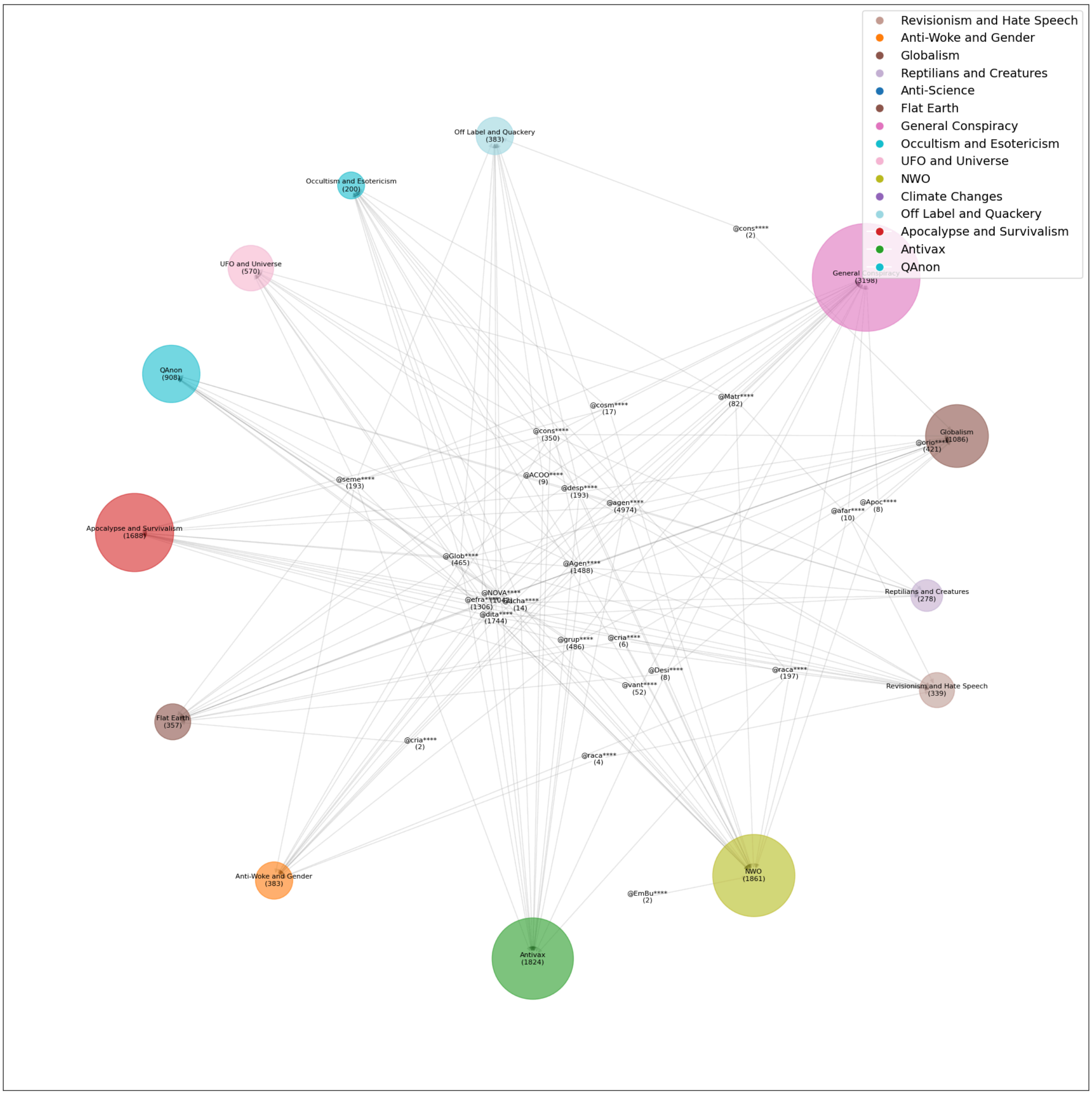

Fonte: Elaboração própria (2024).

Aqui, a figura ilustra como certas comunidades, uma vez exploradas, incentivam a transição dos membros para outros tópicos dentro do ecossistema conspiratório. Grupos que centralizam discussões sobre mudanças climáticas ou anticiência frequentemente guiam os membros para temas correlatos, como Nova Ordem Mundial ou Conspirações Gerais. Esse movimento de saída não apenas diversifica as crenças dos membros, mas também os aprofunda em teorias conspiratórias mais amplas, aumentando o ciclo de radicalização. É interessante notar e cabe como *insight* de achado de como as comunidades de negacionismo de mudanças climáticas e de anticiência parecem atuar como um distribuidor comum para demais temáticas, diferente de outras observações em que há claramente um ou dois temas que recebem mais links, nessas vemos uma distribuição bem hegemônica e equilibrada, significando que essas comunidades negacionistas de mudanças climáticas e anticiência possam atuar como *gatekeepers* de demais teorias da conspiração.

**Figura 04.** Fluxo de links de convites entre comunidades de mudanças climáticas

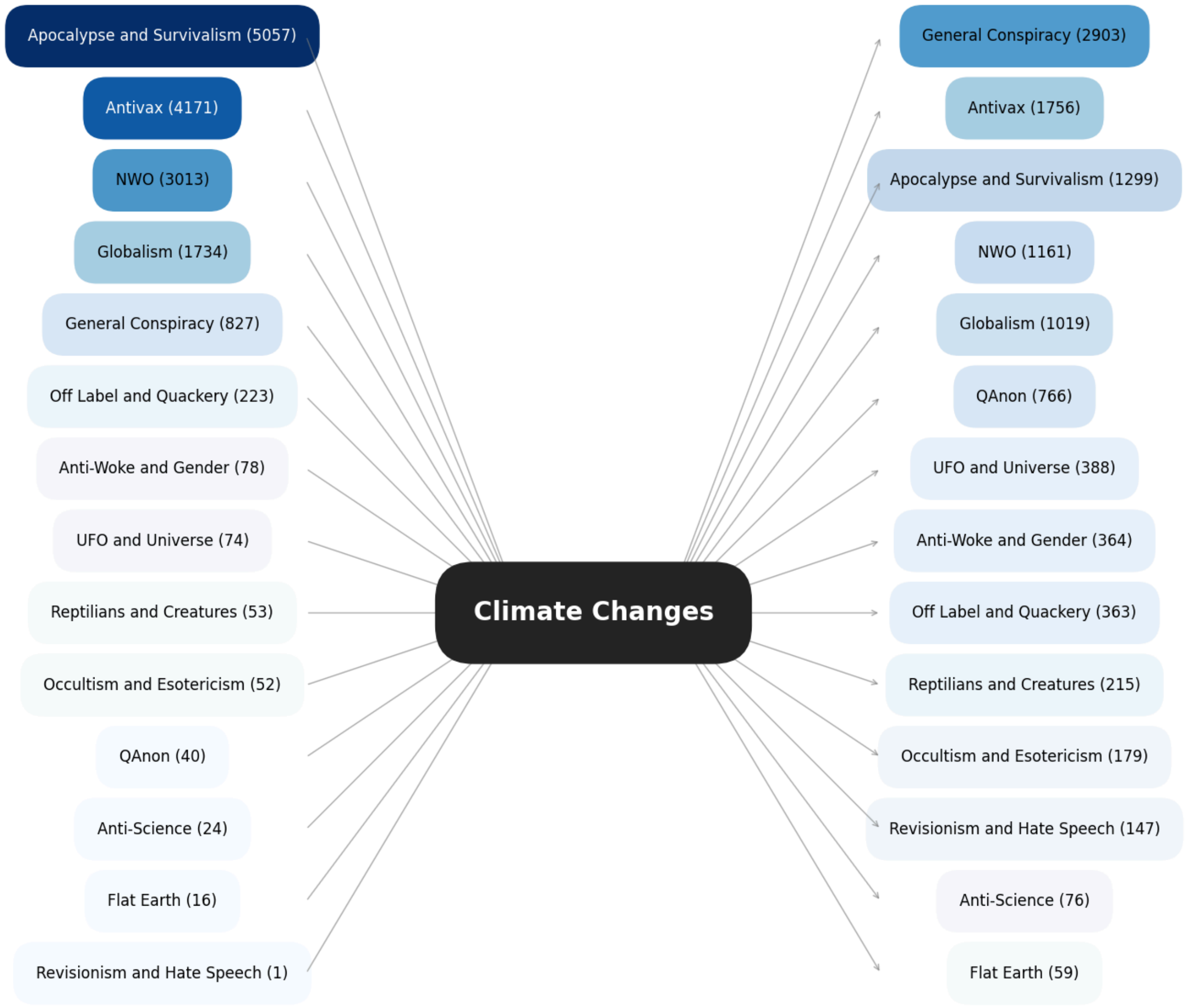

Fonte: Elaboração própria (2024).

A figura demonstra o fluxo de links de convites que partem das comunidades focadas em mudanças climáticas e direcionam seus membros para outras temáticas conspiratórias, assim como o fluxo inverso, onde essas comunidades recebem convites de outras temáticas. À esquerda, observa-se que o maior número de links (5.057) parte das comunidades de Apocalipse e Sobrevivência em direção ao tema de mudanças climáticas, seguido por Antivacinas com 4.171 links e Nova Ordem Mundial com 3.013 links. Esses números indicam que as discussões sobre mudanças climáticas atraem adeptos de narrativas apocalípticas e teorias antivacinas, o que sugere uma interseção significativa entre esses temas. À direita, é notável que as comunidades de mudanças climáticas redirecionam convites de forma bastante equilibrada para uma variedade de outros temas, com destaque para Conspiração Geral (2.903 links), Antivacinas (1.756 links) e Apocalipse e Sobrevivencialismo (1.299 links). Essa distribuição uniforme sugere que as comunidades negacionistas de mudanças climáticas atuam como gatekeepers, distribuindo a desinformação de forma equitativa para várias outras narrativas conspiratórias, fortalecendo sua posição dentro do ecossistema conspiratório.

**Figura 05.** Fluxo de links de convites entre comunidades de anticiência

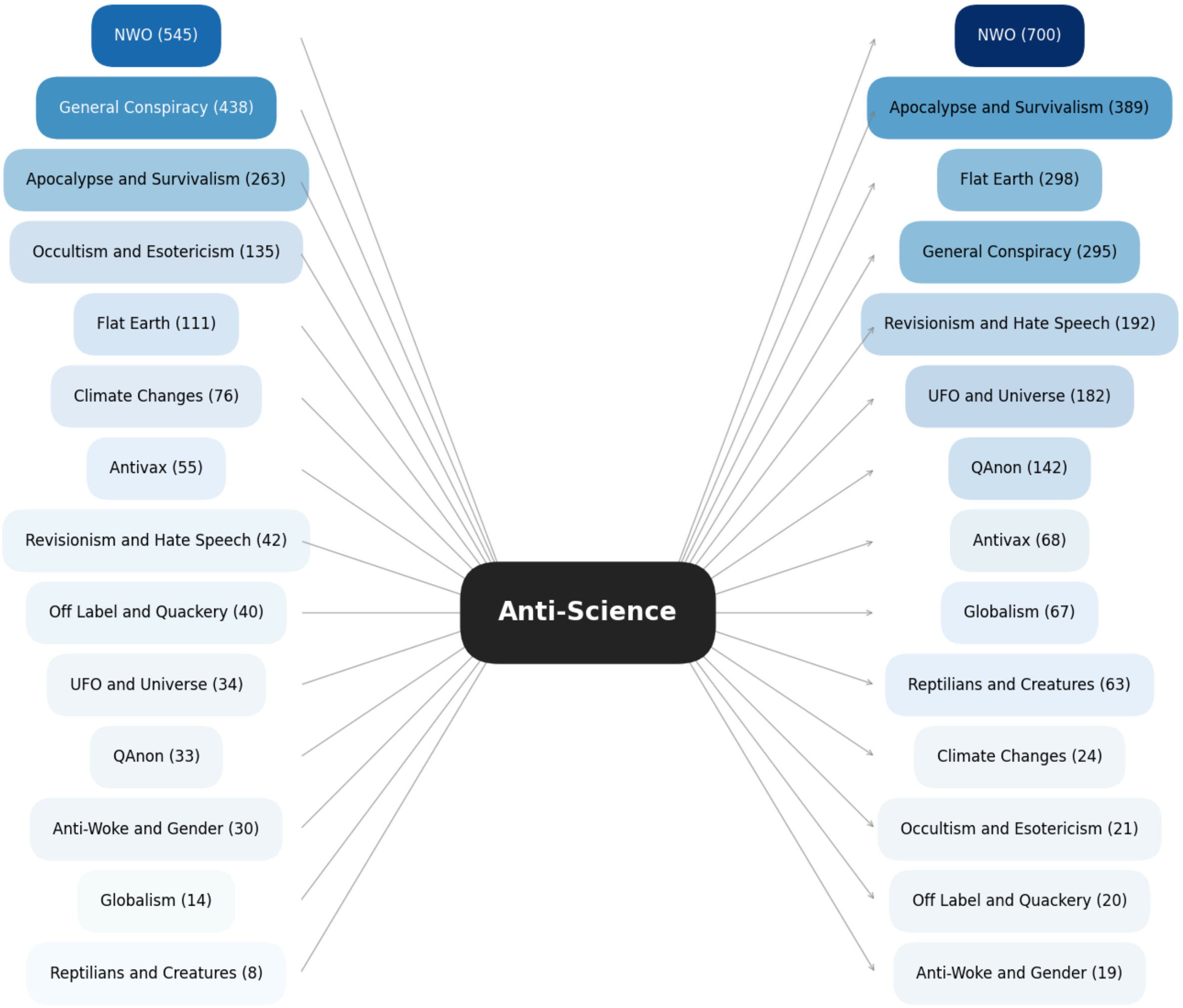

Fonte: Elaboração própria (2024).

Nesta figura, o foco está no fluxo de links de convites que partem das comunidades centradas em anticiência para outras temáticas e vice-versa. À esquerda, a Nova Ordem Mundial (NOM) se destaca com 545 links que conduzem ao tema de anticiência, seguida por Conspiração Geral com 438 links e Apocalipse e Sobrevivencialismo com 263 links. Isso revela que a narrativa anticientífica é atraente para aqueles que já estão imersos em teorias de controle global e sobrevivencialismo apocalíptico. À direita, o tema de anticiência também distribui convites, sendo NOM (700 links), Apocalipse e Sobrevivencialismo (389 links) e Terraplanismo (298 links) os principais destinos. Diferentemente das mudanças climáticas, onde a distribuição de convites é mais equilibrada, aqui vemos uma maior concentração de fluxos para temáticas específicas, o que sugere que as comunidades anticientíficas servem como catalisadores para aprofundar os membros em áreas mais focadas de desinformação, consolidando crenças em teorias bem delineadas como NOM e Terraplanismo.

### 3.2. Séries temporais

As figuras a seguir exploram as séries temporais das comunidades focadas em anticiência e mudanças climáticas ao longo dos anos. Por meio de diferentes representações gráficas, é possível observar como a evolução dessas discussões se comportou ao longo do tempo, revelando padrões de crescimento, declínio e a relação entre esses temas.

**Figura 06.** Gráfico de linhas do período

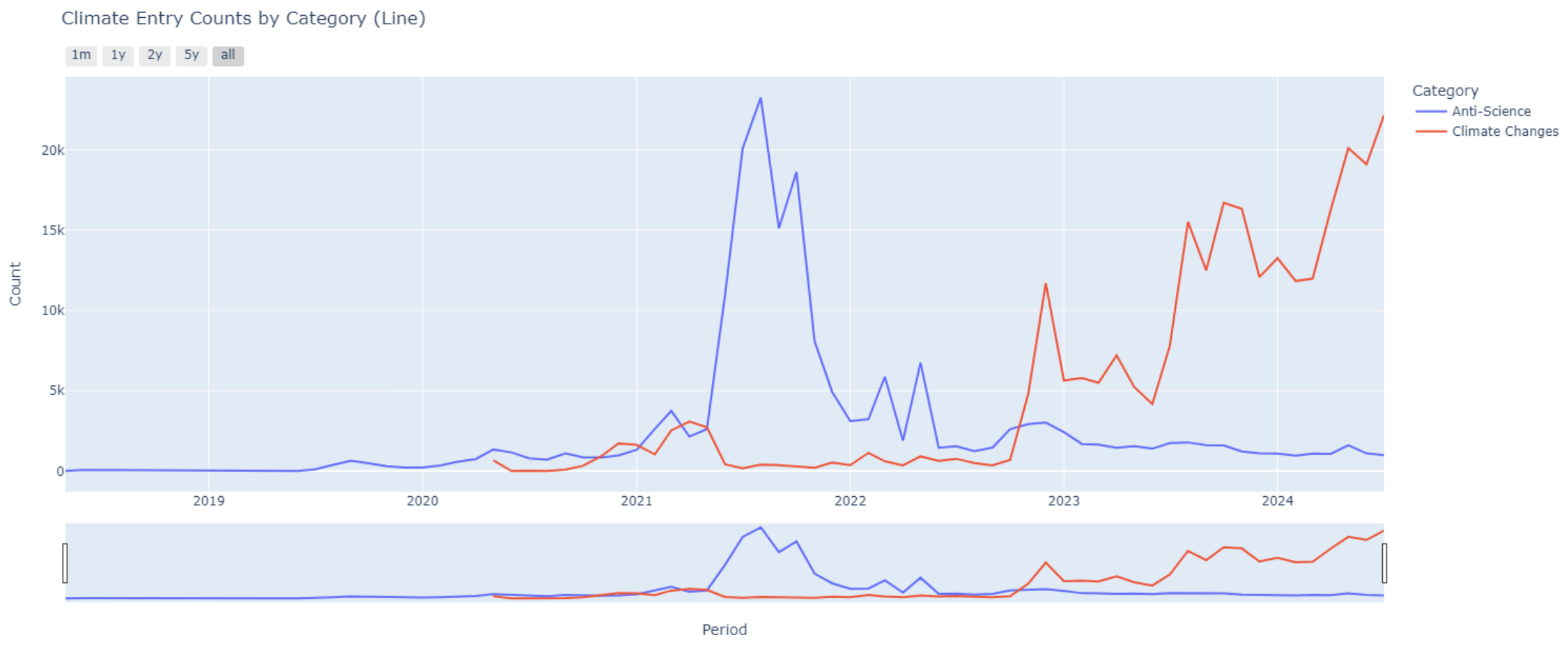

Fonte: Elaboração própria (2024).

O gráfico de linhas do período destaca a evolução temporal das discussões sobre anticiência e mudanças climáticas. Notamos um pico significativo em 2021 para a categoria anticiência, coincidindo com o auge da Pandemia da COVID-19, onde debates sobre a eficácia de vacinas e medidas de saúde pública geraram intensa atividade nessas comunidades. Esse aumento expressivo reflete a polarização e a amplificação de teorias conspiratórias durante momentos de crise global, onde a desinformação encontrou terreno fértil para crescer. Já as discussões sobre mudanças climáticas mostraram um crescimento mais gradual, porém constante, com aumentos notáveis a partir de 2022, quando eventos climáticos extremos e discussões políticas globais, como as conferências climáticas da ONU, impulsionaram as narrativas negacionistas.

[ versão português / english above ]

**Figura 07.** Gráfico de área absoluta do período

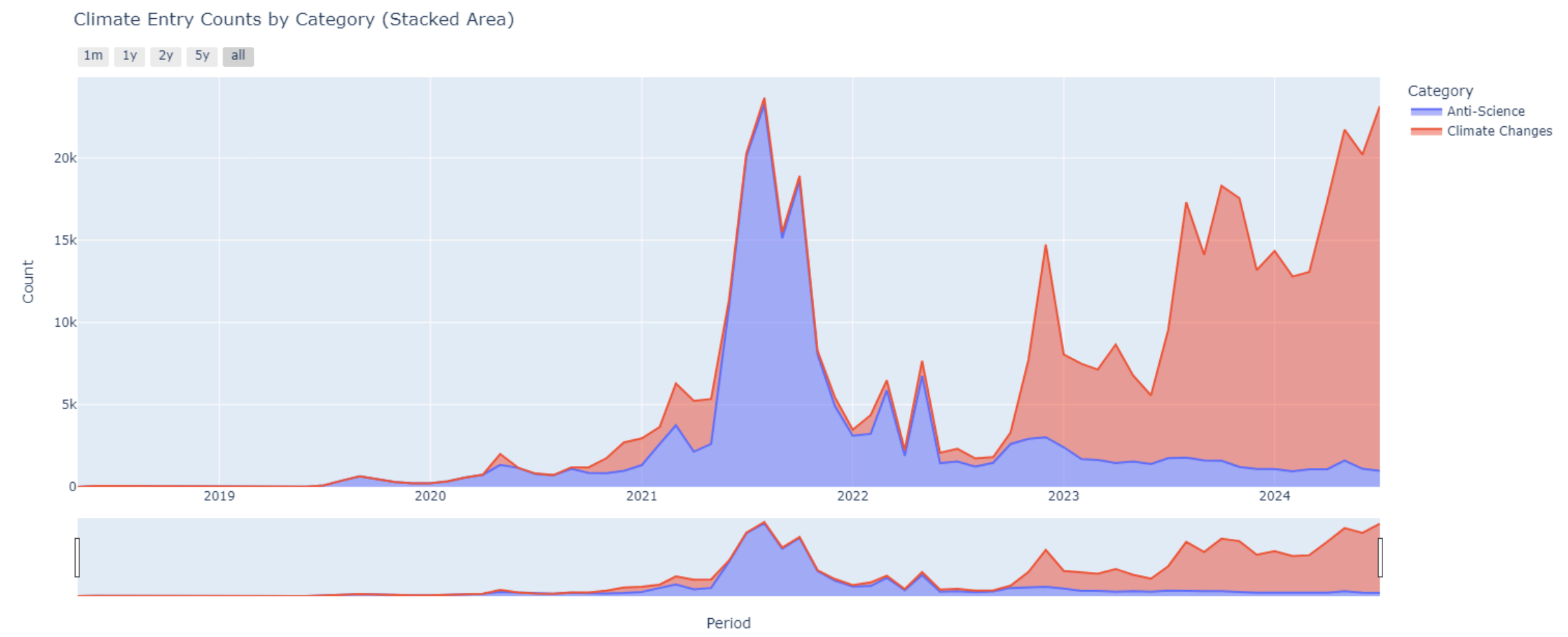

Fonte: Elaboração própria (2024).

No gráfico de área absoluta, a predominância das discussões sobre anticiência em 2021 se torna evidente, ocupando uma grande parte do espaço gráfico durante o pico pandêmico. Isso é seguido por um aumento gradual nas discussões sobre mudanças climáticas, que ganham mais espaço a partir de 2022 e continuam em alta até 2024. Esse padrão sugere que, à medida que a Pandemia foi sendo controlada e novas preocupações globais emergiram, como a crise climática, o foco das comunidades conspiratórias se diversificou. A transição de um domínio absoluto da anticiência para um equilíbrio com as mudanças climáticas reflete a adaptação das narrativas conspiratórias às agendas políticas.

**Figura 08.** Gráfico de área relativa do período

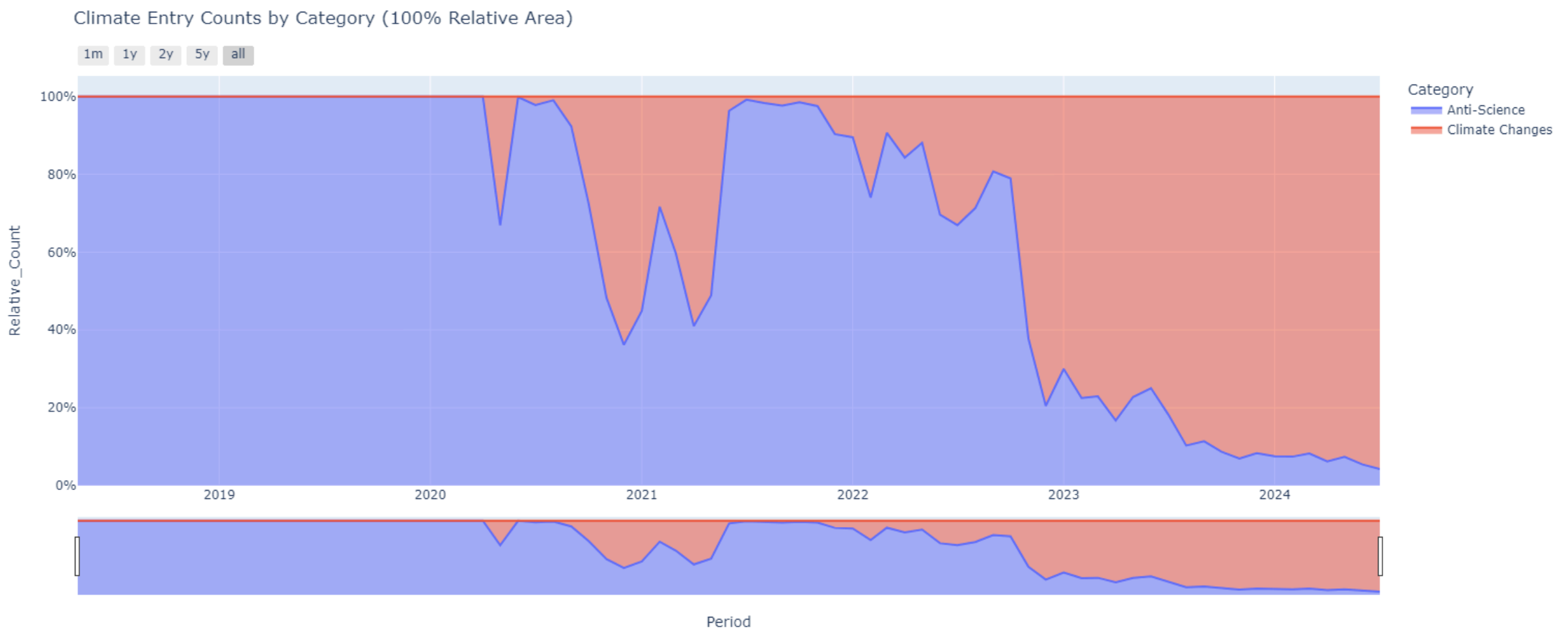

Fonte: Elaboração própria (2024).

O gráfico de área relativa fornece uma perspectiva comparativa entre as duas categorias ao longo do tempo, destacando a variação na dominância de cada tema. Observa-se que, durante 2021, a temática anticiência dominou quase completamente as discussões, refletindo o contexto pandêmico, como já apontado anteriormente. Entretanto, a partir de 2022, há uma inversão gradual, onde as discussões sobre mudanças climáticas começam a ocupar mais espaço relativo, eventualmente superando a anticiência. Esse *shift* sugere que, com o passar do tempo e a redução da urgência pandêmica, as narrativas negacionistas climáticas ganharam proeminência, possivelmente como uma resposta aos crescentes desafios ambientais e às políticas climáticas globais. A mudança na dominância entre as temáticas também reflete a capacidade das comunidades conspiratórias de redirecionar o foco de suas narrativas de acordo com o contexto global.

### 3.3. Análise de conteúdo

As figuras a seguir apresentam uma análise consolidada e temporal das palavras mais recorrentes nas discussões sobre mudanças climáticas e anticiência. Essas nuvens permitem identificar os argumentos e assuntos mais abordados, além das narrativas predominantes, oferecendo uma visão de como certos temas emergem e se transformam ao longo do tempo, refletindo a dinâmica das discussões dentro dessas comunidades.

**Figura 09.** Nuvem de palavras consolidadas de mudanças climáticas e anticiência

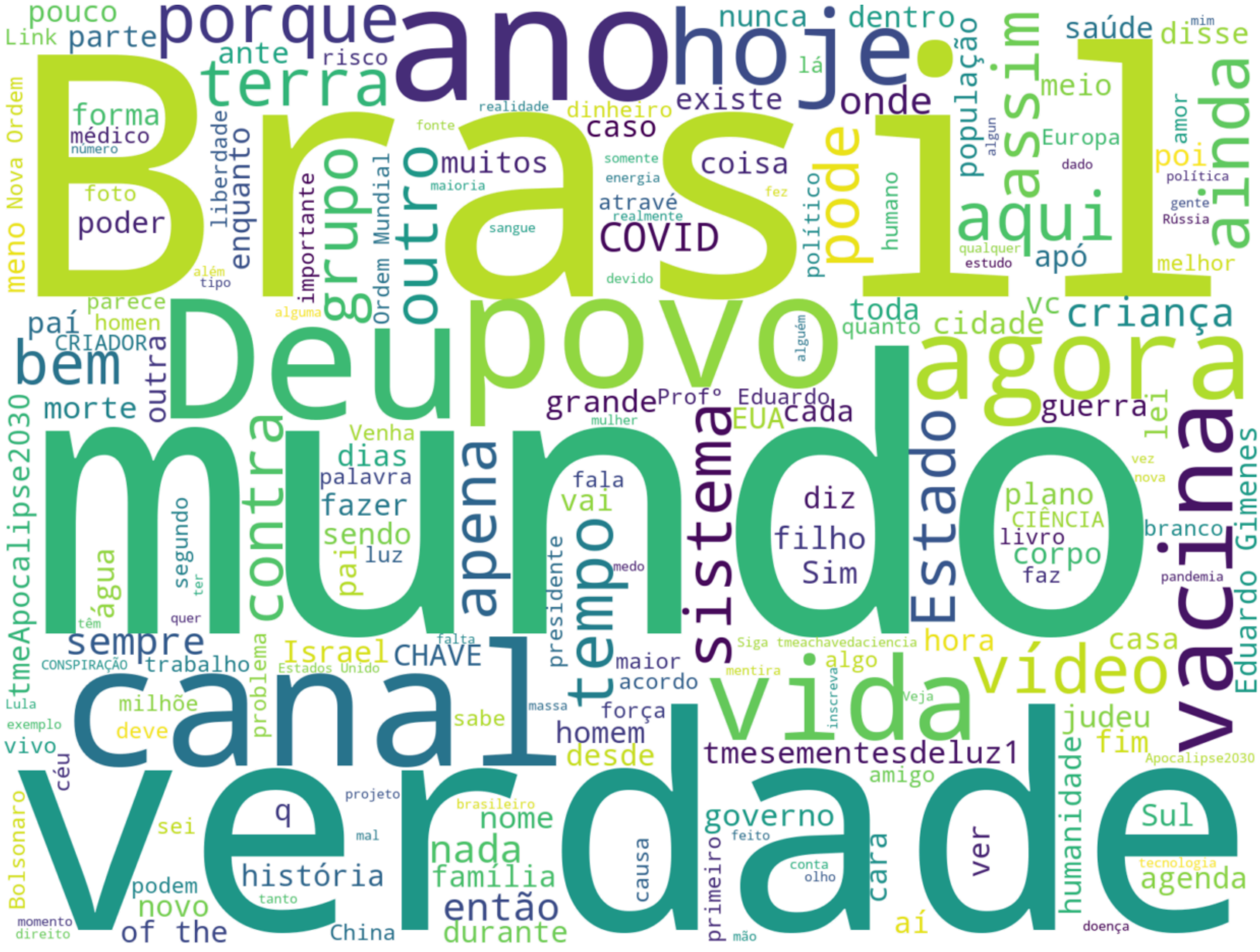

Fonte: Elaboração própria (2024).

A nuvem de palavras consolidada revela a interseção de narrativas entre mudanças climáticas e anticiência. Palavras como “verdade”, “Brasil”, “mundo”, e “povo” destacam-se, e ao observar as publicações podemos notar uma forte presença de discursos que buscam imprimir um viés nacionalista que tentam contestar as narrativas científicas estabelecidas. Termos como “vacina” e “COVID” também aparecem com destaque, refletindo a continuidade das discussões anticientíficas durante e após a Pandemia.

**Quadro 01.** Nuvem de palavras em série temporal de mudanças climáticas

2020

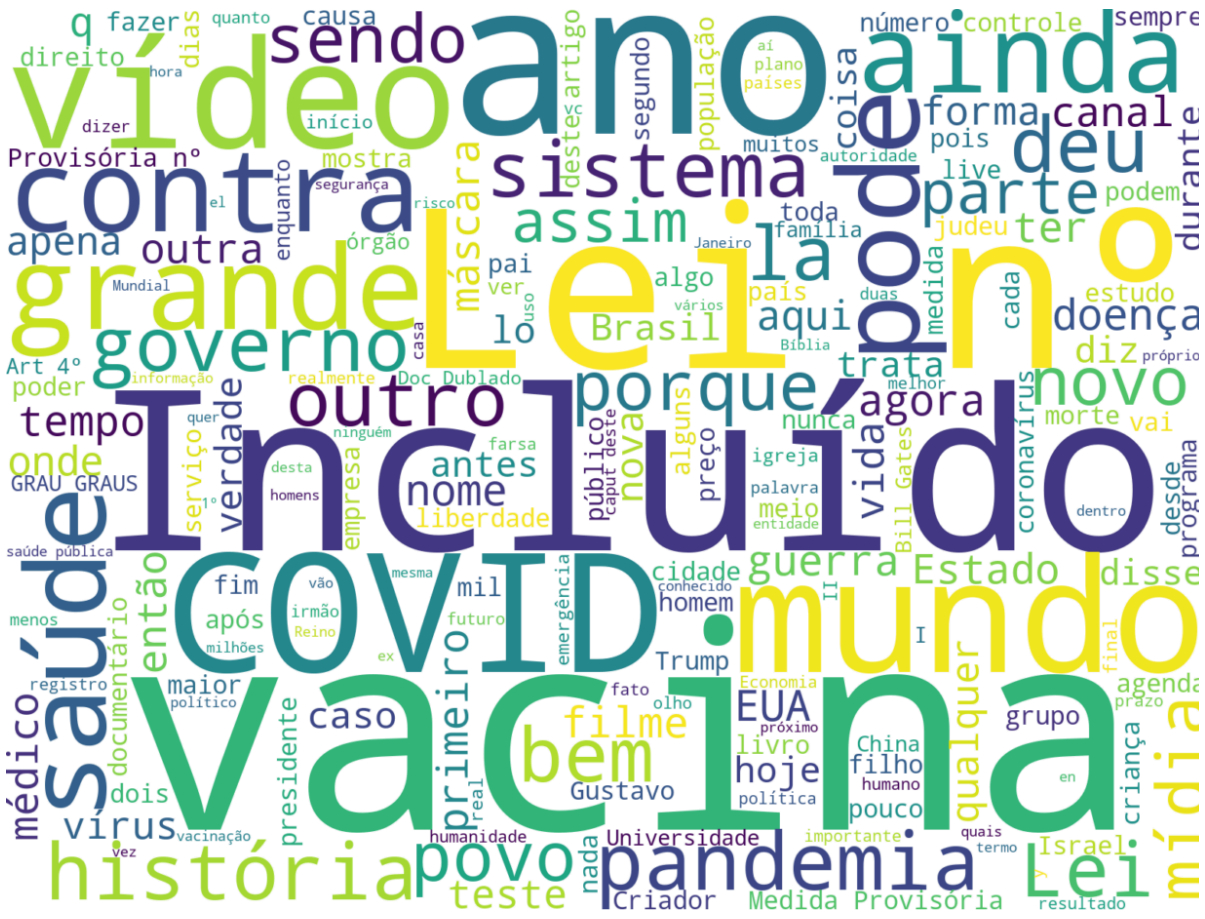

2021

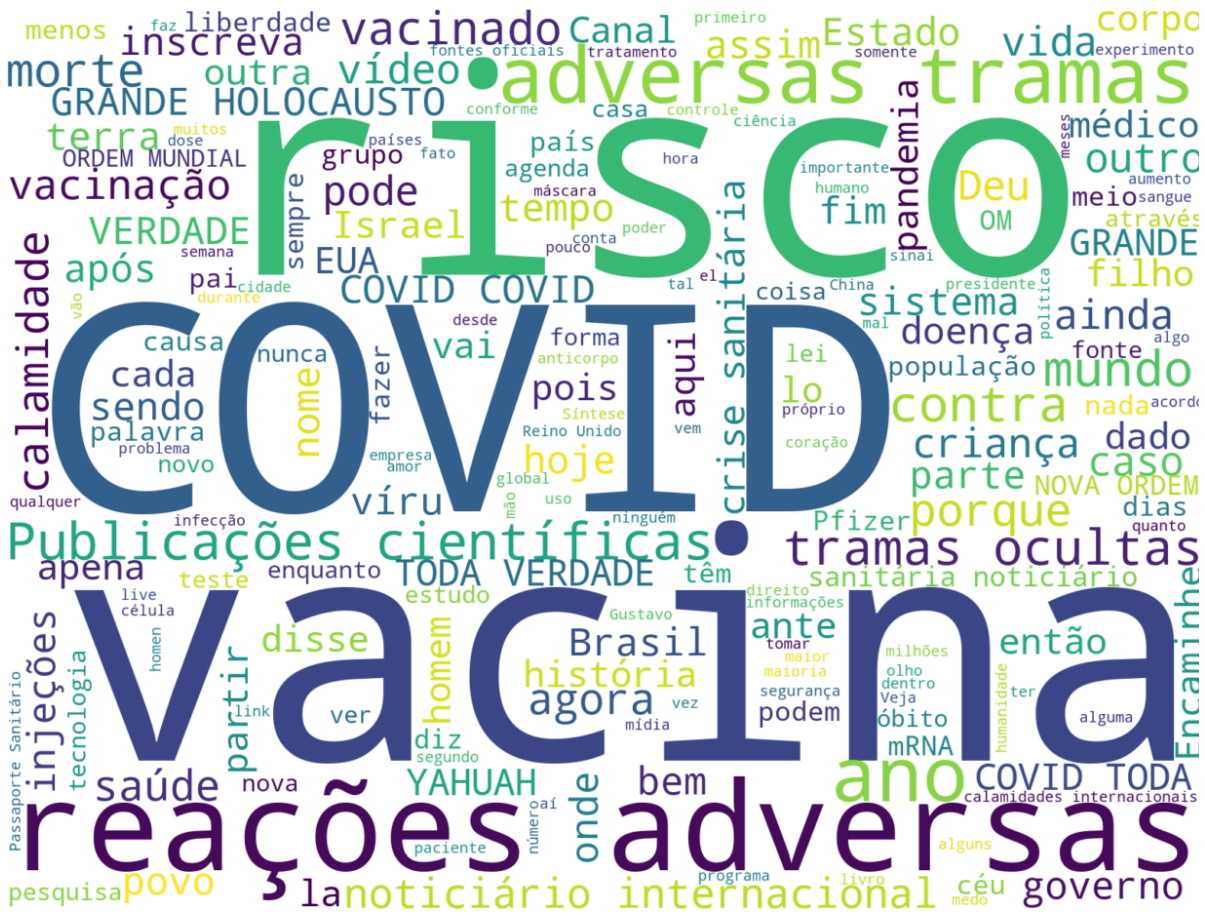

2022

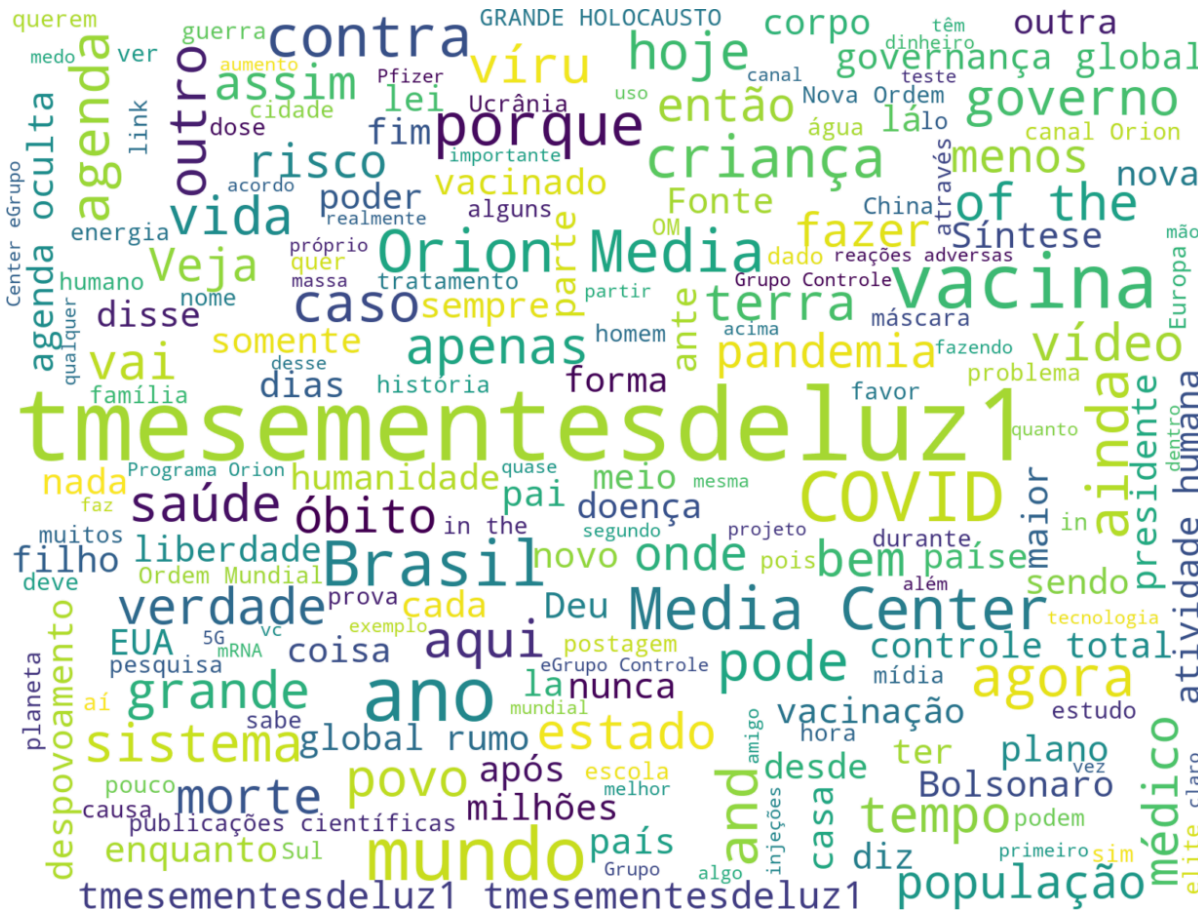

2023

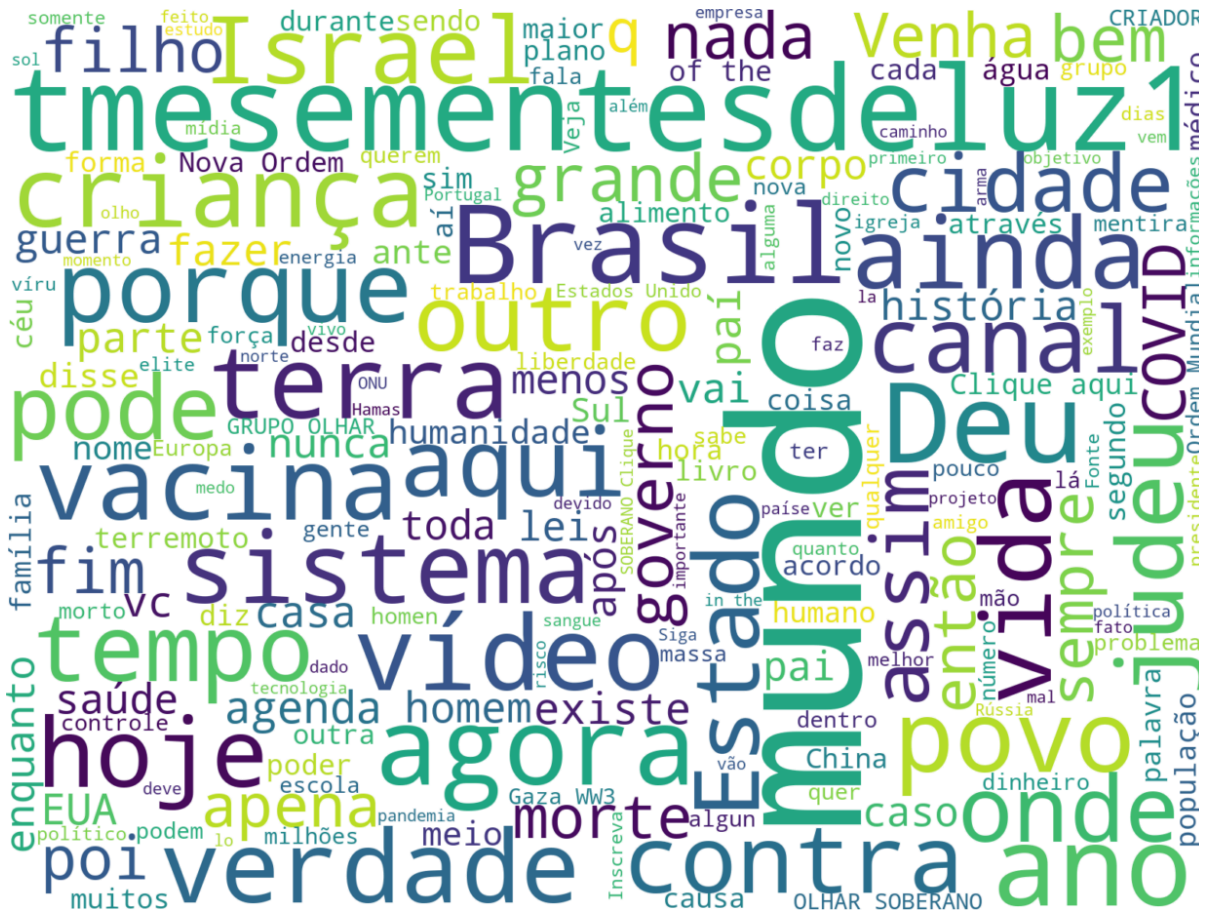

2024

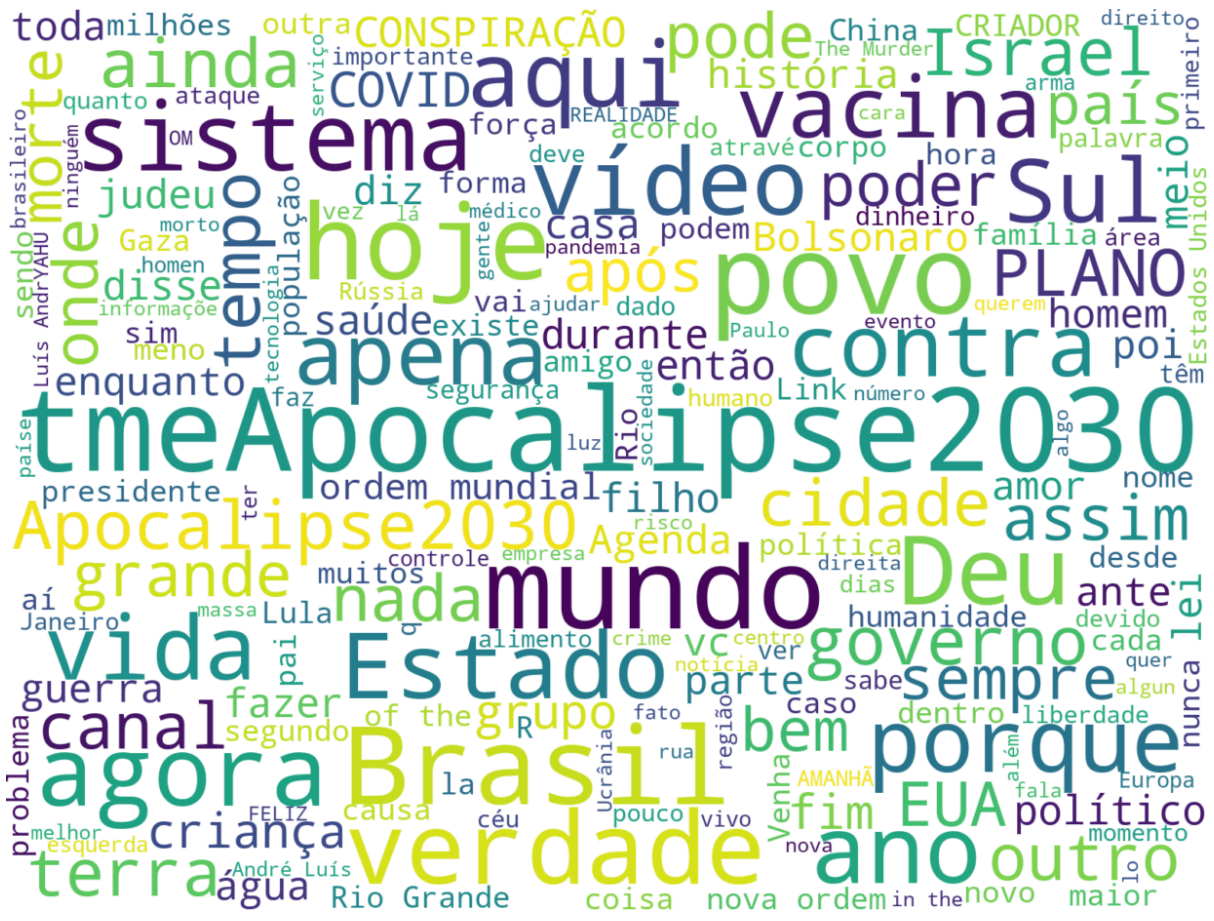

Fonte: Elaboração própria (2024).

O quadro temporal sobre mudanças climáticas mostra uma evolução dos temas discutidos ao longo dos anos. Em 2020, o foco estava claramente nas "vacinas" e "COVID", evidenciando a sobreposição entre as discussões climáticas e pandêmicas. Já em 2021 e 2022, termos como "sistema", "mundo" e "governo" ganharam proeminência, refletindo uma transição para uma discussão mais ampla sobre controle global e desconfiança institucional. Em 2023 e 2024, observamos uma diversificação das narrativas, com a introdução de termos como "Apocalipse2030", além de críticas direcionadas contra a Agenda 2030 da Organização das Nações Unidas (ONU) e os Objetivos e Desenvolvimento Sustentável (ODS), indicando uma conexão crescente entre negacionismo climático e teorias apocalípticas.

**Quadro 02.** Nuvem de palavras em série temporal de anticiência

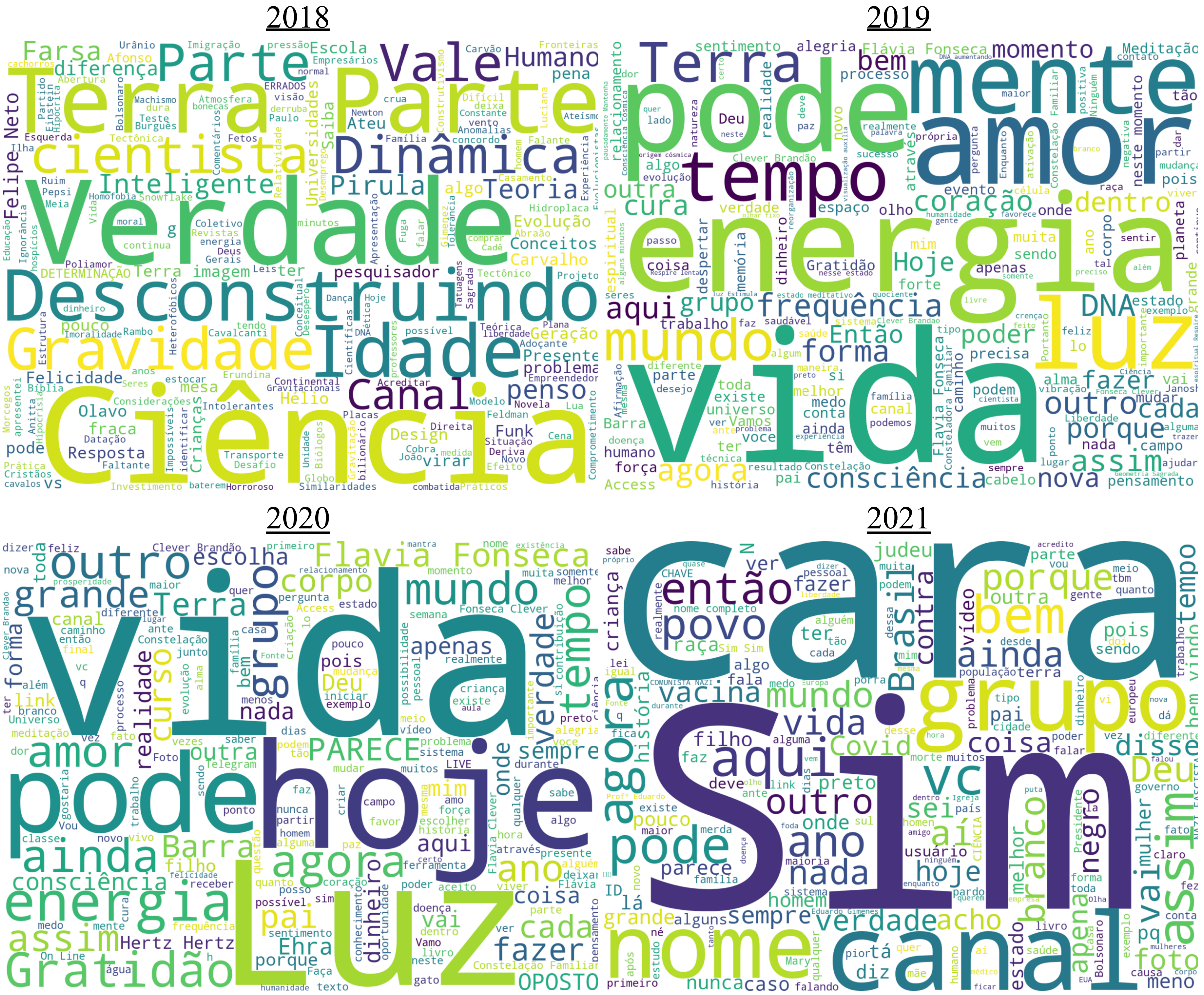

2022

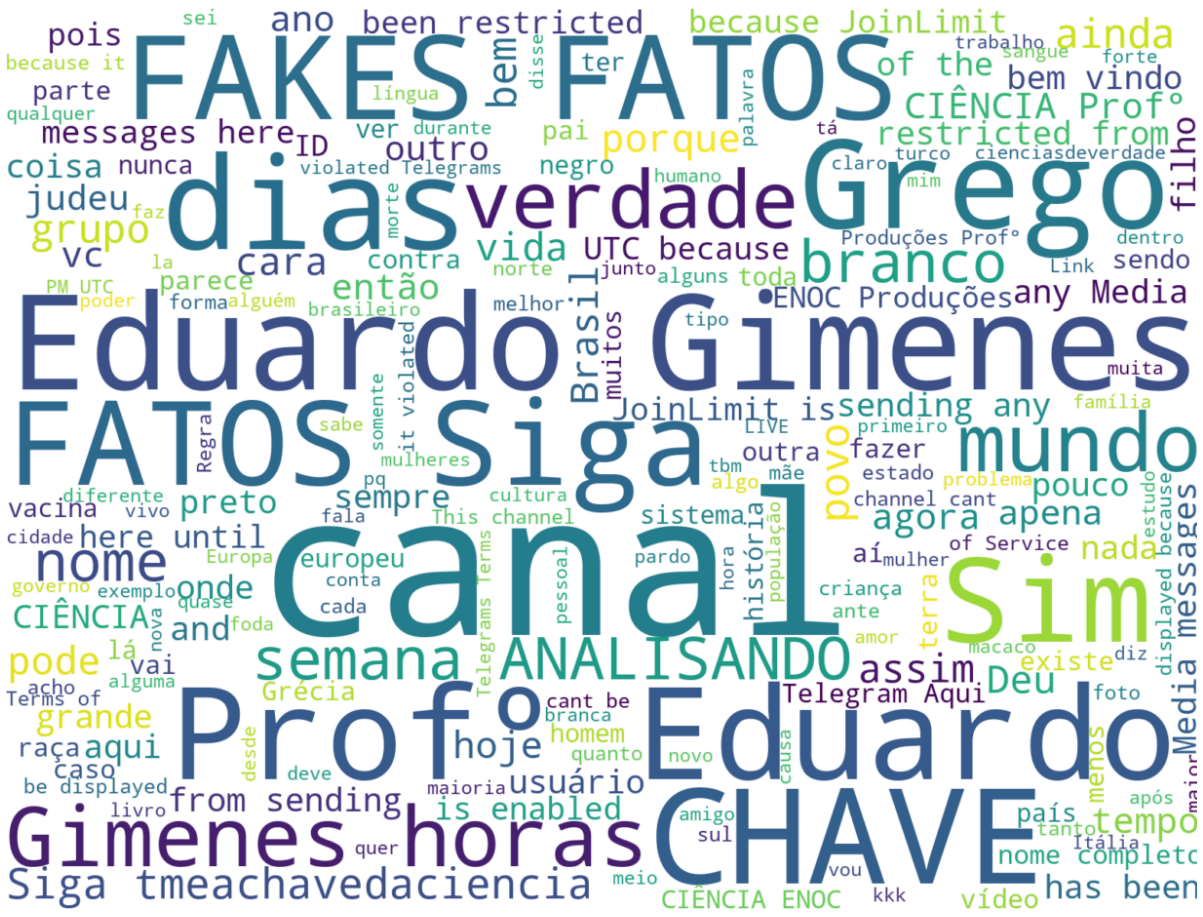

2023

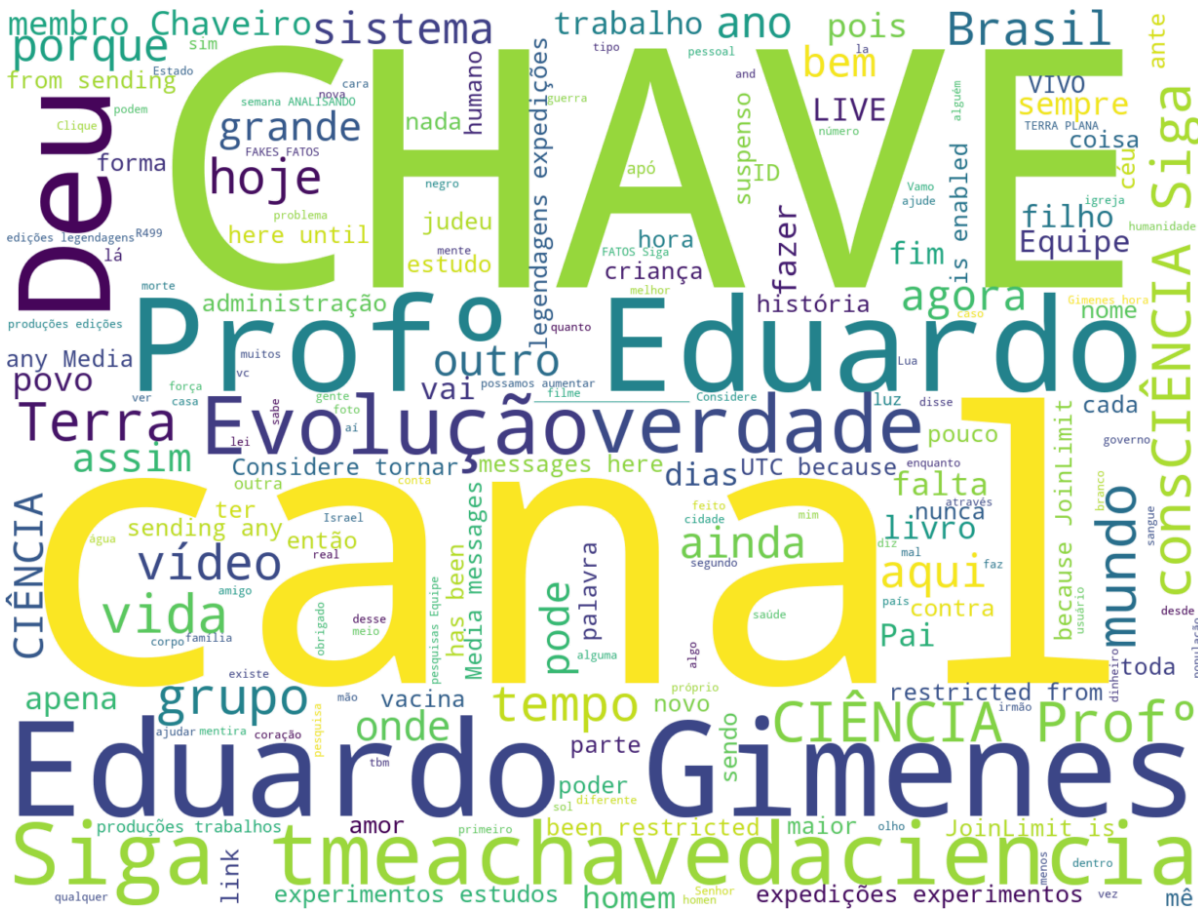

2024

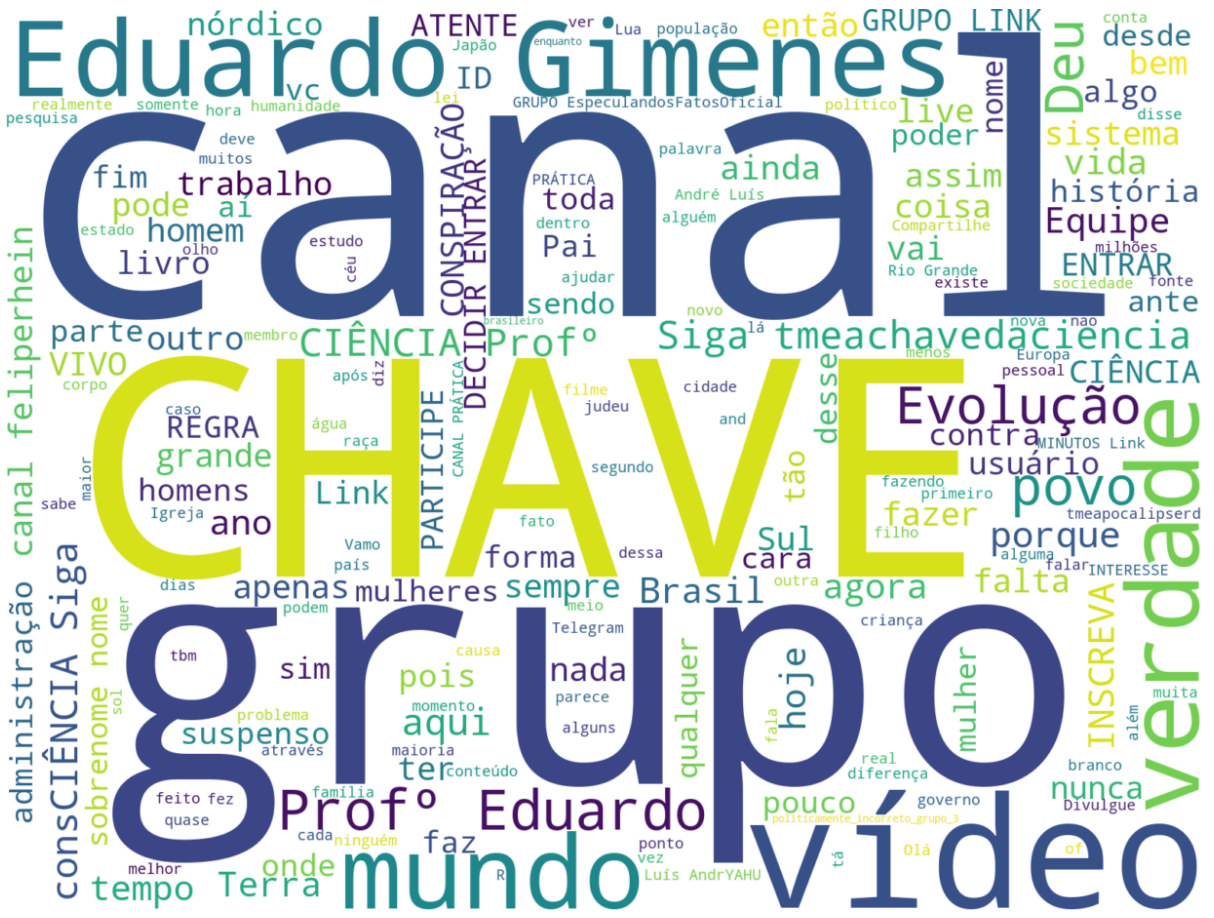

Fonte: Elaboração própria (2024).

Na análise temporal de anticiência, observamos uma consistência no uso de certos termos ao longo dos anos, como “vida”, “canal”, “verdade” e “luz”. Palavras como “Eduardo Gimenes” e “CHAVE” emergem em 2023 e 2024, indicando a influência de figuras ou canais específicos na propagação dessas narrativas. A constância de termos como “vida” e “luz” sugere uma narrativa persistente que mistura esoterismo e ciência, promovendo uma visão de mundo onde a ciência convencional é desacreditada em favor de uma “verdade” alternativa, muitas vezes associada a figuras carismáticas ou influentes dentro dessas comunidades.

### 3.4. Sobreposição de agenda temática

A seguir, são apresentadas figuras que exploram diversas temáticas dentro das comunidades de teorias da conspiração no Telegram, com foco em tópicos como fé e religião, disputas geopolíticas, desinformação vacinal e saúde pública, Nova Ordem Mundial e globalismo, guerras culturais Anti-Woke, e a desinformação em torno da Agenda 2030 da ONU e os ODS. Essas figuras ilustram como diferentes narrativas se interligam, revelando a complexidade das discussões que permeiam essas comunidades. A análise visual dessas conexões permite identificar padrões de interação e sobreposição temática, evidenciando como crenças conspiratórias são reforçadas e disseminadas através de um ecossistema que mistura ciência, religião, política e identidade. Esses gráficos oferecem uma compreensão aprofundada de como temas aparentemente distintos se entrelaçam para criar uma rede coesa de desinformação e resistência a informações científicas e institucionais.

**Figura 10.** Temáticas de fé e religião

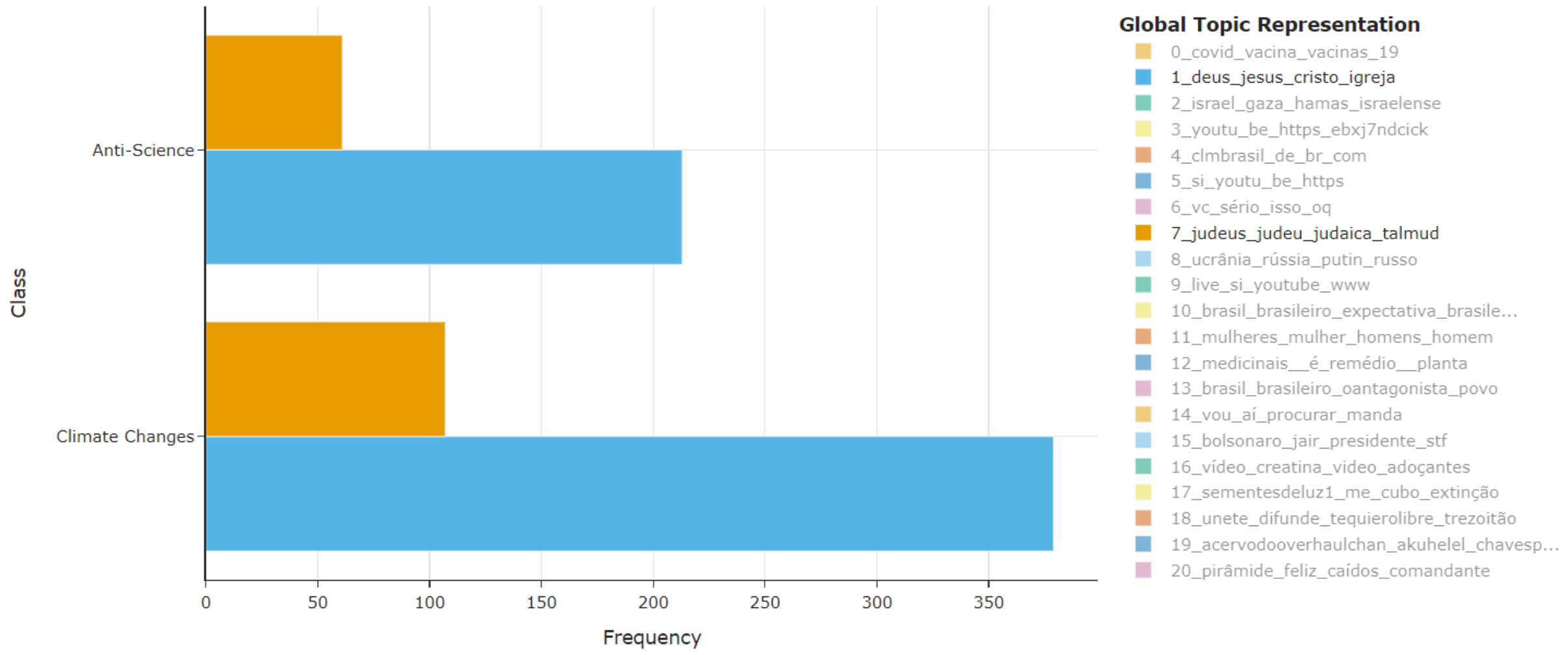

Fonte: Elaboração própria (2024).

Este gráfico ilustra como as temáticas de fé e religião estão interligadas com as discussões sobre mudanças climáticas e anticiência nas comunidades de teorias da conspiração no Telegram. Observa-se uma forte presença de tópicos relacionados a “Deus”, “Jesus”, e “Cristo”, particularmente em discussões sobre anticiência. Em muitos casos observados, versos bíblicos são utilizados para supostamente sustentar argumentos conspiratórios. Essas conexões sugerem que a fé é frequentemente mobilizada para desafiar as narrativas científicas, promovendo uma visão de mundo em que a ciência é vista como contrária ou ameaçadora. Por outro lado, as mudanças climáticas aparecem associadas a tópicos como “Israel” e “judaísmo”, refletindo uma sobreposição com teorias conspiratórias que ligam crises globais a supostos planos de dominação religiosa. Essa interação entre religião e ciência nas discussões revela a profundidade com que crenças pessoais e teorias de conspiração se entrelaçam, fortalecendo a resistência contra informações científicas.

**Figura 11.** Temáticas de disputas geopolíticas

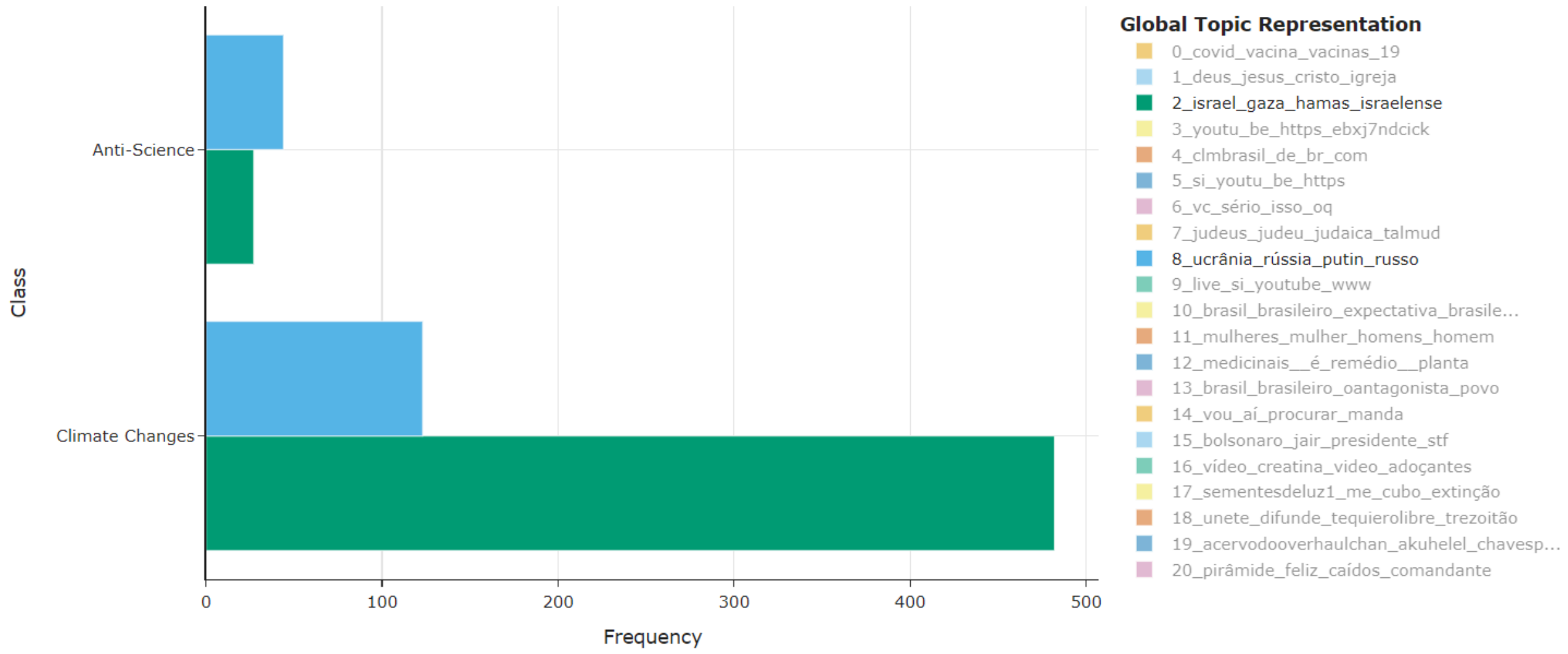

Fonte: Elaboração própria (2024).

Neste gráfico, destacam-se as temáticas de disputas geopolíticas que permeiam as discussões sobre mudanças climáticas e anticiência. As mudanças climáticas, em particular, aparecem fortemente associadas a tópicos como "Ucrânia", "Rússia", e "Putin", evidenciando a narrativa de que eventos climáticos extremos e políticas ambientais são manipulados ou usados como pretexto em conflitos internacionais. A presença significativa desses temas sugere que as discussões climáticas são frequentemente contextualizadas dentro de um quadro geopolítico mais amplo, onde as mudanças climáticas são vistas não apenas como um fenômeno natural, mas como uma ferramenta de controle global. Isso reflete uma estratégia dessas comunidades de inserir questões científicas em debates sobre poder e soberania, ampliando a desconfiança em relação às agendas internacionais e políticas climáticas globais.

**Figura 12.** Temáticas de desinformação vacinal e saúde pública

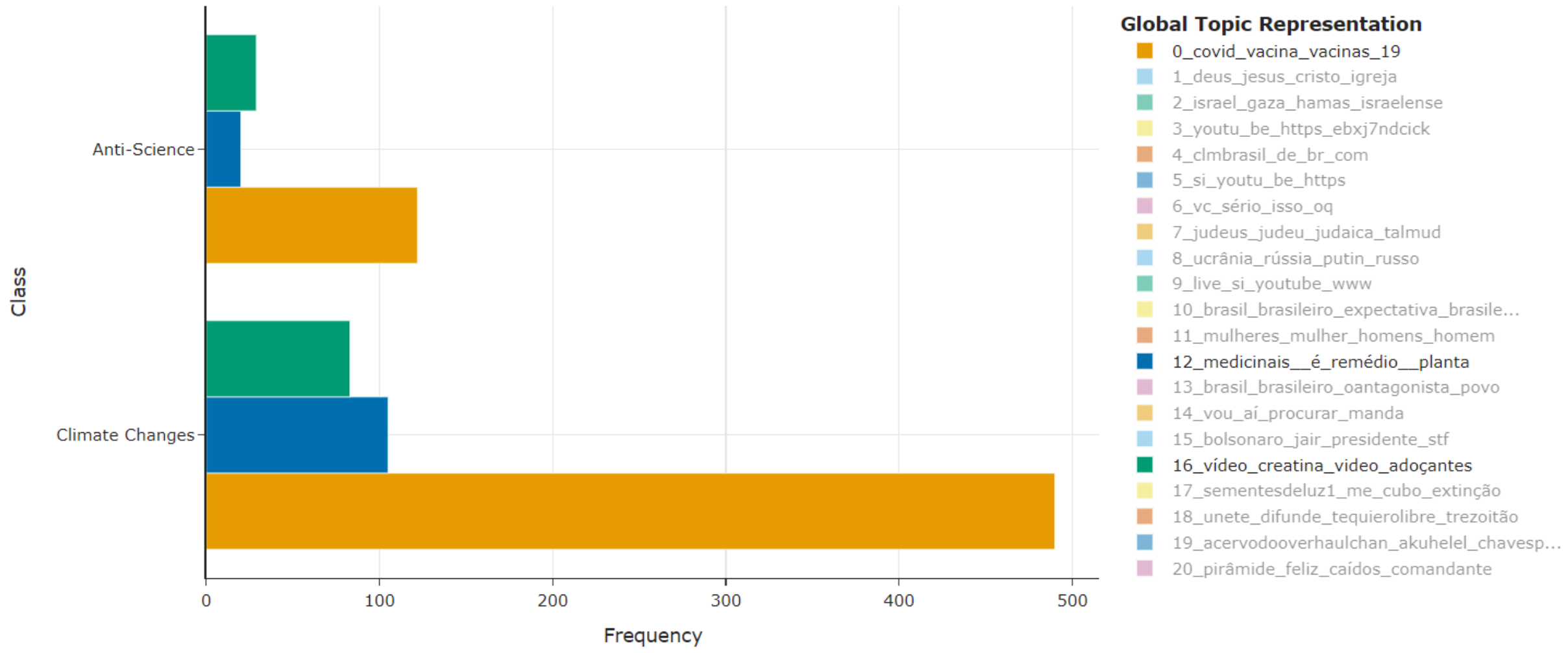

Fonte: Elaboração própria (2024).

Este gráfico examina como as temáticas de desinformação vacinal e saúde pública se entrelaçam com mudanças climáticas e anticiência. A anticiência, em particular, está intimamente ligada a narrativas de desinformação vacinal, com tópicos como "vacinas" e "COVID" dominando as discussões. Isso reforça a ideia de que as comunidades que negam a ciência utilizam a desinformação sobre vacinas como um pilar central em suas campanhas de desinformação. Já as mudanças climáticas estão conectadas a temas de saúde pública através de discussões que questionam a validade das políticas de mitigação e adaptação ao clima, sugerindo que estas políticas fazem parte de um plano maior de controle populacional. A interseção desses tópicos reflete uma estratégia de unificar formas de desinformação em uma narrativa coerente que desafia tanto a ciência quanto às políticas de saúde pública.

**Figura 13.** Temáticas de conspirações de Nova Ordem Mundial e Globalismo

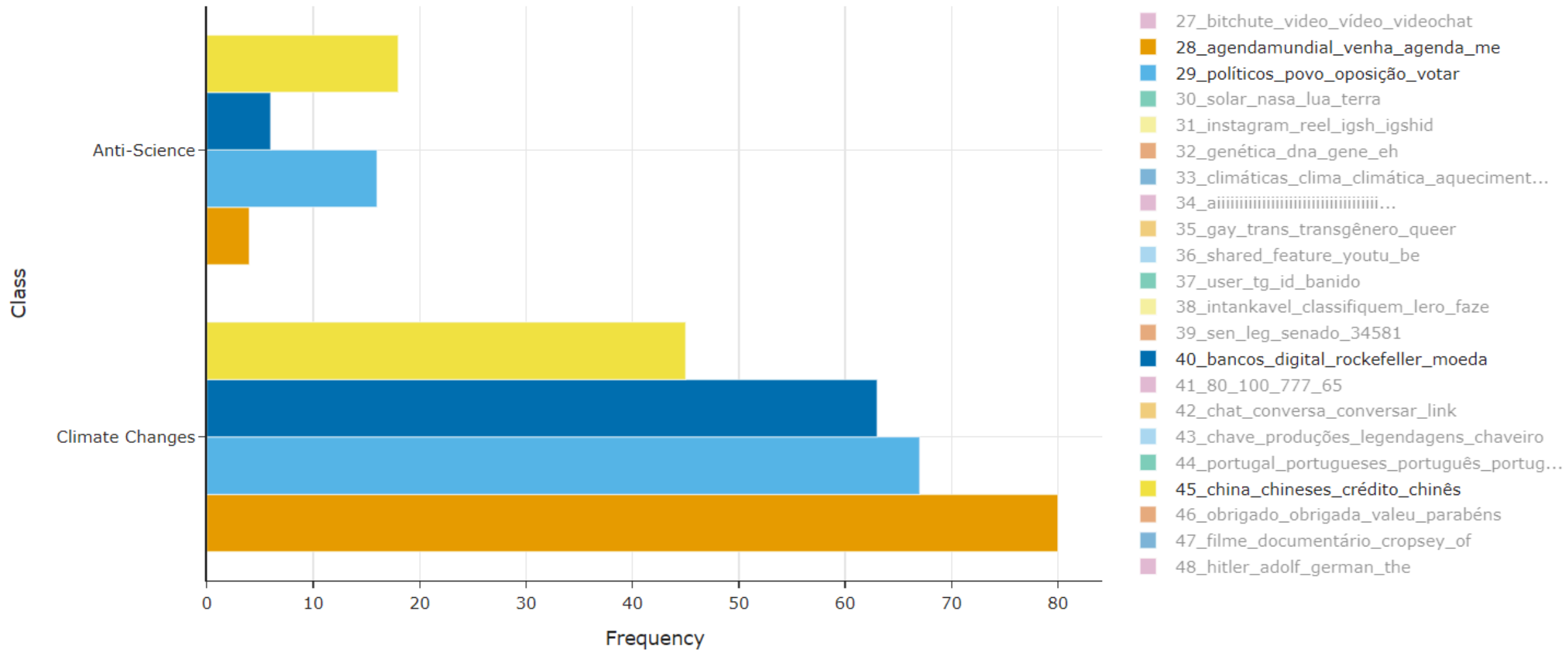

Fonte: Elaboração própria (2024).

O gráfico revela como as temáticas relacionadas à Nova Ordem Mundial e ao Globalismo estão presentes nas discussões sobre mudanças climáticas e anticiência. As mudanças climáticas são frequentemente associadas a tópicos como "agenda mundial", "global" e "moeda digital", indicando que os eventos climáticos são vistos como parte de um plano global para estabelecer um governo único ou controle financeiro global. A anticiência também está ligada a essas teorias, mas de forma menos pronunciada, sugerindo que enquanto a desinformação científica é central, ela é complementada por narrativas mais amplas de conspiração global. Essa sobreposição de temas indica que as discussões climáticas e anticientíficas são moldadas por um medo comum de perda de soberania e controle por elites globais, ampliando o alcance e a persistência dessas narrativas conspiratórias.

**Figura 14.** Temáticas de guerras culturais Anti-Woke

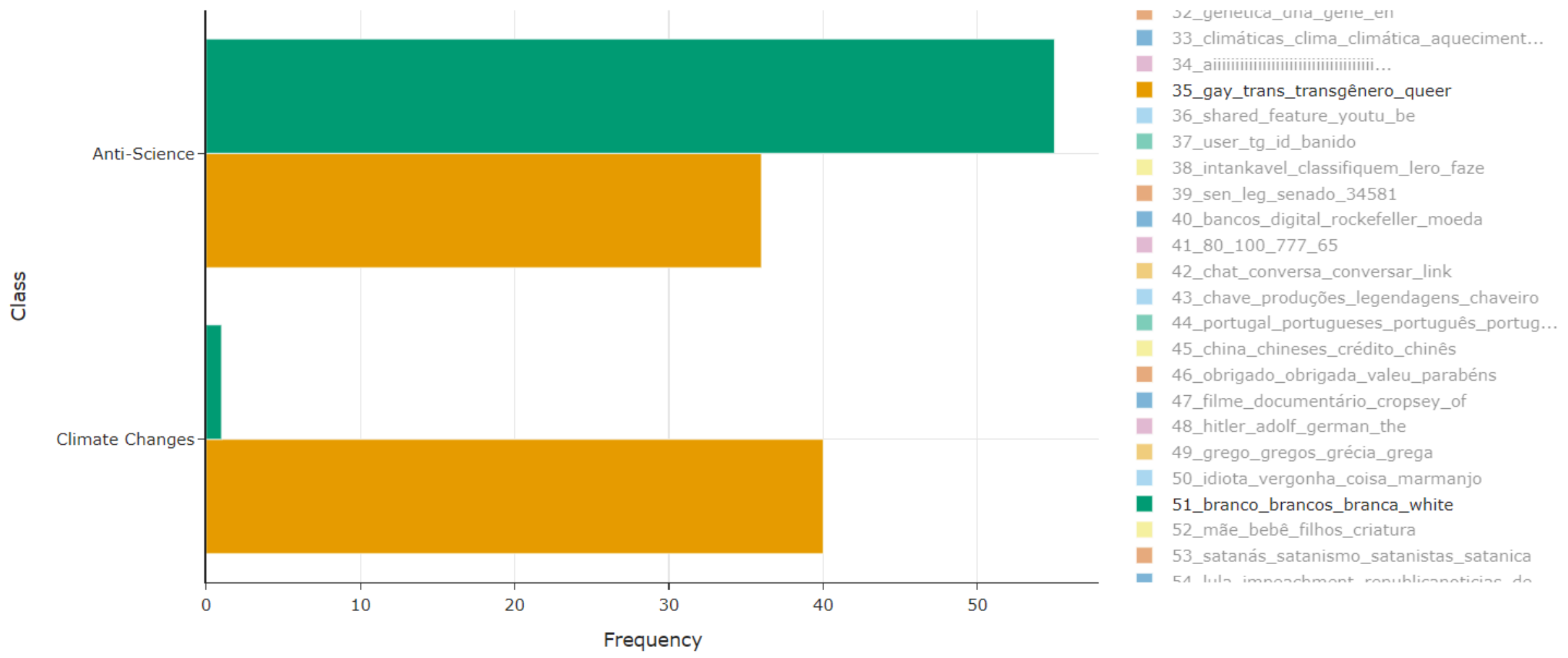

Fonte: Elaboração própria (2024).

Neste gráfico, observamos a intersecção entre as discussões de mudanças climáticas, anticiência, e as guerras culturais Anti-Woke. A anticiência, em particular, está fortemente associada a tópicos como “transgênero”, “gay”, “queer” e “brancos”, refletindo a narrativa de que a ciência moderna e a luta por justiça social são parte de um movimento ideológico que ameaça os valores tradicionais. As mudanças climáticas também aparecem relacionadas a essas guerras culturais, mas de maneira menos intensa, sugerindo que enquanto as discussões climáticas podem ser mobilizadas dentro de debates sobre identidade e cultura, a anticiência é mais diretamente ligada a essas batalhas ideológicas. Mais do que isso, as comunidades de teorias da conspiração sugerem que as mudanças climáticas seriam um plano de dominação que busca masculinizar mulheres e feminilizar homens, por uma suposta agenda anti-Woke. Essa interação destaca como as narrativas conspiratórias utilizam temas culturais e identitários para reforçar a resistência contra mudanças sociais e científicas.

**Figura 15.** Temáticas de desinformação sobre a Agenda 2030 da ONU e os ODS

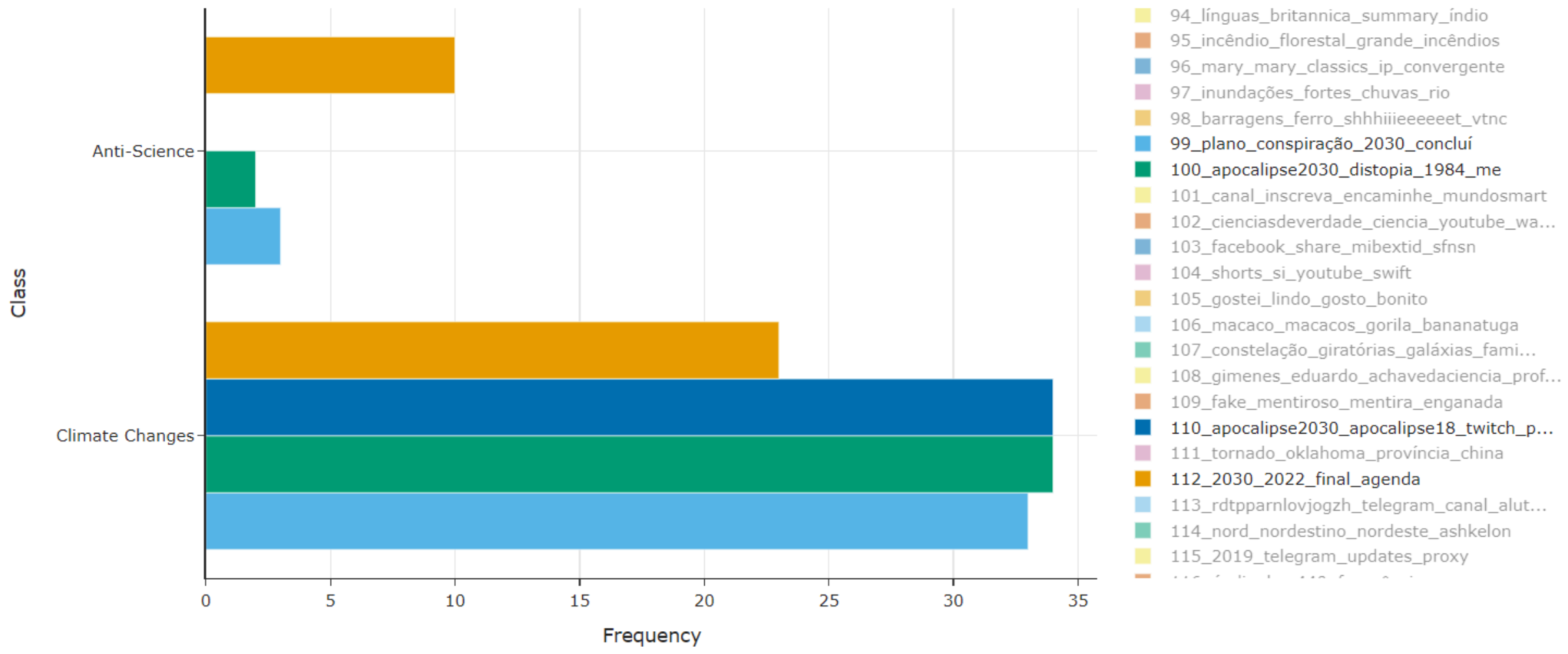

Fonte: Elaboração própria (2024).

Este gráfico explora como as temáticas de desinformação sobre a Agenda 2030 da ONU e os Objetivos de Desenvolvimento Sustentável (ODS) se relacionam com mudanças climáticas e anticiência. As mudanças climáticas, particularmente, estão fortemente ligadas a teorias conspiratórias que descrevem a Agenda 2030 como um plano para supostamente instaurar uma ditadura global ou uma crise fabricada para justificar o controle populacional. Tópicos como "Apocalipse 2030" e "plano conspiração" aparecem com frequência, sugerindo que essas comunidades veem as políticas climáticas internacionais como parte de uma estratégia globalista. A anticiência também está presente, indicando que as discussões científicas são secundárias às narrativas conspiratórias globais. Essa sobreposição de temas revela como a desinformação climática é instrumentalizada para desafiar iniciativas internacionais de sustentabilidade e promover a ideia de uma conspiração global.

## 4. Reflexões e trabalhos futuros

Para responder a pergunta de pesquisa "**como são caracterizadas e articuladas as comunidades de teorias da conspiração brasileiras sobre temáticas de mudanças climáticas e anticiência no Telegram?**", este estudo adotou técnicas espelhadas em uma série de sete publicações que buscam caracterizar e descrever o fenômeno das teorias da conspiração no Telegram, adotando o Brasil como estudo de caso. Após meses de investigação, foi possível extrair um total de 36 comunidades de teorias da conspiração brasileiras no Telegram sobre temáticas de mudanças climáticas e anticiência, estas somando 1.287.938 de conteúdos publicados entre maio de 2018 (primeiras publicações) até agosto de 2024 (realização deste estudo), com 78.252 usuários somados dentre as comunidades.

Foram adotadas quatro abordagens principais: **(i)** Rede, que envolveu a criação de um algoritmo para mapear as conexões entre as comunidades por meio de convites circulados entre grupos e canais; **(ii)** Séries temporais, que utilizou bibliotecas como "Pandas" (McKinney, 2010) e "Plotly" (Plotly Technologies Inc., 2015) para analisar a evolução das publicações e engajamentos ao longo do tempo; **(iii)** Análise de conteúdo, sendo aplicadas técnicas de análise textual para identificar padrões e frequências de palavras nas comunidades ao longo dos semestres; e **(iv)** Sobreposição de agenda temática, que utilizou o modelo BERTopic (Grootendorst, 2020) para agrupar e interpretar grandes volumes de textos, gerando tópicos coerentes a partir das publicações analisadas. A seguir, as principais reflexões são detalhadas, sendo seguidas por sugestões para trabalhos futuros.

### 4.1. Principais reflexões

**Negacionismo climático e anticiência formam uma interseção crítica na propagação da desinformação:** Comunidades de negacionismo climático e anticiência no Telegram brasileiro atuam em sinergia, criando uma rede complexa de desinformação. Essas comunidades interagem de maneira que as narrativas que negam as mudanças climáticas não apenas coexistem com discursos anticientíficos, mas se reforçam mutuamente. Esse fenômeno sugere que essas temáticas conspiratórias se aproveitam de uma desconfiança pré-existente nas instituições científicas, amplificando e espalhando informações falsas. Observa-se um efeito combinado dessas narrativas particularmente potente, pois atua em várias frentes ao mesmo tempo, dificultando o combate à desinformação e ampliando o alcance dessas crenças;

**Mudanças climáticas são alvo de narrativas apocalípticas:** Comunidades que discutem Apocalipse e Sobrevivência direcionaram 5.057 links para grupos de mudanças climáticas, demonstrando que essas teorias apocalípticas são uma porta de entrada significativa para o negacionismo climático. Essas narrativas apocalípticas frequentemente associam eventos climáticos extremos a previsões de catástrofes iminentes, que são então usadas para questionar a validade científica das mudanças climáticas. Ao promover essa visão, essas comunidades conseguem capturar a atenção de pessoas que já estão predispostas a acreditar em teorias apocalípticas, e redirecioná-las para uma rejeição mais ampla da ciência climática, exacerbando a polarização em torno deste tema;

**Comunidades de anticiência funcionam como distribuidores de desinformação para diversas teorias conspiratórias:** Os links que partem dessas comunidades são distribuídos de forma equilibrada para outros temas, como Nova Ordem Mundial e Globalismo, sugerindo que anticiência atua como um gatekeeper central. Essas comunidades não se limitam a uma única narrativa conspiratória, mas sim servem como pontos de convergência para várias teorias, onde os membros são expostos a uma ampla gama de desinformação. Isso cria um ambiente onde diferentes narrativas conspiratórias podem se reforçar mutuamente, criando uma teia complexa e coesa de desinformação que é difícil de desmantelar, pois atua em múltiplos níveis e temas simultaneamente;

**Crescimento significativo de discussões anticientíficas durante a Pandemia da COVID-19:** Houve um pico notável em 2021 nas discussões anticientíficas, impulsionado principalmente pela desinformação sobre vacinas e outras medidas de saúde pública. Durante esse período, as comunidades anticientíficas no Telegram se expandiram rapidamente, aproveitando-se do medo e da incerteza gerados pela Pandemia para disseminar informações falsas. Esse crescimento é indicativo de como momentos de crise global são explorados por essas comunidades para fortalecer e ampliar sua base de seguidores. A explosão dessas discussões revela não apenas o aumento na quantidade de conteúdo, mas também a intensificação das crenças anticientíficas, que passaram a se misturar com outras teorias conspiratórias, formando uma rede ainda mais robusta de desinformação;

**Comunidades negacionistas das mudanças climáticas atraem membros de outras teorias conspiratórias:** Grupos focados em globalismo e Nova Ordem Mundial enviaram 3.013 e 2.903 links, respectivamente, para comunidades de mudanças climáticas, indicando uma convergência entre essas narrativas conspiratórias. Isso sugere que, para os adeptos dessas teorias, as mudanças climáticas são vistas não apenas como um fenômeno natural, mas como parte de um plano maior de controle global. Essa interseção entre diferentes teorias permite que as comunidades negacionistas climáticas se expandam e integrem novas ideias, atraindo membros de outros grupos conspiratórios que compartilham uma visão cética ou hostil em relação às autoridades e às explicações científicas tradicionais. Assim, as mudanças climáticas se tornam mais um campo de batalha na luta contra uma suposta dominação global;

**A desinformação sobre mudanças climáticas está fortemente vinculada a disputas geopolíticas:** Termos como “Ucrânia” e “Rússia” são frequentemente associados a debates climáticos, sugerindo que a crise climática é enquadrada como parte de uma estratégia de manipulação geopolítica. Nesses debates, eventos climáticos extremos e políticas ambientais são retratados como ferramentas utilizadas por potências globais para influenciar e controlar outras nações. Essa narrativa é particularmente poderosa porque se aproveita de tensões geopolíticas reais, adicionando uma camada de desconfiança e conspiracionismo às discussões sobre mudanças climáticas. O resultado é uma visão distorcida das questões ambientais, onde cada evento climático é visto não como um sintoma de uma crise global, mas como um instrumento de poder usado por nações ou elites globais para avançar agendas;

**Narrativas anticientíficas integram crenças esotéricas e conspiratórias:** Comunidades que promovem esoterismo e ocultismo estão frequentemente ligadas às discussões anticientíficas, reforçando a ideia de uma "verdade alternativa" que desafia a ciência convencional. Essas comunidades oferecem um espaço onde as fronteiras entre ciência, espiritualidade e teoria da conspiração se tornam borradas. A ciência é frequentemente retratada como uma força opressora que esconde a “verdadeira” natureza da realidade, que é supostamente revelada por meio de conhecimentos esotéricos. Isso não apenas fortalece as crenças dentro dessas comunidades, mas também cria uma narrativa coesa que integra diferentes formas de desinformação, tornando-as mais atraentes e convincentes para aqueles que já estão predispostos a questionar as explicações científicas convencionais;

**Discussões sobre mudanças climáticas evoluem para temas relacionados à dominação global:** A partir de 2022, os discursos climáticos passaram a incluir mais referências a um suposto controle global, conectando-se com teorias da Nova Ordem Mundial e outras conspirações. Essa evolução reflete como as narrativas sobre mudanças climáticas são moldadas para se alinhar com uma visão mais ampla de conspirações globais, onde eventos climáticos são vistos como pretextos para implementar políticas autoritárias e centralizadas. Essa abordagem ressoa com aqueles que já estão desconfiados das elites globais e das organizações internacionais, reforçando a ideia de que as mudanças climáticas são uma invenção ou exagero criado para justificar uma perda de liberdade e soberania;

**Comunidades anticientíficas promovem desinformação sobre saúde pública com base em teorias conspiratórias:** A desinformação sobre vacinas é central nas discussões anticientíficas, sendo usada como pilar para desafiar as políticas de saúde pública e reforçar outras narrativas de conspiração. Nessas comunidades, as vacinas são frequentemente apresentadas não como uma medida de proteção, mas como um instrumento de controle ou dano intencional. Essa visão distorcida se alimenta de medos e incertezas já presentes na sociedade, ampliando a resistência às políticas de saúde pública e criando uma base para rejeitar outras iniciativas científicas. A partir dessas discussões, surgem narrativas que se entrelaçam com teorias de controle global, onde a ciência é vista como uma ferramenta nas mãos de elites para manipular e controlar a população;

**A Agenda 2030 da ONU é retratada como uma conspiração para controle populacional:** Comunidades negacionistas das mudanças climáticas e anticientíficas frequentemente associam as iniciativas da ONU a teorias conspiratórias que alegam um plano de dominação global. Dentro dessas narrativas, a Agenda 2030 e os Objetivos de Desenvolvimento Sustentável (ODS) são apresentados não como um esforço para mitigar as mudanças climáticas e promover o desenvolvimento sustentável, mas como uma fachada para uma agenda mais sinistra de controle populacional e perda de soberania nacional. Essa visão distorcida é alimentada por medos de um governo mundial e é amplamente difundida entre os membros dessas comunidades, fortalecendo a resistência contra as políticas internacionais e minando os esforços globais para enfrentar desafios ambientais e sociais.

### 4.2. Trabalhos futuros

Considerando as descobertas detalhadas neste estudo, várias direções de pesquisa emergem como promissoras para explorar mais profundamente as dinâmicas do negacionismo climático e do movimento anticiência. Uma área particularmente rica para investigação é a análise de como a convergência entre narrativas apocalípticas e o negacionismo climático contribui para a formação de comunidades altamente polarizadas e resistentes à mudança de opinião. Estudos futuros poderiam se concentrar em examinar a evolução dessas narrativas, especialmente em resposta a eventos climáticos extremos, e como essas comunidades utilizam crises globais para reforçar e legitimar suas crenças entre seus membros e além.

Outra direção relevante seria a exploração do papel das disputas geopolíticas na propagação de desinformação sobre mudanças climáticas. Futuras pesquisas poderiam investigar como essas conexões são estabelecidas, ampliadas, e como elas afetam a percepção pública sobre mudanças climáticas em diferentes contextos regionais e sociopolíticos. Além disso, uma análise mais aprofundada poderia examinar o impacto dessas narrativas geopolíticas na formulação e implementação de políticas ambientais.

Além disso, a interação entre esoterismo e anticiência merece uma atenção especial e detalhada. Estudos poderiam focar em como as comunidades que promovem crenças esotéricas e ocultistas contribuem de maneira significativa para a disseminação de desinformação sobre mudanças climáticas e sobre a ciência de modo geral. Compreender a psicologia subjacente dessas comunidades, as motivações que levam à adesão a essas crenças, e como elas se mantêm coesas e atraentes para novos membros, pode fornecer *insights* cruciais para o desenvolvimento de estratégias mais eficazes de combate à desinformação, especificamente adaptadas para esses ambientes.

Futuras pesquisas também poderiam explorar mais profundamente a evolução das discussões sobre mudanças climáticas que estão cada vez mais se alinhando com teorias de controle global. Essa tendência de conectar eventos climáticos a narrativas de dominação global e perda de liberdade individual pode ter implicações significativas não apenas para a formulação de políticas públicas, mas também para a comunicação científica em geral. Investigar como essas narrativas se desenvolvem, se consolidam e ganham força em diferentes comunidades digitais poderia oferecer *insights* valiosos para interromper a disseminação de desinformação antes que ela se torne enraizada e influente em larga escala.

Por fim, uma análise mais detalhada das narrativas que associam a Agenda 2030 da ONU a teorias conspiratórias seria fundamental para compreender como essas ideias são disseminadas e amplificadas nas comunidades de negacionismo climático e anticiência. Estudos poderiam explorar como essa narrativa específica é construída, propagada e reforçada dentro dessas comunidades, além de avaliar o impacto potencial dessas crenças na implementação de políticas globais de sustentabilidade. Identificar os principais pontos de entrada dessas narrativas, entender as motivações por trás de sua disseminação, e desenvolver abordagens inovadoras para neutralizá-las, pode ser crucial para proteger o progresso em questões climáticas e ambientais, bem como para assegurar a adesão a acordos e políticas internacionais essenciais para o futuro do planeta.

## 5. Referências

## 6. Biografia do autor

**Ergon Cugler de Moraes Silva** possui mestrado em Administração Pública e Governo (FGV), MBA pós-graduação em Ciência de Dados e Análise (USP) e bacharelado em Gestão de Políticas Públicas (USP). Ele está associado ao Núcleo de Estudos da Burocracia (NEB FGV), colabora com o Observatório Interdisciplinar de Políticas Públicas (OIPP USP), com o Grupo de Estudos em Tecnologia e Inovações na Gestão Pública (GETIP USP), com o Monitor de Debate Político no Meio Digital (Monitor USP) e com o Grupo de Trabalho sobre Estratégia, Dados e Soberania do Grupo de

[ versão português / english above ]

Estudo e Pesquisa sobre Segurança Internacional do Instituto de Relações Internacionais da Universidade de Brasília (GEPSI UnB). É também pesquisador no Instituto Brasileiro de Informação em Ciência e Tecnologia (IBICT), onde trabalha para o Governo Federal em estratégias contra a desinformação. Brasília, Distrito Federal, Brasil. Site: https://ergoncugler.com/.